\documentclass[12pt]{article}
\usepackage{epsf}
\usepackage{graphicx}
\usepackage{amsfonts}
\usepackage[mathscr]{eucal}
\usepackage{booktabs}

\def\href#1#2{#2}   
\newif\ifdraft
\draftfalse	  
\reversemarginpar   
\makeatletter
\let\mlabel=\label
\let\adkendequation=\endequation%
\def\endequation{\adkendequation\adklabel\global\@ignoretrue}
\let\adkendeqnarray=\endeqnarray%
\def\endeqnarray{\adkendeqnarray\adklabel\global\@ignoretrue}
\newbox\marglabbox
\def\adklabel{\ifvoid\marglabbox\else\marginpar{\unhbox\marglabbox}\fi}
\def\label#1{\ifdraft\ifmmode%
  \global\setbox\marglabbox=\hbox{\hfill\fbox{\tiny\verb*~#1~}}%
  \else\ifinner\else\marginpar{\hfill\fbox{\tiny\verb*~#1~}}%
  \fi\fi\fi \mlabel{#1}}
\makeatother
\ifdraft%
\fi
\def\bb{\mathbb}
\def\eusm{\mathscr}
\font\twelvefrak=eufm10 scaled 1200
\font\tenfrak=eufm10
  \newfam\frakfam
  
  \textfont\frakfam=\twelvefrak
  \scriptfont\frakfam=\tenfrak
  \scriptscriptfont\frakfam=\scriptfont\frakfam
\def\sqr#1#2{{\vcenter{\hrule height.#2pt
   \hbox{\vrule width.#2pt height#1pt \kern#1pt
      \vrule width.#2pt}
   \hrule height.#2pt}}}

\def\bsqr#1#2{{\vrule width #1pt height#2pt}}
\def\bsquare{{\mathchoice\bsqr66\bsqr66\bsqr33\bsqr33}}
\def\badbreak{\penalty1000}

\def\sgn{\mathop{\rm sgn}}                  
\def\rational#1#2{{\mathchoice{\textstyle{#1\over#2}}%
  {\scriptstyle{#1\over#2}}{\scriptscriptstyle{#1\over#2}}{#1/#2}}}
\def\half{\rational12}			    
\def\quarter{\rational14}		    
\def\R{{\bb R}}				    
\newcommand{\gfive}{\gamma_{5}}             
\newcommand{\cP}{{\cal P}}                  
\def\rpc{x}                                 
\def\Xg{X}                                  
\def\Fg{{\eusm X}}                          
\def\Fset{{\bb X}}                          
\def\xd{P}                                  
\def\plp{{\Gamma}}                          
\def\cop{{C}}                               
\def\pr{{\cP}}                              
\def\df{{\cP}_f}                            
\def\db{{\cP}_b}                            
\def\dr{{\cP}_r}                            
\def\dop{{\xd}_r}                           
\def\da{{\plp}_r}                           
\def\Fgr{{\Fg}_r}                           
\def\dopset{{\bb P}}                        

\begin{document}


\begin{center}
{\Large{\bf The Analysis of Space-Time Structure in QCD}} \\
\vspace*{.16in}
{\Large{\bf Vacuum II: Dynamics of Polarization and}} \\
\vspace*{.16in}
{\Large{\bf Absolute X--Distribution}} \\
\vspace*{.25in}
{\large{Andrei Alexandru$^1$, Terrence Draper$^2$, Ivan Horv\'ath$^2$ and Thomas Streuer$^3$}}\\
\vspace*{.24in}
$^1$The George Washington University, Washington, DC, USA\\
$^2$University of Kentucky, Lexington, KY, USA\\
$^3$University of Regensburg, Regensburg, Germany


\end{center}

\vspace*{0.10in}

\begin{abstract}

  \noindent
  We propose a framework for quantitative evaluation of dynamical tendency 
  for polarization in arbitrary random variable that can be decomposed 
  into a pair of orthogonal subspaces. The method uses measures based on comparisons 
  of given dynamics to its counterpart with statistically independent components. 
  The formalism of previously considered $\Xg$--distributions is used to express 
  the aforementioned comparisons, in effect putting the former approach on 
  solid footing. Our analysis leads to definition of a suitable correlation 
  coefficient with clear statistical meaning. We apply the method to the dynamics 
  induced by pure--glue lattice QCD in local left--right components of overlap Dirac 
  eigenmodes. 
  It is found that, in finite physical volume, there exists a non--zero physical scale 
  in the spectrum of eigenvalues such that eigenmodes at smaller (fixed) eigenvalues 
  exhibit convex $\Xg$--distribution (positive correlation),   
  while at larger eigenvalues the distribution is concave (negative correlation). 
  This {\em chiral polarization scale} 
  thus separates a regime where dynamics enhances chirality relative to statistical 
  independence from a regime where it suppresses it, and gives an objective definition
  to the notion of ``low'' and ``high'' Dirac eigenmode. We propose to investigate 
  whether the polarization scale remains non--zero in the infinite volume limit, 
  in which case it would represent a new kind of low energy scale in QCD.
  
\end{abstract}

\vspace*{0.10in}


\section{Introduction}

In physics it is sometimes useful to decompose a multi--component variable/quantity 
with respect to a pair of orthogonal subspaces spanning its range. Examples relevant for 
the area of application that we target here, namely QCD and its vacuum structure, include 
the decomposition of a bispinor into left--right chirality components, and 
the decomposition of the gauge field strength tensor into self--dual and 
anti--self--dual parts. Physical reasons 
for considering such decompositions usually have to do with the fact that the dynamics of 
the variable involves (or is expected to involve) a specific type of relation among such 
suitably defined projections. The expected correlations can then be tested in a relevant 
physical experiment or evaluated in theoretical calculations. 

The property of polarization, namely the tendency for asymmetric participation of the two 
subspaces in the preferred values of the variable, is frequently of interest in the above 
context. While the qualitative meaning of the concept is obvious, the ways to characterize 
and quantify the degree of polarization in the dynamics are highly non--unique {\em a priori}. 
One typically operates under the implicit assumption that the physical context 
uniquely ``selects the tools'', i.e.\ that the choice of the polarization characteristics
is driven by the physical nature of the quantity at hand. There are situations however,
the study of QCD vacuum structure being one of them, where we may be interested 
in polarization characteristics that pertain to dynamics alone and are not conditioned 
on which specific physical situation the dynamics is modelling.
Our goal in this paper is to develop a framework to describe and quantify such 
{\em dynamical tendency for polarization}. In addition to being free of kinematic effects,
we also require that the framework provides a possibility for high level of detail. 
In other words, we are not only interested in a suitable average polarization parameter,
but rather want the aspects of polarization to be separated from other content of 
the dynamics.


To construct our formalism, we will build upon the $\Xg$--distribution proposal of 
Ref.~\cite{Hor01A} which grew from similar motivations albeit with a specific
aim to characterize the local chirality of QCD Dirac eigenmodes. 
The $\Xg$--distribution approach is based on the following logic. 
Consider the left--right decomposition $\psi = \psi_L + \psi_R$ of the local value 
in the eigenmode, and some function $\Fg\equiv\Fg(\psi_L,\psi_R)$ that reflects 
the degree of chiral polarization in $\psi$. Since $\Fg$ is defined at every space--time point, 
it is a local characteristic providing the information on changes in the left--right 
polarization within the mode. The equilibrium ensemble of QCD gauge configurations induces 
the distribution of associated $\Fg$--values for chosen sets of eigenmodes.\footnote{In the 
early works, additional constraints to define sub-ensembles of $\Fg$--values, such as 
restrictions to space--time regions with strong fields, were frequently used.} 
In other words, denoting by $\Xg$ an independent variable whose domain is in the range 
of $\Fg$, there exists a function $\xd(\Xg)$, namely the probability distribution of 
the polarization characteristic specified by $\Fg$. The $\Xg$--distribution $\xd(\Xg)$ 
provides rather detailed information about the chiral polarization properties of 
the selected group of modes. 

While the concept of $\Xg$--distributions is suitable for our purposes, the issue of 
``dynamical versus kinematical'' has not been explicitly addressed in that framework. 
The channel through which kinematics enters here is the function $\Fg$ since
there is a large freedom in choosing measures that qualitatively reflect chiral polarization. 
Rather than analyzing this freedom, the related work was carried out with the implicit
rationale that the proper fixed choice of this {\em polarization function} can be made.
For example, the original definition in Ref.~\cite{Hor01A} involved $\Fg$ linear in 
the polar angle of point $(|\psi_L|, |\psi_R|)$ from the ``left--right plane'' 
(chiral orientation parameter). This was used in most of the early follow--up 
works~\cite{Followup} with some departures~\cite{Blu01A,Gat02A}. However, especially in 
the case of Dirac eigenmodes, it 
is not entirely clear which physical consideration should fix the selection. Thus,
focusing exclusively on the dynamics, in the sense discussed above, provides a meaningful 
direction to proceed in.

With the above discussion in mind, we wish to propose a construct that (1) provides a detailed
differential information on polarization modeled after the $\Xg$--distribution but (2)
doesn't rely on a fixed polarization function $\Fg$, and (3) reflects only the dynamics.
Note that while the involvement of kinematics proceeds via the polarization function $\Fg$, 
requirement (2) only represents a necessary condition for this characteristic to be considered 
purely dynamical.
Indeed, we view the concept of ``dynamical'' in this context as entailing a specific
meaning of ``correlational'' or, in more general terms, based on some form of comparison 
to statistical independence. This is why the requirements (2) and (3) are listed
separately and need to be fulfilled simultaneously.

To satisfy these requirements, we first reinterpret the $\Xg$--distribution as 
a relative measure.
In other words, we show that one can view $\xd(\Xg)$ not only as an object assigned
to given dynamics and the polarization function, but also as an object comparing this dynamics
to another dynamics associated with the polarization function in question.
We develop a formalism that allows us to switch freely between the two interpretations which,
in technical terms, allows us to construct an $\Xg$--distribution for arbitrary pair 
of dynamics. Equipped with this tool, we then proceed to define a truly dynamical 
characteristic as a relative $\Xg$--distribution assigned to dynamics of interest 
and the related one where left and right components are chosen independently of one another 
from corresponding marginal distributions. Note that the latter is the needed point 
of comparison representing statistical independence. The crucial point, implicitly
embedded in this procedure, is that it is independent of the polarization function used 
to carry it out. In other words, while the implementation must proceed via some 
polarization function $\Fg$, the final result doesn't depend on this choice. 
We thus refer to the above dynamical characteristic as an {\em absolute $\Xg$--distribution} 
in order to distinguish it from the original one that can only be 
viewed in a relative sense.\footnote{One can also refer to this construct as 
the {\em dynamical $\Xg$--distribution}, explicitly emphasizing the fact that its purpose is 
to remove any kinematical content from the description of polarization.}

In terms of requirement (3), one can draw a parallel between our construction and 
the definition of the Pearson correlation coefficient for random variables $q_1$, $q_2$ 
governed by a joint probability distribution $\cP(q_1,q_2)$, namely 
$r \equiv (\langle q_1 q_2\rangle - \langle q_1 \rangle \langle q_2 \rangle)/
(\sigma_1 \sigma_2)$, with $\sigma_1$, $\sigma_2$ being the corresponding standard 
deviations. Computing just $\langle q_1 q_2\rangle$ will not convey any dynamical
information (it won't tell us anything about tendencies of the relationship between 
$q_1$ and $q_2$), nor does this value have any specific statistical meaning. 
However, after the subtraction of ``uncorrelated relationship'' 
$\langle q_1 \rangle \langle q_2 \rangle$, the resulting covariance acquires
a qualitative dynamical content. Finally, the normalization via standard deviations, 
ensuring that $-1\le r \le 1$, gives the coefficient a well-defined statistical meaning, 
and thus a quantitative dynamical significance. In effect, the dynamical construct
that we propose implements the ``subtraction of statistical independence'' at the level 
of distributions which, among other things, has an added bonus that proper normalizations 
are automatically in place.

With absolute $\Xg$--distribution at hand, one can naturally define an averaged quantity 
based on it, and this role is played by the {\em correlation coefficient of polarization} 
(CCP). CCP has a clear statistical meaning in that it is linearly related to the probability 
that the sample chosen from the distribution governing the dynamics is more polarized than 
the sample chosen from its counterpart with statistically independent components. 
One utility of CCP is that it naturally ranks (orders) possible dynamics with respect 
to the polarization. In particular, larger CCP means larger dynamical tendency 
for polarization. Moreover, the positive CCP signals that the dynamics enhances the polarization 
relative to statistical independence, while the negative value means that polarization is being 
dynamically suppressed. Due to its absolute nature, we propose CCP as a basic 
``figure of merit'' quantifying polarization properties inherent to dynamics.

As a first application of the above techniques, we analyze local chiral properties of 
low--lying QCD Dirac eigenmodes. It was already the motivation of Ref.~\cite{Hor01A} that local 
properties of these modes are expected to be connected to low--energy features of QCD, 
especially to those related to spontaneous chiral symmetry breaking. However, our goals in this 
investigation are quite different from those of Ref.~\cite{Hor01A}. Rather than checking 
the consistency of assumptions for certain models, as we did in Ref.~\cite{Dra04A} which 
contains the preliminary discussion of techniques developed here, our aim is to build 
a model--independent information on relevant properties of these modes in line with 
the {\em bottom--up} approach to QCD vacuum structure as formulated in Ref.~\cite{Hor06B}. 
Indeed, it is the purpose of this series~\cite{Hor04A} to contribute to developing the set 
of techniques that can facilitate these goals.

The summary of our main initial findings on the chiral polarization properties of 
low--lying overlap Dirac eigenmodes are as follows. The absolute $\Xg$--distributions 
of the lowest modes are convex on their domain, while the $\Xg$--distributions of higher 
modes are concave. This implies that the lowest modes exhibit a dynamical tendency for chirality 
(positive CCP) while the higher modes dynamically suppress it (negative CCP). 
At fixed volume and lattice spacing, there exists a transition point $\Lambda_T(V,a)$ 
in the spectrum where the absolute $\Xg$--distribution is strictly flat, and the polarization
properties are indistinguishable from statistical independence. Based on four lattice spacings,
our data supports the proposition that this scale remains finite in the continuum limit,
i.e.\ $\lim_{a \to 0} \Lambda_T(V,a) = \Lambda_T(V) > 0$ for $V<\infty$. The appearance
of this sharply--defined new feature in the QCD Dirac spectrum could be relevant 
for understanding the mechanism of spontaneous chiral symmetry breaking.

The structure of the paper is as follows. We start by discussing an explicit example of 
the absolute $\Xg$--distribution calculation in Sec.~\ref{sec:primer}. Our goal here
is to provide a concise introduction to the method with sufficient practical and conceptual
information for it to be directly applied to the problem of interest here, namely 
the study of local chiral properties in overlap Dirac eigenmodes. This application and 
corresponding results 
are then described in Sec.~\ref{sec:qcd}. In Appendix~\ref{sec:general} we describe 
in detail the theoretical structure, the meaning and the scope of absolute polarization
methods proposed here. This is a rather extensive topic of its own, and the section 
can be read independently of the QCD content. Indeed, this discussion is carried out 
in a completely generic setting, emphasizing the universal applicability of the formalism 
to any area where the concept of polarization plays relevant role.

\section{Absolute $\Xg$--Distribution -- a Primer}
\label{sec:primer}

\begin{figure}[t]
   \centering
   \includegraphics[width=15truecm,angle=0]{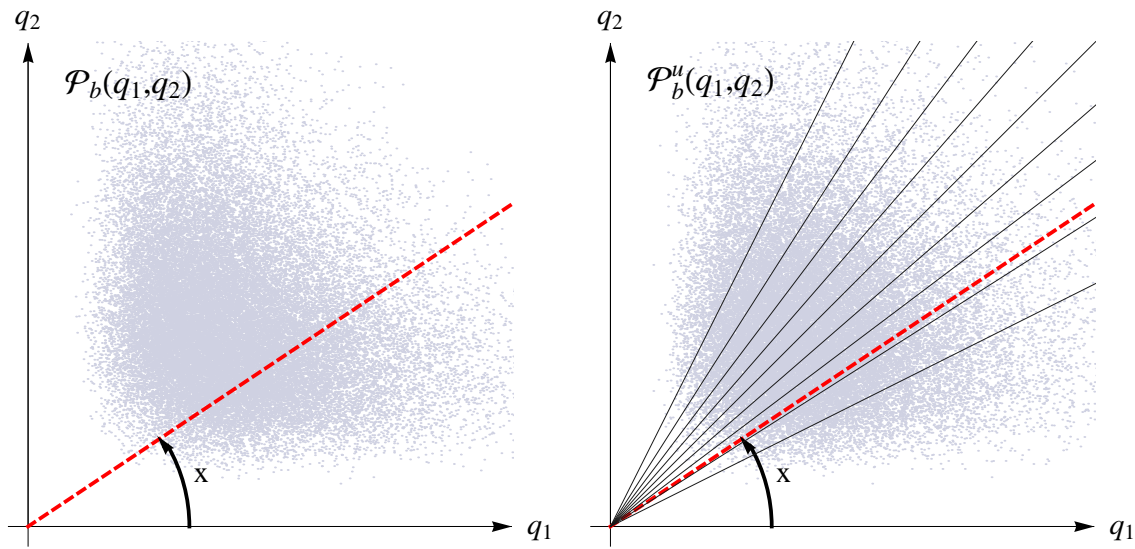}
   \caption{Left: Dynamics governing the polarization properties of two lowest modes 
            for ensemble $E_{1}$.
            Right: The associated ``uncorrelated'' dynamics; solid gray lines separate 
            the quadrant into 10 sectors, each containing 10\% of the population.}
   \label{fig:dyn_E1}
\end{figure}

Given that our first application of absolute polarization methods targets chiral properties 
of Dirac eigenmodes, we will use the specifics of that setting to introduce the approach.
For definiteness, consider the set of lattice Dirac eigenmodes comprised of lowest 
two (non--zero) conjugate pairs on gauge backgrounds from ensemble $E_1$ of 
Table~\ref{tab:ensembles}.\footnote{Details concerning the parameters of lattice gauge 
ensembles and lattice Dirac operator are given in Sec.~\ref{sec:qcd} and are not important
for this discussion.} 
Collecting all local values $\psi$ from these modes into single pool, let us
consider the associated collection of two--component objects $(q_1,q_2)$, namely 
the magnitudes of left and right components of $\psi$
\begin{equation}
   (q_1, q_2) \,\equiv\, (\,|\psi_L|,|\psi_R|\,)  \qquad\quad
    \psi_L \equiv \half \, ( 1-\gfive )\, \psi   \qquad
    \psi_R \equiv \half \, ( 1+\gfive )\, \psi   
    \label{eq:n10}  
\end{equation}
The distribution of these ``samples'' is shown on the left side of 
Fig.~\ref{fig:dyn_E1}, and statistically approximates the underlying {\em base} probability 
distribution $\db(q_1,q_2)$ describing the limit of infinite gauge statistics.

One can assess a degree of polarization in a given sample by its polar angle 
$\varphi \in [0,\pi/2]$ which is convenient to symmetrize into
{\em reference polarization coordinate}, namely~\cite{Hor01A}
\begin{equation}
    \rpc \equiv \frac{4}{\pi} \, \varphi \,-\,1    \qquad \quad
    \rpc \in [-1,1]
    \label{eq:n20}
\end{equation}
Note that strictly left--polarized and strictly right--polarized samples are 
characterized by values $\rpc=-1$ and $\rpc=+1$ respectively. 
The probability distribution $\dop(\rpc)$ of polarization coordinate over 
the population of samples is sometimes invoked as a detailed measure 
characterizing chiral polarization properties of Dirac modes in question, 
and is usually referred to as the (reference) $\Xg$--distribution~\cite{Hor01A}. 
Our goal is to define a construct of similar nature which however focuses on 
the relation of the dynamics in question to the ``uncorrelated'' one 
where left and right components of $\psi=(\psi_L,\psi_R)$ are paired with one 
another at random from available pool of possibilities. Indeed, such comparative
polarization measure could then be viewed as {\em dynamical}. Note that 
the needed point of comparison (the population of statistically independent
components) associated with our exemplary dynamics is shown on the 
right side of Fig.~\ref{fig:dyn_E1}. It approximates the underlying 
distribution $\db^u(q_1,q_2)\equiv p(q_1) p(q_2)$ with $p(q)$
being the distribution of a single component in original $\db(q_1,q_2)$. 
This uncorrelated dynamics carries its own reference 
$\Xg$--distribution $\dop^u(\rpc)$. 

\begin{figure}[t]
   \centering
   \includegraphics[width=15truecm,angle=0]{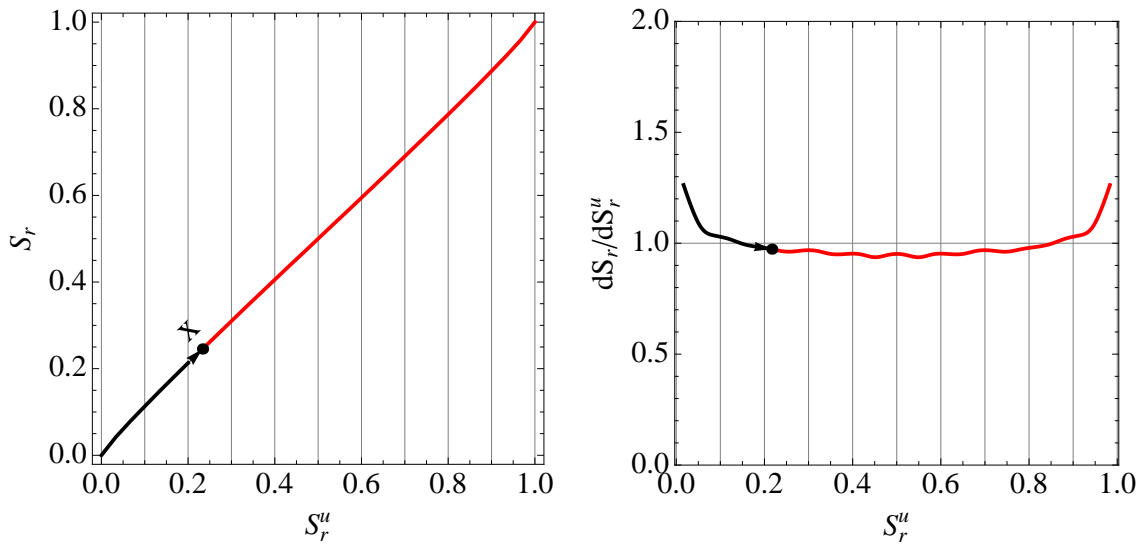} 
   \vskip -0.5 truecm
   \caption{Steps in the construction of absolute $\Xg$--distribution for data shown in 
            Fig.~\ref{fig:dyn_E1}. See the discussion in the text.}
   \label{fig:sr_vs_sru}
\end{figure}

To obtain the polarization measure of the above kind, 
we perform a differential comparison of polarization in 
$\db(q_1,q_2)$ relative to $\db^u(q_1,q_2)$ as follows. A ray passing through
the origin can be specified by its reference polarization coordinate $\rpc$, 
as shown in Fig.~\ref{fig:dyn_E1}. Determine the fraction of population contained 
between the $q_{1}$-axis and this ray, separately for $\db$ and $\db^{u}$. These 
fractions represent the cumulative probability functions $S_{r}(\rpc)$ and 
$S_{r}^{u}(\rpc)$ associated with $\dop(\rpc)$ and $\dop^u(\rpc)$ respectively.
Eliminating $\rpc$, every line is represented by a point in the cumulative 
distribution plane $(S_r^u, S_r)$ with the set of all such points forming a graph 
starting at $(0,0)$ and ending at $(1,1)$. If the the two situations (``dynamics'')
in question had identical polarizations, the graph would be a straight line. 
On the left side of Fig.~\ref{fig:sr_vs_sru} we plot the result of this construction 
for data displayed in Fig.~\ref{fig:dyn_E1}. One can see that the correlated and 
uncorrelated dynamics have very similar polarizations since the graph is close 
to being linear. 
To obtain a differential comparison, and to better see the differences, we compute 
the slope of the cumulative polarization graph and show it on the right side of 
Fig.~\ref{fig:sr_vs_sru}. Since this dependence represents a probability distribution 
in variable $S_{r}^{u}$, we can see that the correlated case exhibits a small 
excess of probability with respect to the uncorrelated one (the horizontal line) near 
the extremal values. Noting that the extremal values $S_r^u=0,1$ actually correspond 
to strictly polarized cases $\rpc=-1,1$ respectively, the observed qualitative behavior 
of the slope graph 
in fact conveys the excess of polarization in dynamics described by $\db(q_1,q_2)$ 
relative to dynamics described by $\db^u(q_1,q_2)$. To standardize this polarization
characteristic, we finally convert the distribution in $S_r^u$ into distribution 
in rescaled polarization--like variable $S_r^u \rightarrow 2 S_{r}^{u}-1 \equiv \Xg$ 
whose domain is $[-1,1]$. The resulting construct $P_A(X)$ will be referred to as the 
{\em absolute $\Xg$--distribution}.  
\smallskip

\noindent We now wish to make several remarks regarding the above method.
\smallskip

\begin{figure}[t]
\begin{center}
    \centerline{
    \hskip 0.10in
    \includegraphics[width=8.8truecm,angle=0]{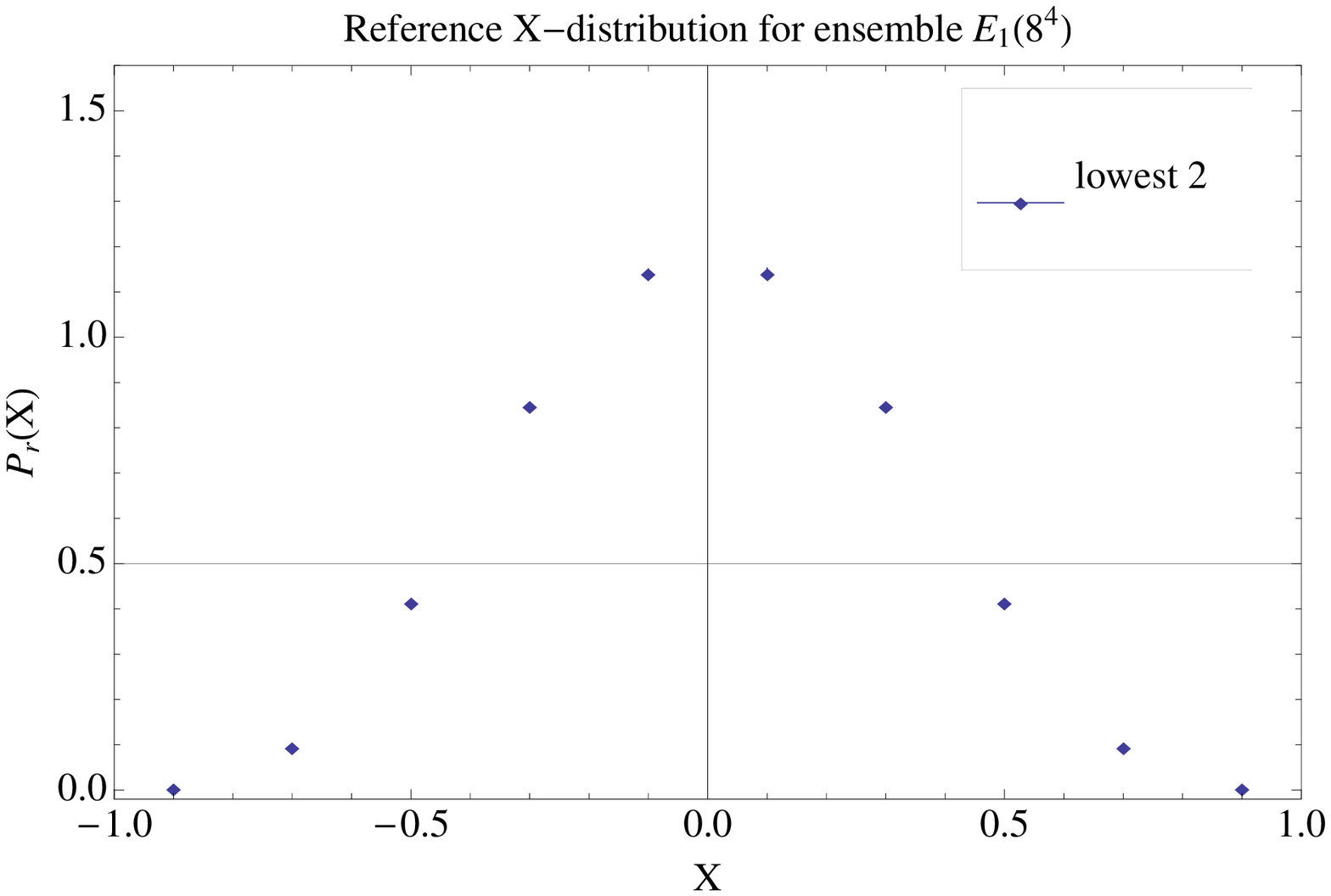}
    \hskip -0.28in
    \includegraphics[width=8.8truecm,angle=0]{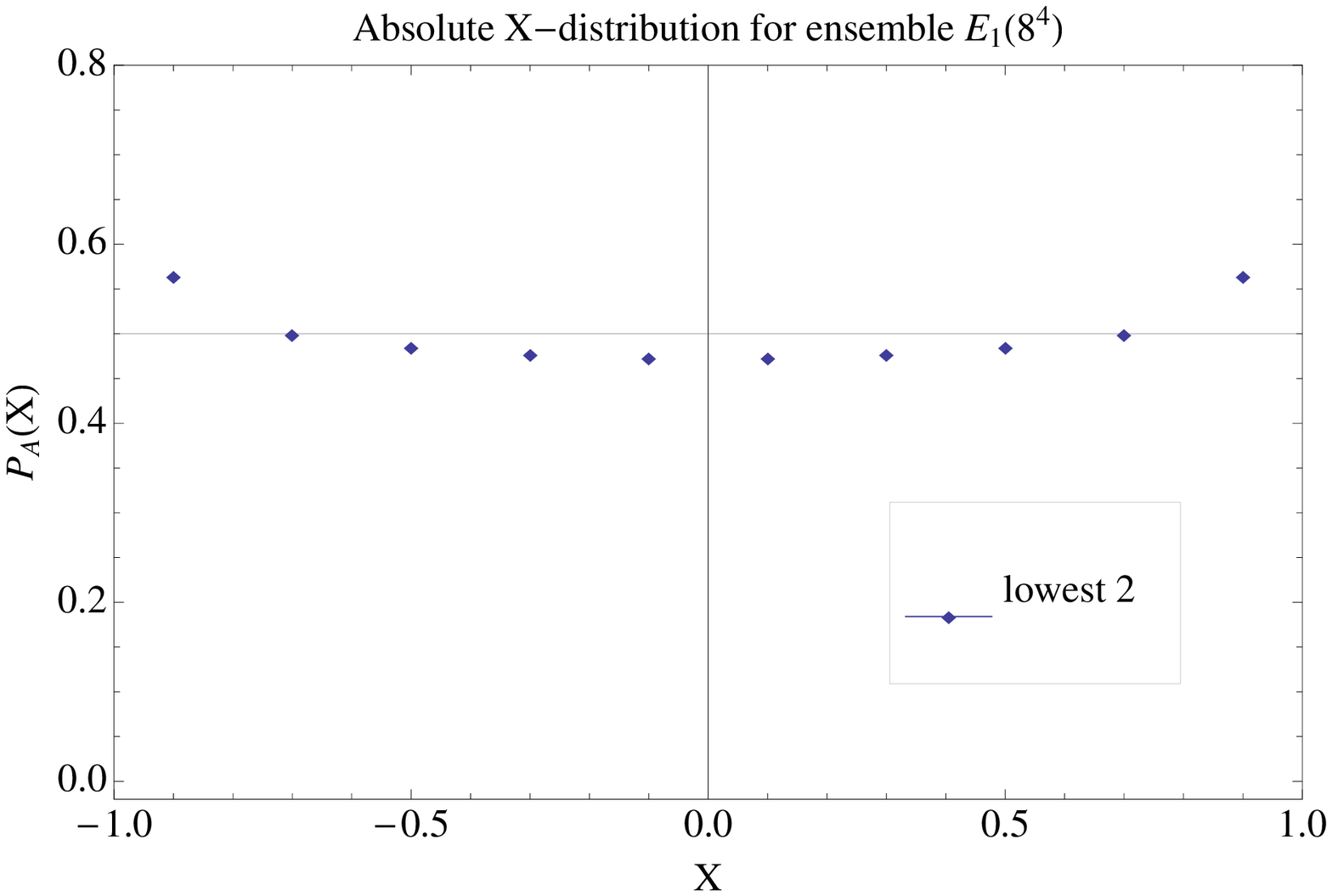}     
     }
     \vskip -0.00in
     \caption{Reference $\Xg$--distribution (left) and absolute 
     $\Xg$--distribution (right) for data shown in Fig.~\ref{fig:dyn_E1} 
     with resolution of 10 bins.}
     \vskip -0.3in 
     \label{fig:E1_10bins}
\end{center}
\end{figure}

\noindent {\em (i)}
Performing a practical calculation of absolute $\Xg$--distribution with uniform 
resolution in $\Xg$ requires that we evaluate cumulative probabilities at values 
$x_i$ that split the uncorrelated population into $N_b$ bins of equal size. This 
is indicated on the right hand side of Fig.~\ref{fig:dyn_E1} for $N_b=10$ by the rays 
enclosing 10\% of population each, and similarly by vertical lines 
in Fig.~\ref{fig:sr_vs_sru}. The corresponding reference distribution 
$\dop(\Xg \equiv \rpc)$ and absolute $\Xg$--distribution $P_A(\Xg)$ computed with 
this resolution are shown in Fig.~\ref{fig:E1_10bins}.
\smallskip

\noindent {\em (ii)}
As one can readily inspect, the above geometric 
construction would lead to the same result had we decided to quantify the degree 
of chiral polarization in $\psi$ by 
some arbitrary {\em polarization function} $\Fg(\rpc)$ rather than by the default
choice $\Fg(\rpc)=\rpc$. Indeed, such different possibilities correspond to different 
parametric representations of the same curve in the $(S_{r}^{u},S_{r})$ plane, and
this information drops out of the calculation. The associated 
{\em reparametrization invariance} is significant since it expresses the fact that 
the proposed description of polarization is independent of the ``language'' 
(or ``reference frame'') chosen to represent the concept at the sample level. 
This is necessary for the description to be considered purely dynamical and 
reflects the absolute nature of this construction. 
\smallskip

\noindent {\em (iii)}
The definition of absolute $\Xg$--distribution given above is equivalent 
to constructing the $\Xg$--distribution of Ref.~\cite{Hor01A}, but associated 
with polarization function
\begin{equation}
   \Fg_A(\rpc) \equiv 2 \int_{-1}^\rpc d y \, \dop^u(y) - 1
   \label{eq:n25}   
\end{equation}
This makes the fact that the absolute $\Xg$--distribution represents a valid 
polarization measure more explicit. In fact, depending on the dynamics 
$\db(q_1,q_2)$ in question, the procedure adjusts the polarization characteristic 
so that it directly measures deviations relative to uncorrelated dynamics. 
In this language, the reparametrization invariance of absolute $\Xg$--distribution 
follows from the fact that $\Fg_A$ is invariant under the choice of reference 
polarization coordinate as one can directly verify.
\smallskip

\noindent {\em (iv)}
We wish to emphasize that the strong dependence on the choice of the polarization
function (dependence on ``kinematics'') provided a major motivation for extending
the old $\Xg$--distribution approach in the direction described here. Indeed,
as discussed extensively in Appendix~\ref{sec:general}, even qualitative 
conclusions can be entirely misleading if one relies only on the $\Xg$--distribution 
associated with fixed polarization function. This is illustrated in 
Fig.~\ref{fig:arbitrary} where a suitable choice of polarization function can either
produce seemingly clear chiral double--peaking or what appears as completely 
non--chiral single peak.
\smallskip
 
\noindent{\em (v)}
In addition to detailed description of polarization contained in absolute 
$\Xg$--distribution, we also define a single figure of merit reflecting the overall 
dynamical tendency for polarization, namely
\begin{equation}
   C_A \,\equiv \,
    2\,\int_{-1}^{1} d \Xg \, |\Xg| \, P_A(\Xg) \,-\,1
    \label{eq:n30}
\end{equation}
The positive value of this {\em correlation coefficient of polarization} 
$C_A \in [-1,1]$ indicates that the dynamics enhances polarization relative 
to randomness, while the negative value implies its dynamical suppression.
Indeed, as one can check (see Appendix~\ref{sec:general}), $C_A$ is linearly
related to the probability that the sample chosen from $\db(q_1,q_2)$
is more polarized than the sample (independently) chosen from 
$\db^u(q_1,q_2)$, leading to the above interpretation.

The elements described in this section are at the core of the general dynamical
polarization method we are proposing in this work. Formal definitions and detailed 
discussion of the logical structure involved are given in Appendix~\ref{sec:general}.
While the concise overview given in this section is sufficient to apply the method
to the case of QCD Dirac eigenmodes in what follows, we emphasize that the content
of Appendix~\ref{sec:general} is important not only for deeper understanding 
the scope and the meaning of the approach, but also for its practical application 
in case of singular polarization dynamics.

\section{QCD Dirac Eigenmodes}
\label{sec:qcd}

We will now apply the absolute polarization methods to the case of low--lying QCD Dirac 
eigenmodes in a more systematic manner. As argued extensively e.g.~in Ref.~\cite{Hor01A}, 
the properties of these modes are expected to encode many important features of QCD vacuum. 
Their direct connection to spontaneous chiral symmetry breaking, 
beyond the Banks--Casher relation~\cite{Ban80A}, is particularly anticipated.

\newcommand\fm{\mathop{\rm fm}}
\begin{table}[b]
   \centering
   \begin{tabular}{@{} ccccccccc @{}} 
      \toprule
      Ensemble    & Size & $N_{\rm config}$  & Volume & Lattice Spacing & 
      $\Lambda_{\rm LOW}^{\rm MAX}$ & $\Lambda_{\rm LOW}^{\rm AVE}$ & 
      $\Lambda_{\rm HIGH}^{\rm MIN}$ & $\Lambda_{\rm HIGH}^{\rm AVE}$\\
      \midrule
     $E_{1}$ &  $8^{4}$ & 100 & $(1.32\fm)^{4}$ & $0.165 \fm$ &
     449 & 226 & 1956 & 1980\\
     $E_{2}$ & $12^{4}$ & 97 & $(1.32\fm)^{4}$ & $0.110 \fm$ &
     407 & 169 & 1711 & 1735\\
     $E_{3}$ & $16^{4}$ & 99 & $(1.32\fm)^{4}$ & $0.0825 \fm$ &
     304 & 142 & 1513 & 1553\\
     $E_{4}$ & $24^{4}$ & 96 & $(1.32\fm)^{4}$ & $0.055 \fm$ &
     344 & 136 & 1338 & 1366\\
     $E_{5}$ & $16^{4}$ & 99 & $(1.76\fm)^{4}$ & $0.110 \fm$ &
     162 & 58 & 1087 & 1123\\
      \bottomrule
   \end{tabular}
   \caption{The summary of five ensembles used in overlap eigenmode calculations. The right 
    side
    of the table describes some properties of the spectra (in MeV) with 
    $\Lambda_{\rm LOW}^{\rm AVE}$ denoting the average magnitude of lowest near--zero 
    eigenvalue over the ensemble and $\Lambda_{\rm HIGH}^{\rm AVE}$ denoting the same
    for highest eigenvalue. $\Lambda_{\rm LOW}^{\rm MAX}$ is the magnitude of the maximal
    lowest eigenvalue, and $\Lambda_{\rm HIGH}^{\rm MIN}$ the magnitude of the minimal
    highest eigenvalue.}
   \label{tab:ensembles}
\end{table}

We perform our exploratory calculations in the context of pure glue lattice QCD with 
Iwasaki gauge action~\cite{Iwasaki}. The left side of Table~\ref{tab:ensembles} 
summarizes the parameters of our five ensembles. Lattice scale has been determined from 
the string tension following the methods and results of Ref.~\cite{Oka99A}. 
Ensembles $E_1$--$E_4$ have fixed physical volume, and are intended for investigation
of the continuum limit. Ensemble $E_5$ has larger physical volume and serves mainly
to check whether our qualitative conclusions remain stable in that regard.

Low--lying eigenmodes of the overlap Dirac operator~\cite{Neu98BA} were calculated
on gauge backgrounds from the above ensembles. Specifically, we used the massless
overlap operator with Wilson kernel, the negative mass parameter $\rho=26/19$ 
($\kappa=0.19$), and Wilson parameter $r=1$. The eigenmodes and eigenvalues were 
computed using our own implementation of the implicitly restarted Arnoldi algorithm 
with deflation~\cite{ira}. The zero modes and about 50 pairs of lowest near--zero 
modes \footnote{The exact number of computed near--zero mode pairs depends on 
the number of zero modes (topological charge) of a given configuration. 
If $N_0$ is the number of zero modes then there were $55-N_0$ pairs computed.}
with complex--conjugated eigenvalues were obtained for each configuration. 
The right side of Table~\ref{tab:ensembles} compiles the information on spectral 
boundaries of computed regions for each ensemble (see the caption). Note that since 
zero modes are exactly chiral, it is the near--zero modes that are of interest 
here. They are in fact the relevant eigenmodes to consider in the context of 
spontaneous chiral symmetry breaking.

\subsection{Reference and Absolute $\Xg$--Distributions}

In order to get the basic idea about the dynamical tendency for chirality in 
the obtained eigenmodes, we first focus on the lowest non--zero and highest computed 
modes. For this purpose, we collect into the ``lowest'' eigenmode group the two
lowest near--zero modes for each configuration from the given gauge ensemble. Since 
the eigenmodes of conjugate pairs have identical chiral properties, we actually
include one representative from each of the two lowest conjugate pairs. Similarly,
the ``highest'' eigenmode group contains eigenmodes from the highest two pairs. 
The collection of all local values of $(q_1,q_2)$ (magnitudes of left and right 
components) in the grouped eigenmodes represent the dynamics $\db(q_1,q_2)$ that we 
are investigating.

\begin{figure}
\begin{center}
    \centerline{
    \hskip 0.08in
    \includegraphics[width=8.4truecm,angle=0]{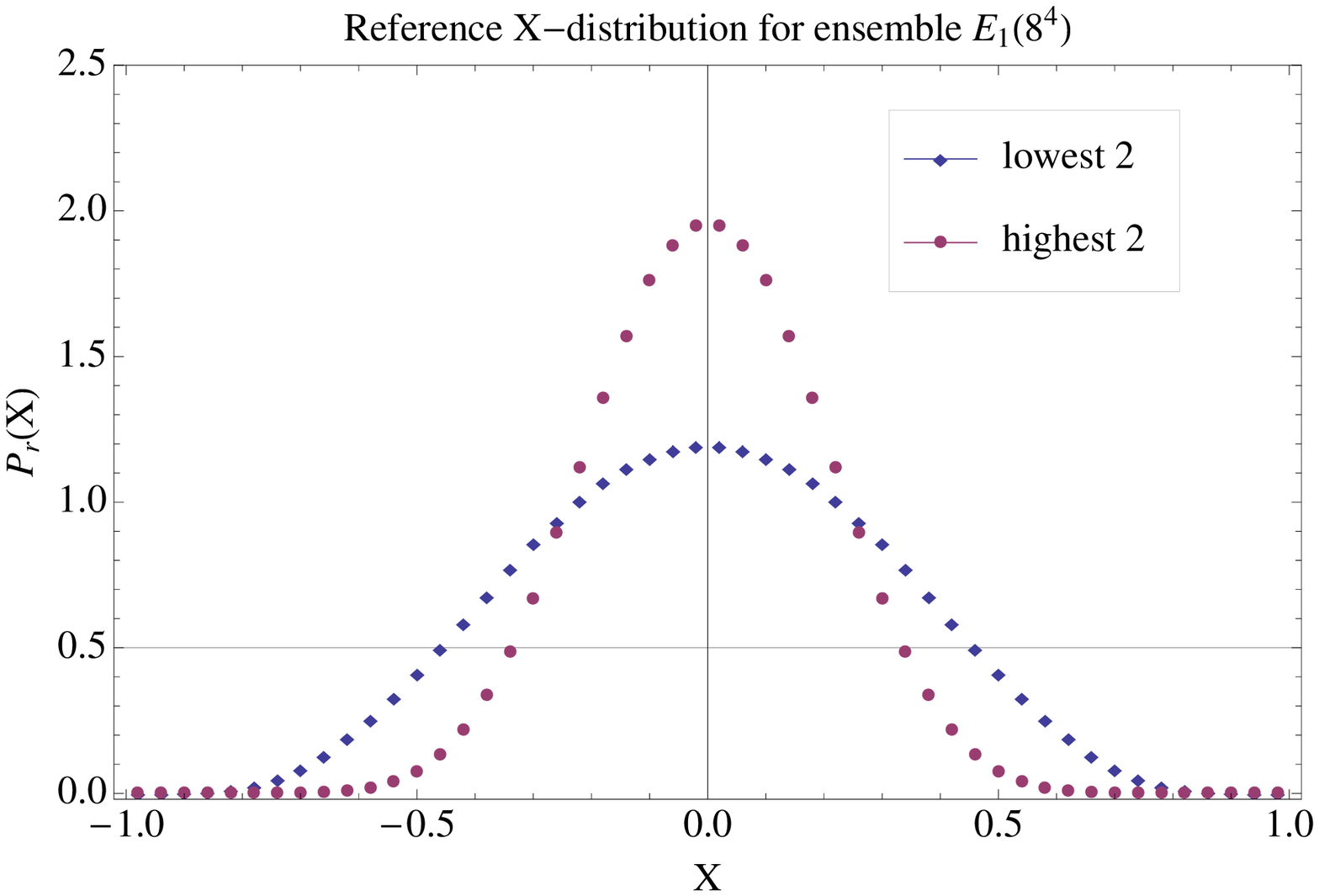}
    \hskip -0.1in
    \includegraphics[width=8.4truecm,angle=0]{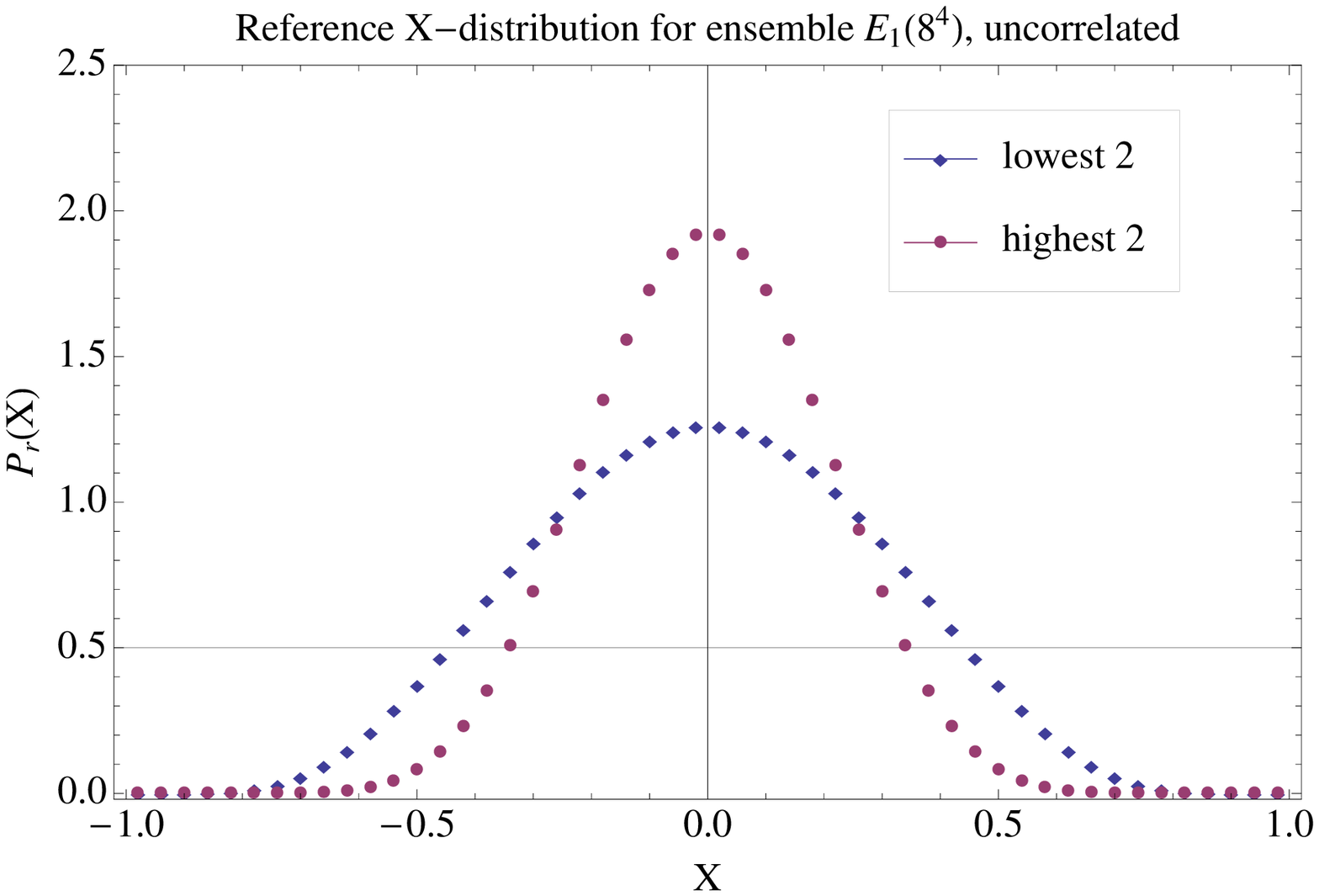}     
     }
     \vskip -0.00in
    \centerline{
    \hskip 0.08in
    \includegraphics[width=8.4truecm,angle=0]{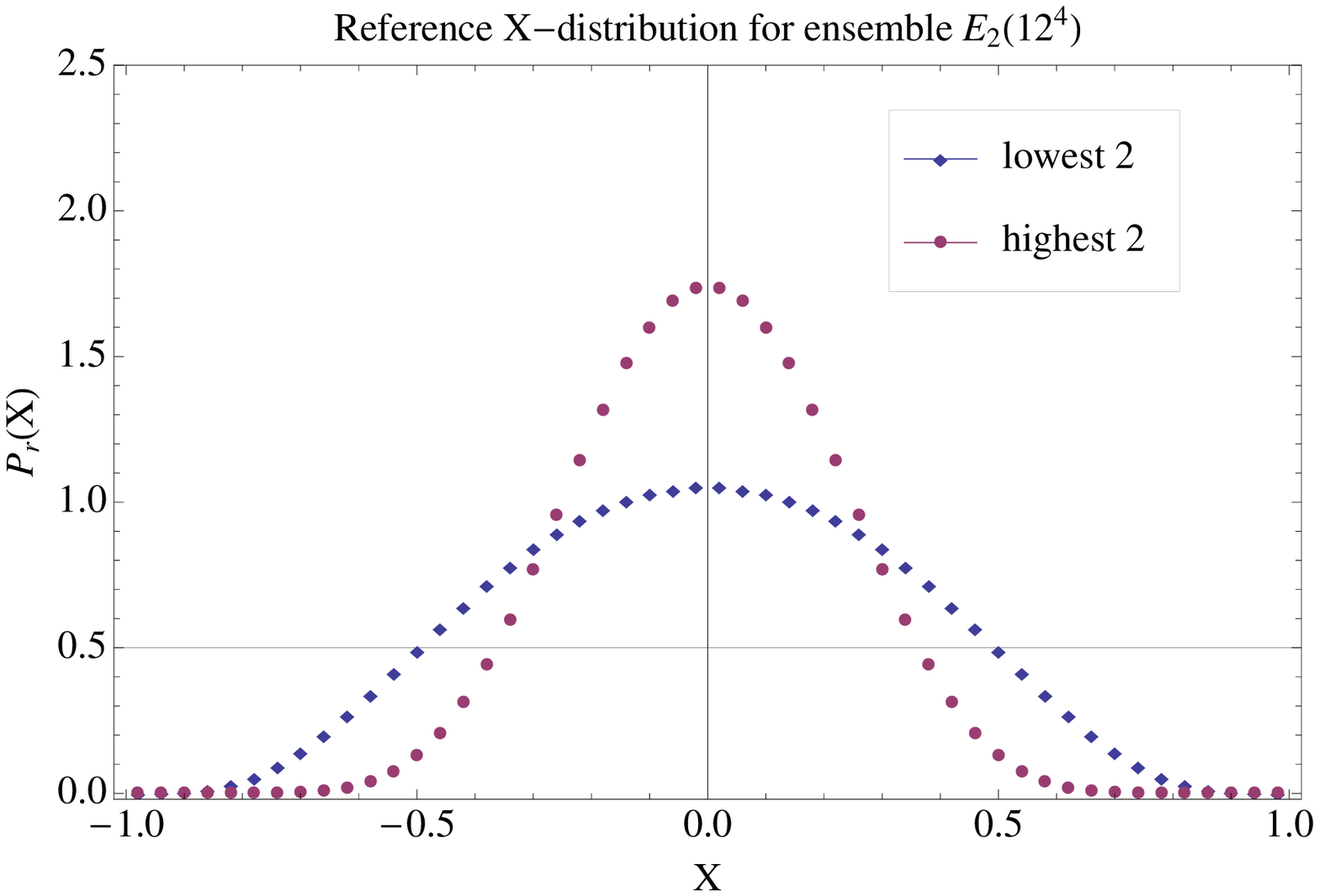}
    \hskip -0.1in
    \includegraphics[width=8.4truecm,angle=0]{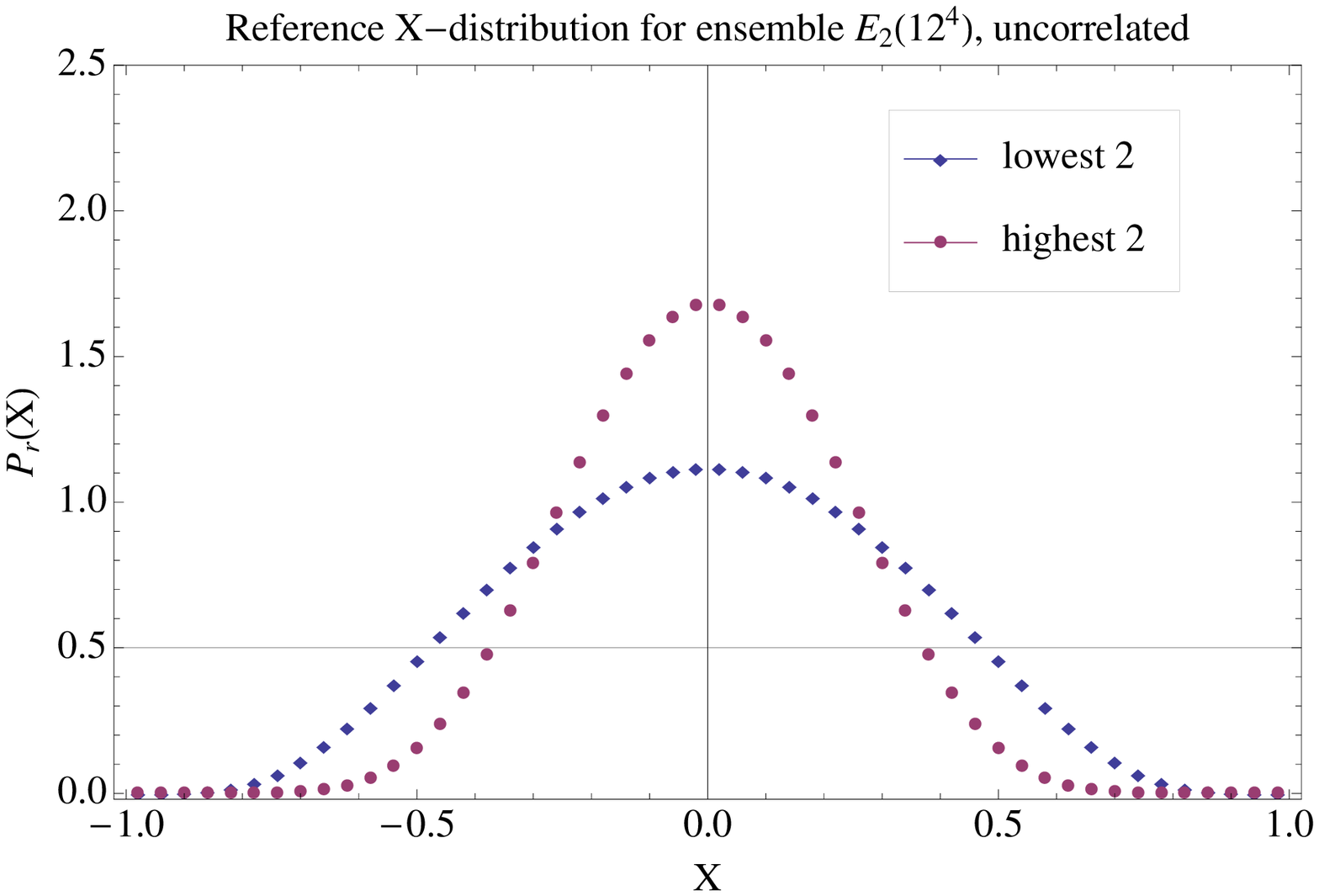}     
     }
     \vskip -0.00in
    \centerline{
    \hskip 0.08in
    \includegraphics[width=8.4truecm,angle=0]{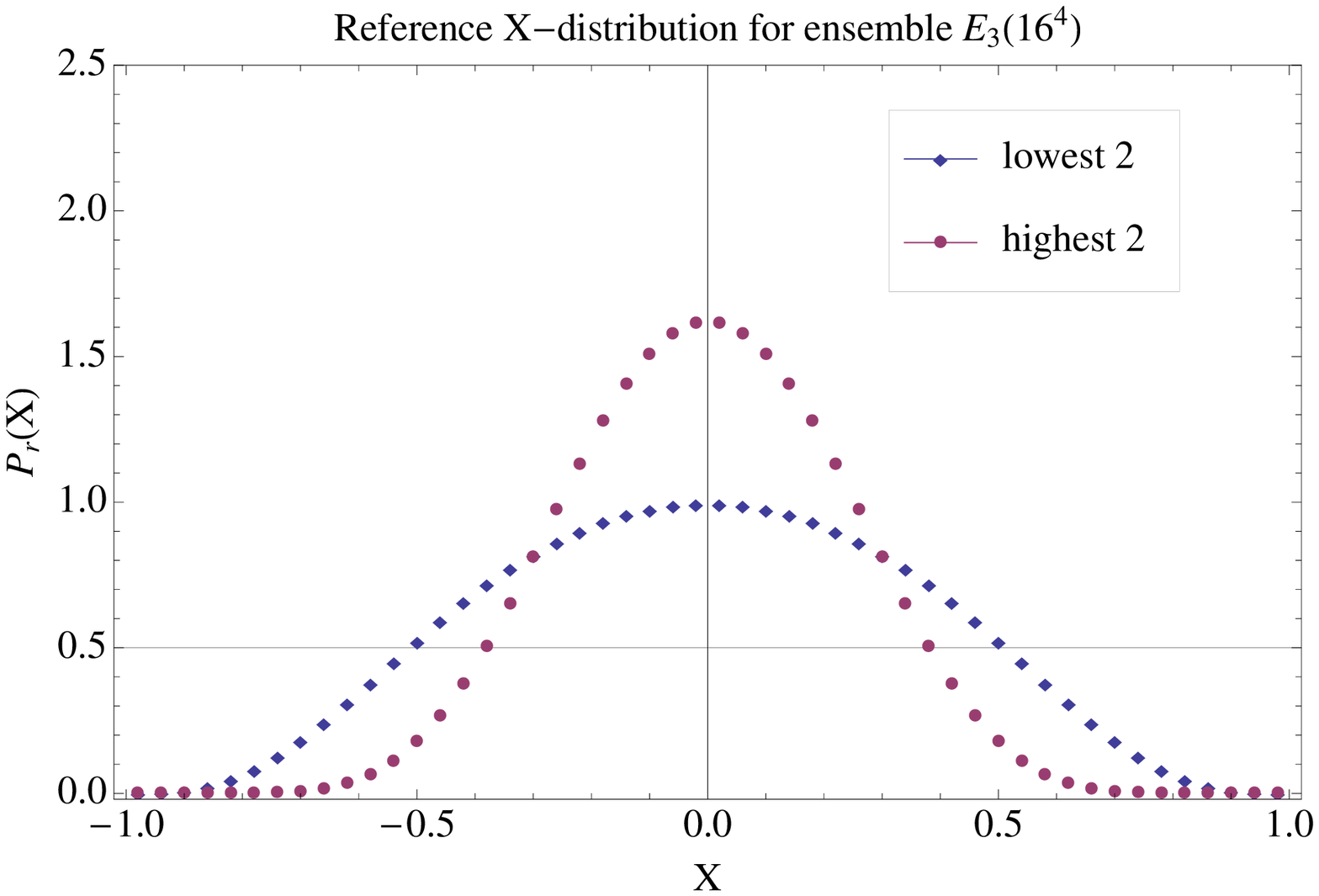}
    \hskip -0.1in
    \includegraphics[width=8.4truecm,angle=0]{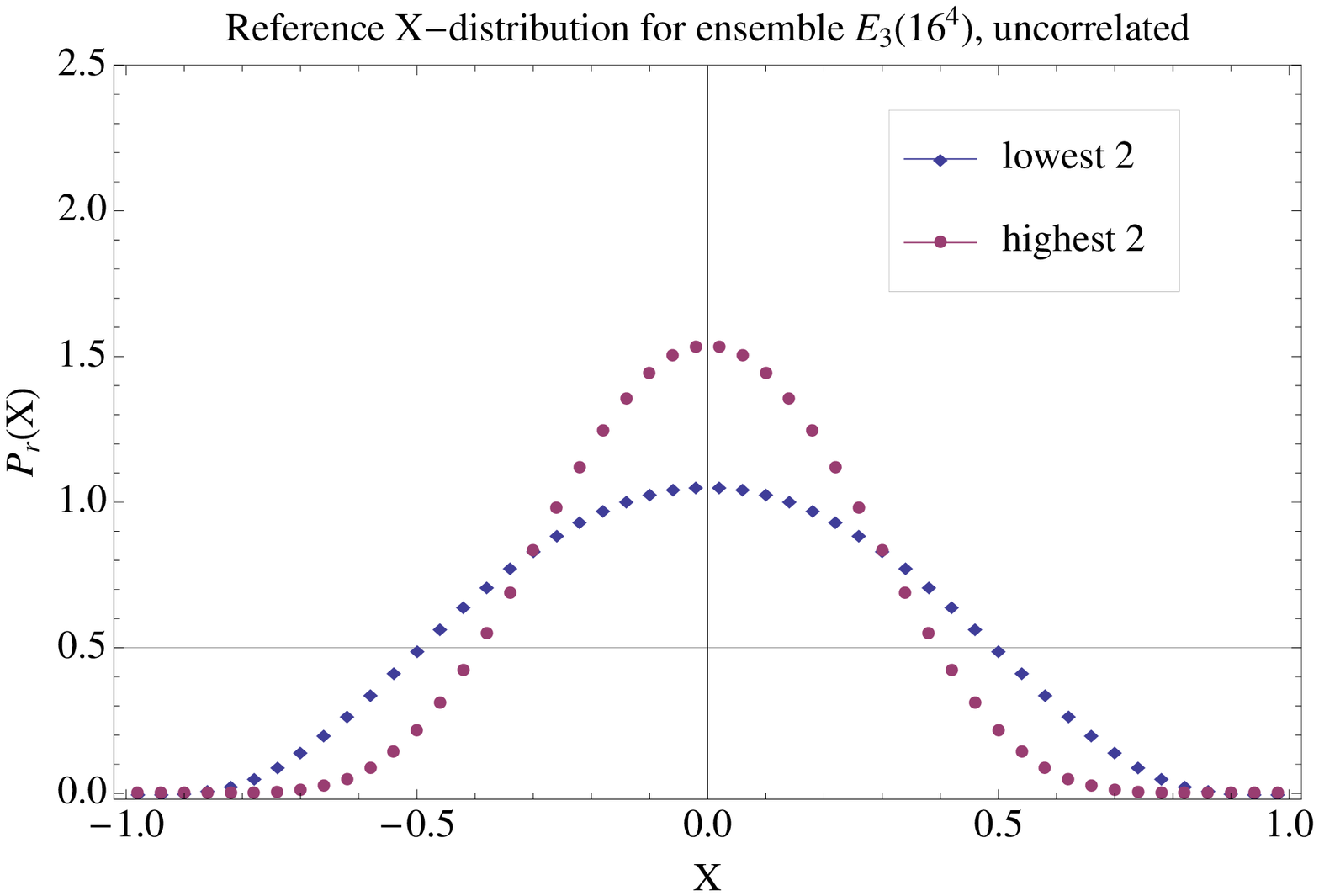}     
     }
     \vskip -0.00in
    \centerline{
    \hskip 0.08in
    \includegraphics[width=8.4truecm,angle=0]{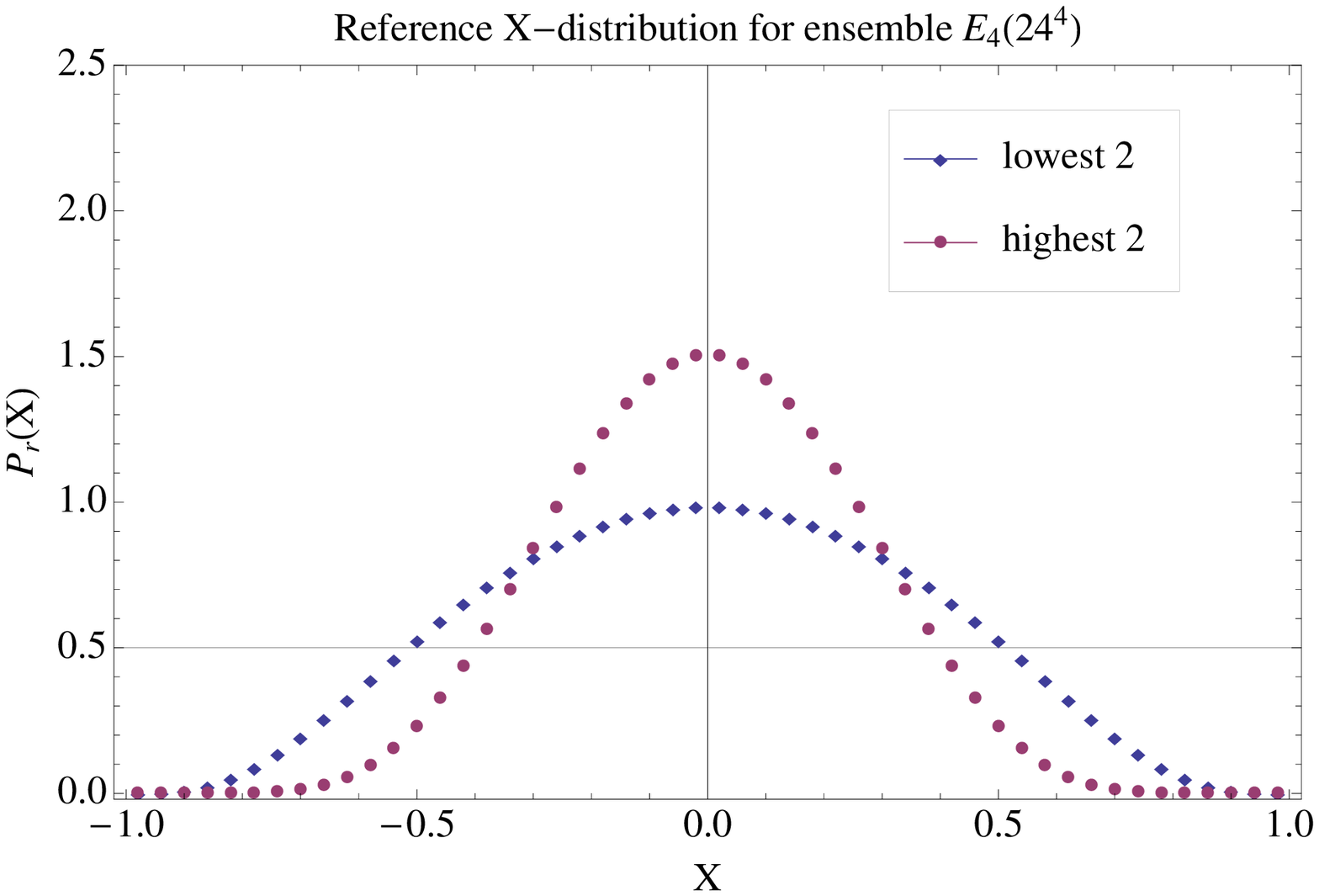}
    \hskip -0.1in
    \includegraphics[width=8.4truecm,angle=0]{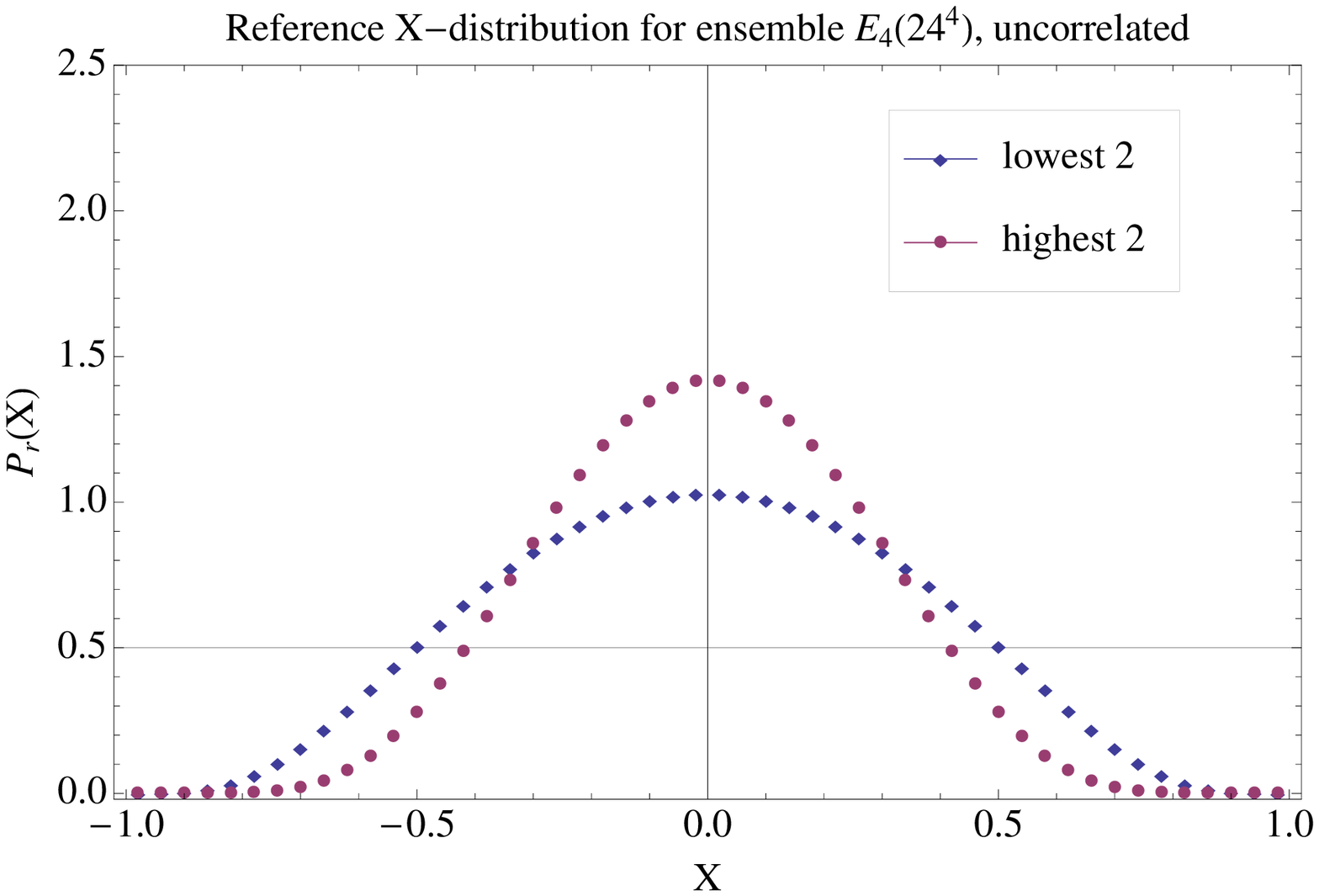}     
     }
     \vskip -0.10in
     \caption{Reference $\Xg$--distributions for ``lowest'' and ``highest'' modes in 
     ensembles $E_1$--$E_4$ are shown in the left column. The associated distributions 
     with statistically independent left--right components (``uncorrelated'') are shown
     in the right column.}
     \vskip -0.1in 
     \label{fig:pr-low-high}
\end{center}
\end{figure}

On the left side of Fig.~\ref{fig:pr-low-high} we show the reference 
$\Xg$--distributions, or reference polarization dynamics, for ``lowest'' and 
``highest'' eigenmodes in ensembles $E_1$--$E_4$. On the right side of the same
figure are the associated $\Xg$--distributions for the dynamics with 
statistically independent left--right components. A detailed description of 
the method used to obtain this ``randomized'' or ``uncorrelated'' dynamics
is given in Appendix~\ref{app:implement}, as are the other technical details related 
to the construction of absolute $\Xg$--distributions. 

Looking at the left side of Fig.~\ref{fig:pr-low-high}, the most obvious feature 
is the narrowing of the distributions going from low to higher modes. This is as 
expected from many studies of standard $\Xg$--distributions even though we should 
point out that, contrary to most of such studies, here we are not making 
any restrictions based on magnitudes of $\psi(x)$. Rather, all local values 
are part of the dynamics, as they should be.

\begin{figure}
\begin{center}
    \centerline{
    \hskip 0.08in
    \includegraphics[width=8.4truecm,angle=0]{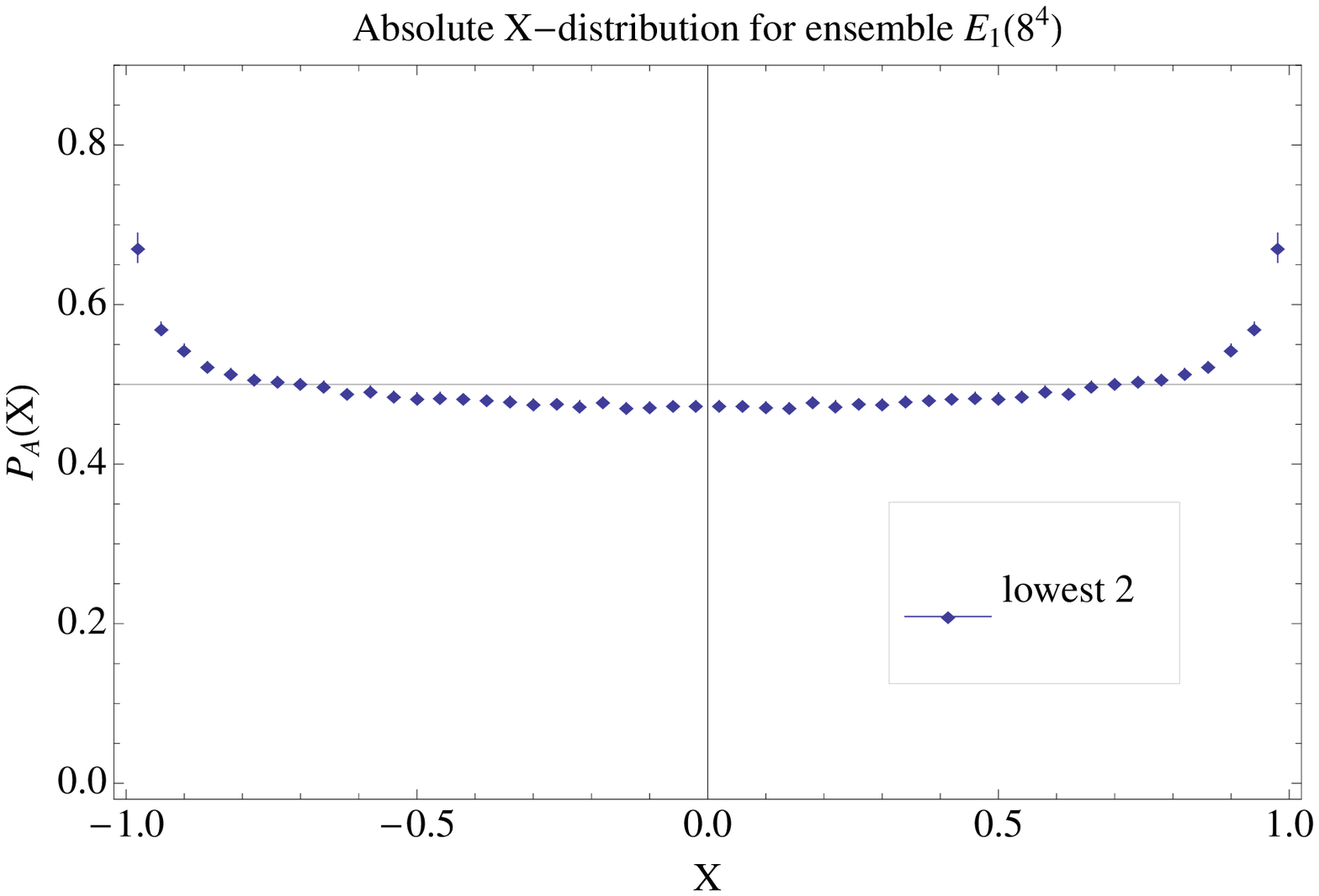}
    \hskip -0.1in
    \includegraphics[width=8.4truecm,angle=0]{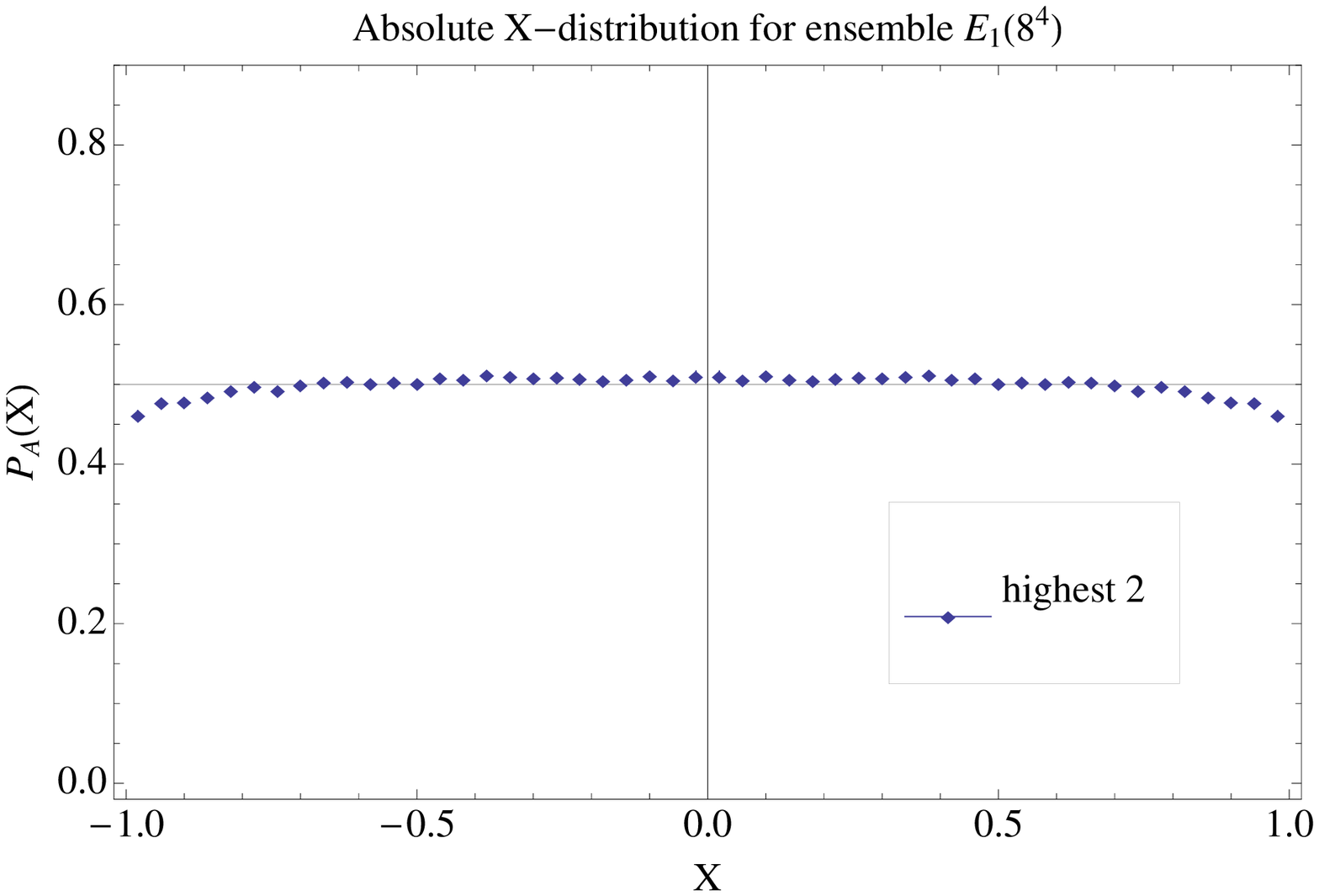}     
     }
     \vskip -0.00in
    \centerline{
    \hskip 0.08in
    \includegraphics[width=8.4truecm,angle=0]{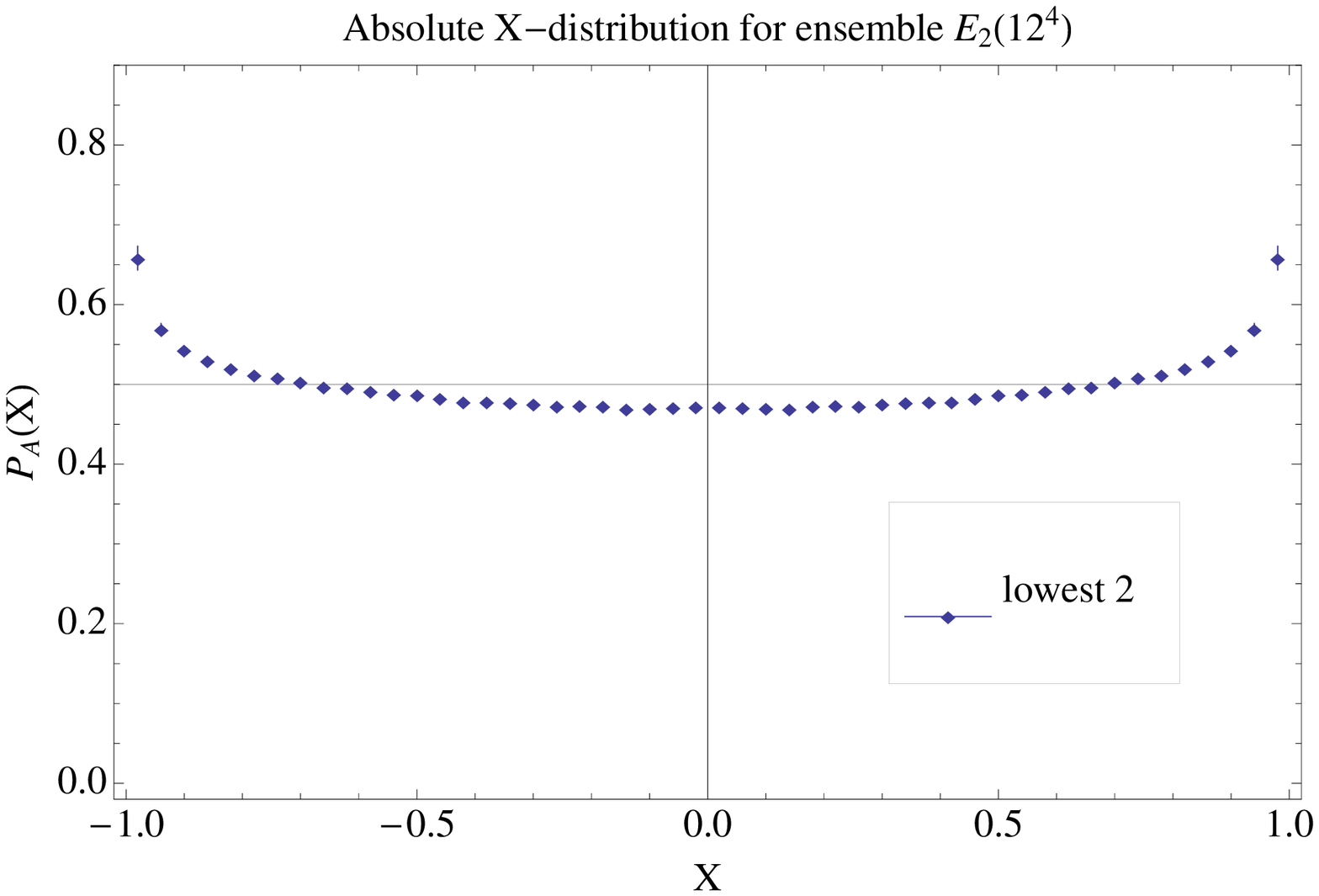}
    \hskip -0.1in
    \includegraphics[width=8.4truecm,angle=0]{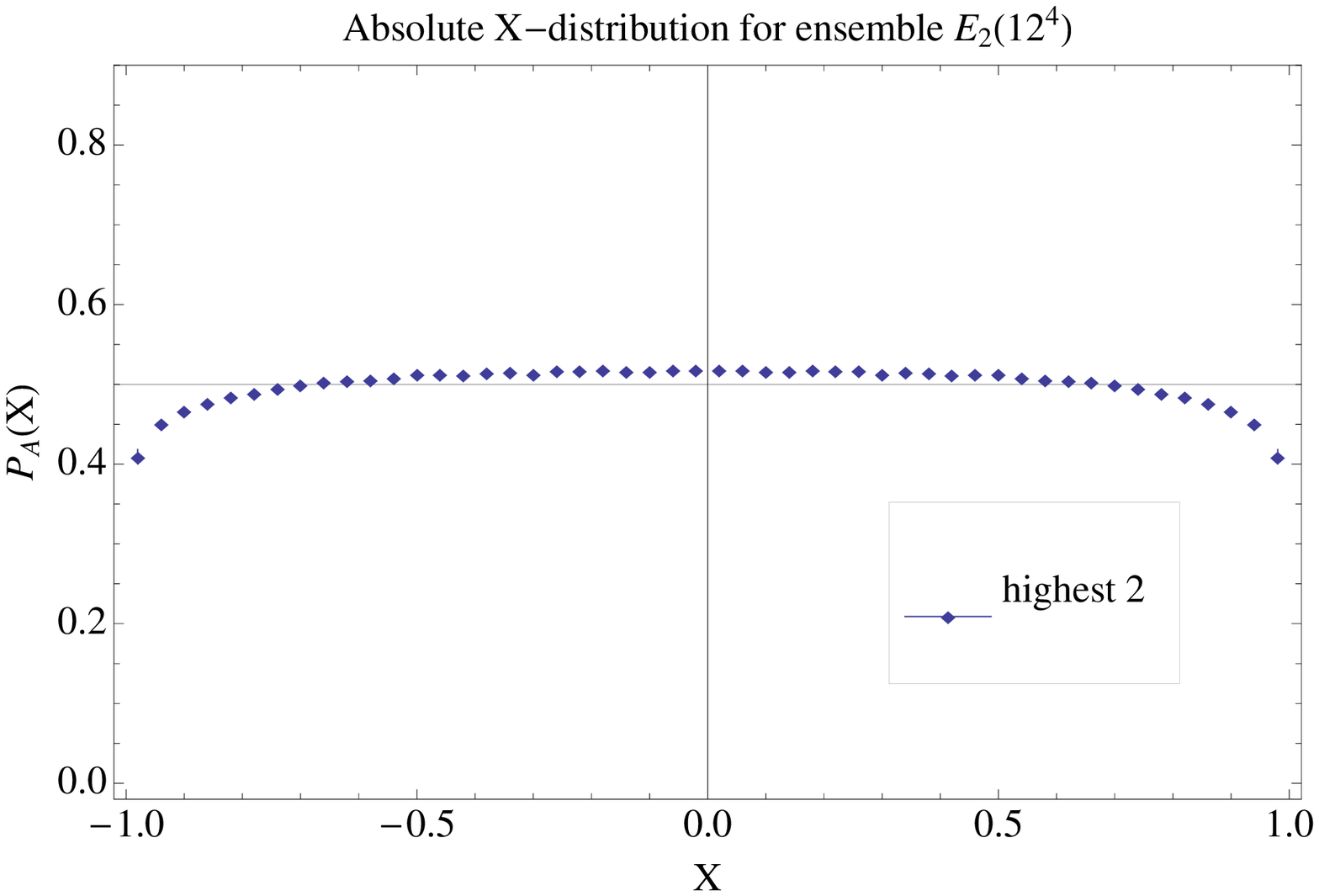}     
     }
     \vskip -0.00in
    \centerline{
    \hskip 0.08in
    \includegraphics[width=8.4truecm,angle=0]{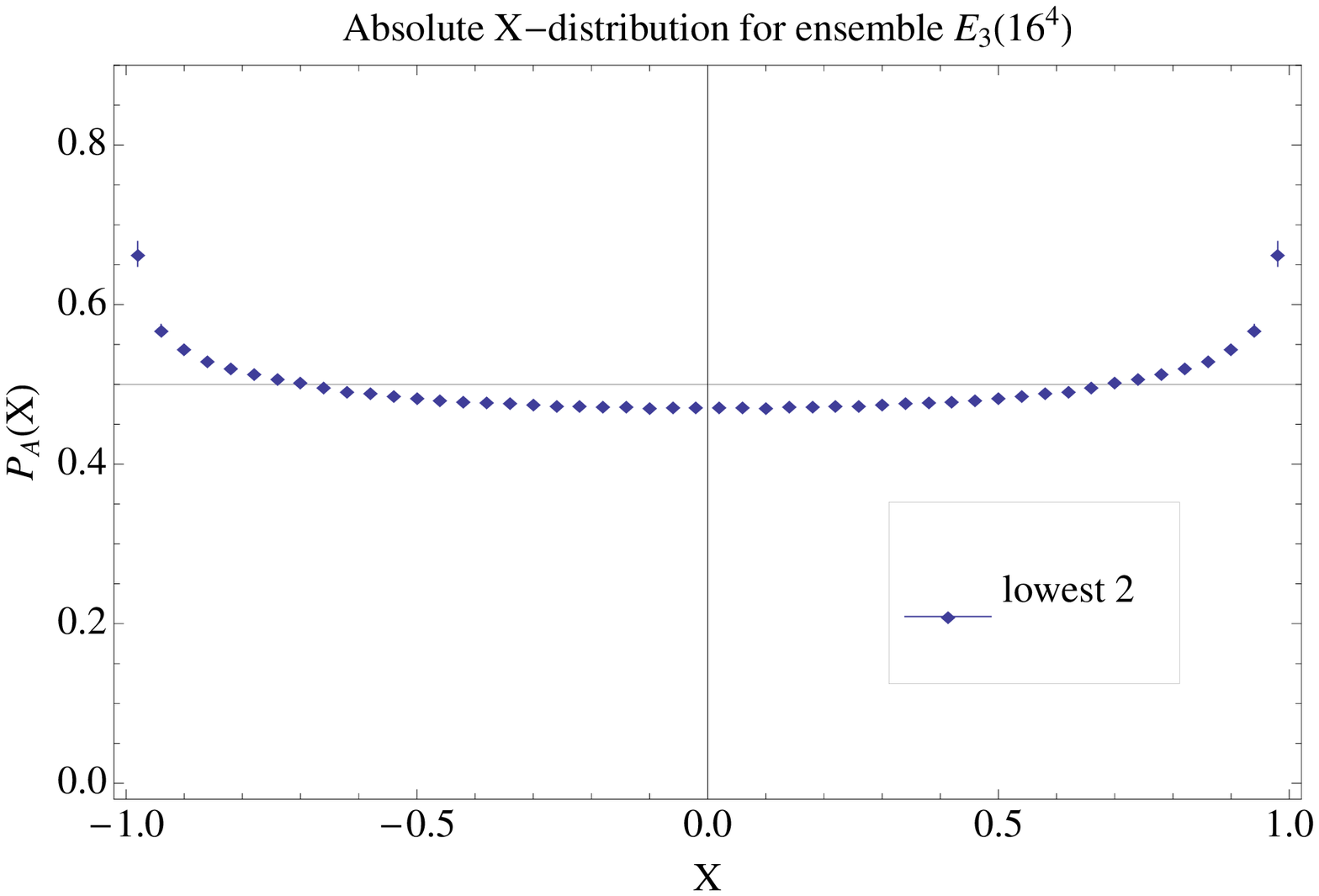}
    \hskip -0.1in
    \includegraphics[width=8.4truecm,angle=0]{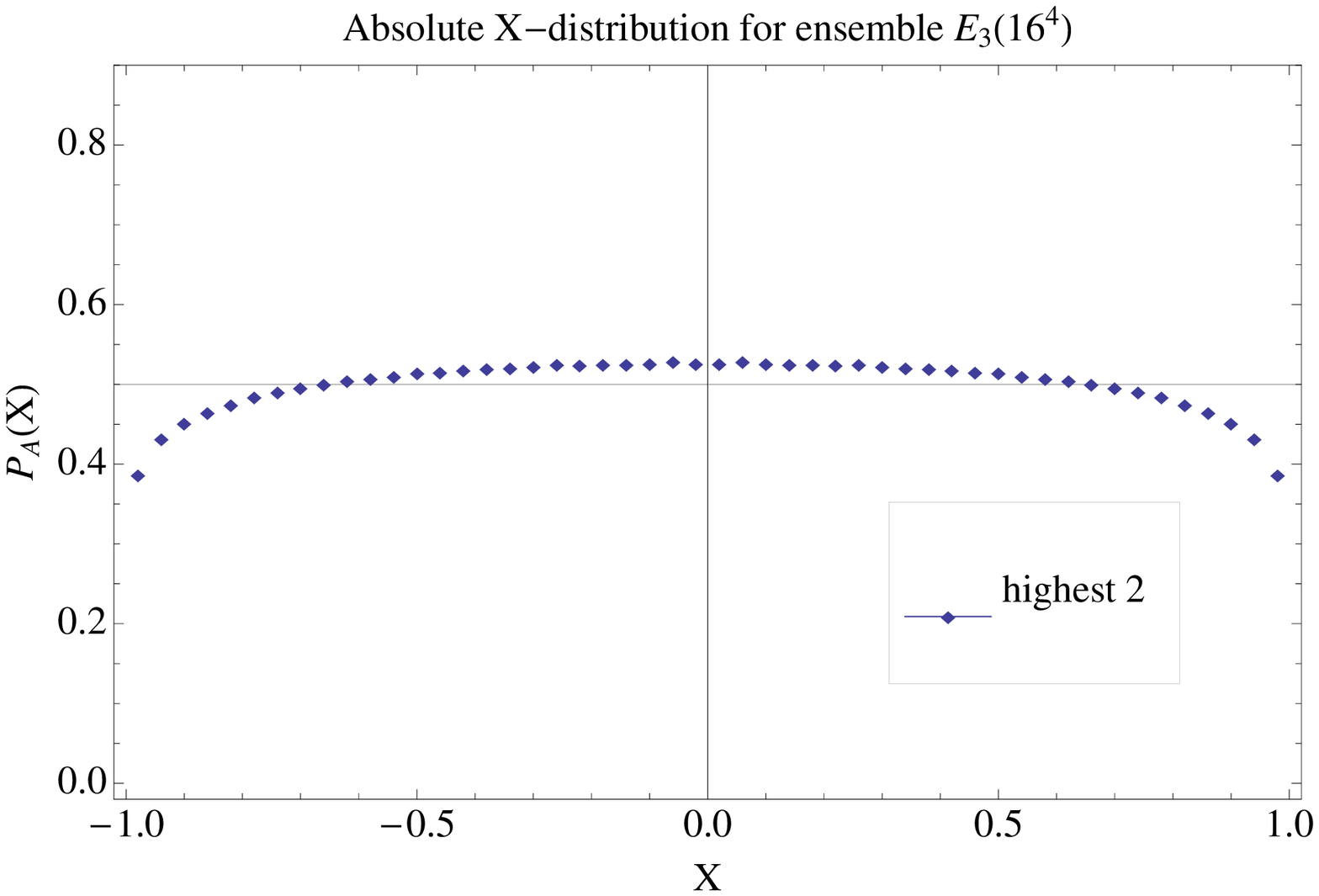}     
     }
     \vskip -0.00in
    \centerline{
    \hskip 0.08in
    \includegraphics[width=8.4truecm,angle=0]{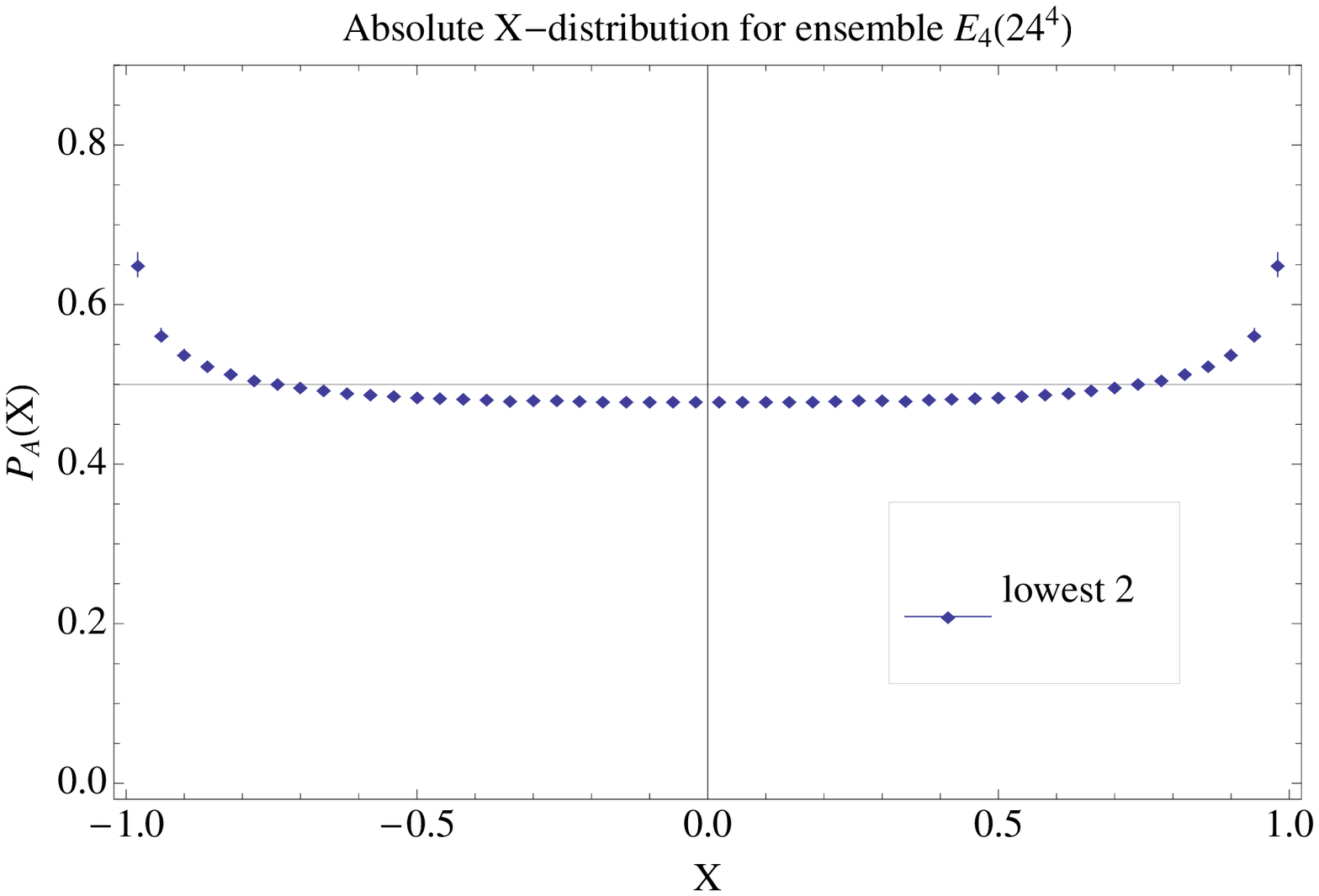}
    \hskip -0.1in
    \includegraphics[width=8.4truecm,angle=0]{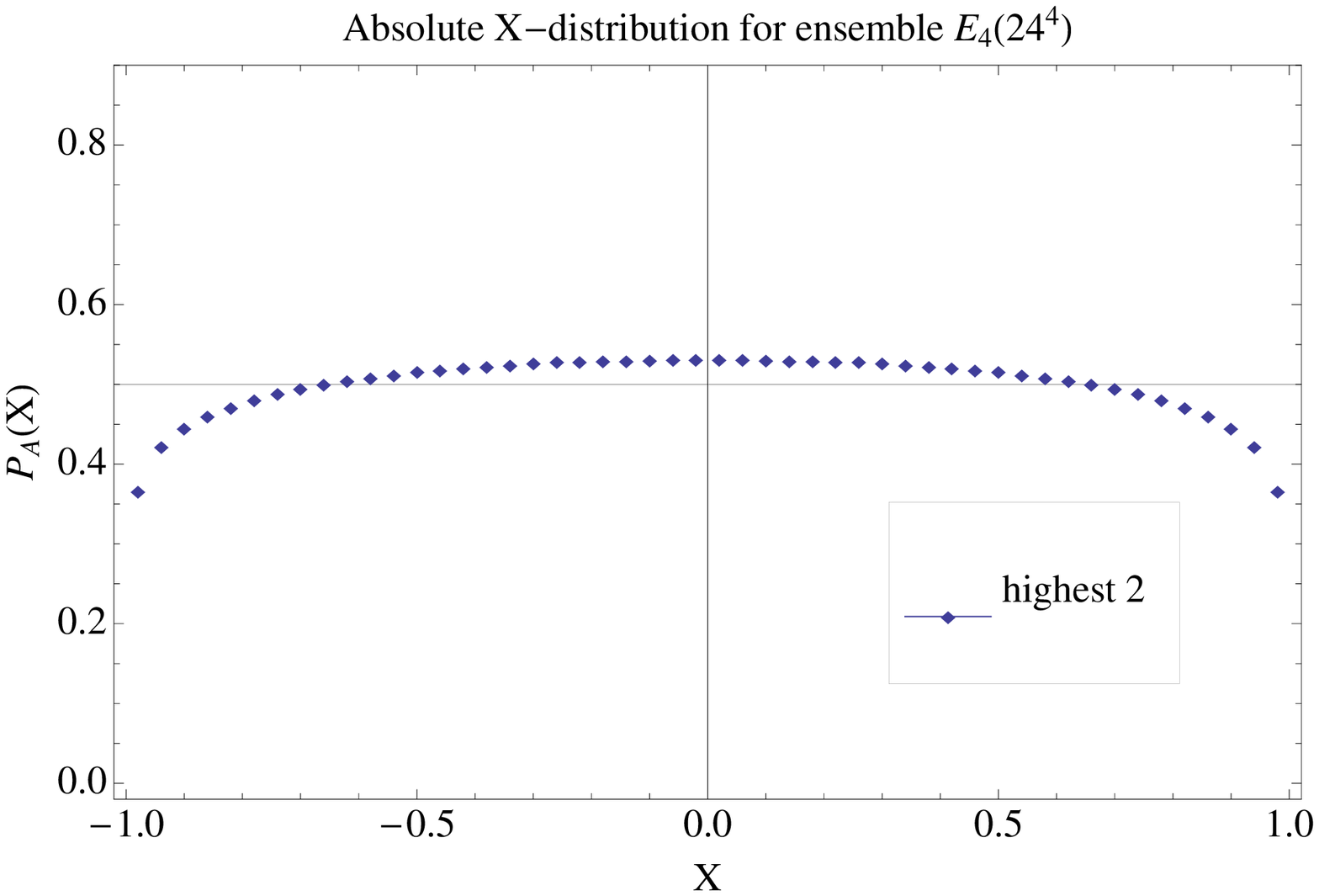}     
     }
    \vskip -0.10in
    \caption{Absolute $\Xg$--distributions for lowest (left column) and highest (right column) 
     two pairs of modes for ensembles $E_1$--$E_4$. Lattice spacing is decreasing from top to 
     down.}
    \vskip -0.1in 
    \label{fig:pa-low-high}
\end{center}
\end{figure}

An interesting aspect of Fig.~\ref{fig:pr-low-high} is the comparison of left 
and right columns, i.e.\ the comparison of pure kinematics (right side) to kinematics
combined with dynamics (left side). As one can see, the two are barely distinguishable
implying that the qualitative appearance of reference $\Xg$--distributions is almost 
entirely driven by kinematics. In order to expose the dynamical polarization tendencies,
we construct the corresponding absolute $\Xg$--distributions that effectively remove
the features due to kinematics. The resulting distributions are shown in 
Fig.~\ref{fig:pa-low-high} with lowest modes on the left and highest modes 
on the right. 

One can easily interpret the absolute $\Xg$--distributions since the uncorrelated
dynamics is represented by the uniform distribution shown as a solid horizontal
line. Thus, for the lowest modes, there is an enhancement of samples with largest
polarization relative to statistical independence, and the corresponding suppression
of samples with low polarization. In fact, one can read from the graph complete
differential information on how much enhancement/suppression there is for arbitrary
segment of uncorrelated population. The correlation coefficient of polarization (CCP)
is obviously positive for the lowest modes, and the dynamics governing them exhibits 
the dynamical tendency for polarization. As can be inspected from the right column,
this situation reverses for the highest computed modes in all lattice gauge 
ensembles. Indeed, in that case the dynamics suppresses samples with largest 
polarization and CCP is negative.

\subsection{Convexity of Absolute $\Xg$--Distributions}

A characteristic feature of absolute $\Xg$--distributions shown in Fig.~\ref{fig:pa-low-high},
as well as of all the other ones that we computed from our ensembles, is that they are 
in some sense the simplest they can be. For example, while there doesn't seem to be a reason 
{\em a~priori} excluding multiple intersections of $\xd_A(\Xg)$ with uniform distribution on each 
``chiral half'' (interval $[-1,0]$ and $[0,1]$), we have generically seen just one. Note that
these crossings define segments in the sample space with dynamically enhanced/suppressed chirality.
Moreover, consistently with the above, the absolute $\Xg$--distributions for QCD 
eigenmodes seem to be either purely convex or purely concave on their domain $[-1,1]$. 
This implies that they are also strictly monotonic on each chiral half.

To give a well-defined meaning to these kinds of observations, one should specify more precisely
the dynamics of chiral pairs in question. In other words, one should specify how are 
the associated populations defined in terms of selected Dirac eigenvectors so that 
the corresponding characteristics are both physically meaningful and have proper continuum 
limits. In that regard, things simplify for lattice QCD in the infinite volume since
one can assign an $\Xg$--distribution to an individual eigenmode in that case, at least
for equilibrium gauge backgrounds. Moreover, under standard assumptions, the local properties 
of such modes will not depend on the specific equilibrium configuration in question. Thus, in 
the infinite volume, the eigenvalue $\lambda$ is the only label distinguishing the local 
behavior of Dirac eigenmodes in the underlying gauge theory. It is a common practice that 
the above is also mimicked in the finite volume, except that the gauge ensemble average has 
to be performed in that case. In other words, one defines properties of eigenmodes at scale 
$\Lambda$ by inserting $\delta(\Lambda -|\lambda|)$ into the expression for ensemble average 
of an observable involving a spectral sum over individual eigenmodes $\psi_\lambda$.  

Here we will follow the above logic and all the conclusions regarding the polarization properties 
of Dirac eigenmodes will implicitly refer to dynamics of modes at fixed $\Lambda$ in the above
sense.\footnote{Note that the practical implementation of this, i.e.~the selection of appropriate 
eigenmodes with only finite gauge ensembles at our disposal, can vary depending on how 
much data is actually available. For example, our selection of ``lowest'' modes in the previous 
section could be viewed as a procedure to determine the behavior of eigenmodes at $\Lambda=0$ in 
the continuum limit.} With that in mind we suggest the following for further investigation.

\medskip
\noindent {\em \underline{Proposition 1}\,(Convexity):$\;$ 
           The absolute $\Xg$--distributions of low--lying overlap Dirac modes in SU(3) pure 
           glue lattice gauge theory are strictly convex, strictly concave or constant.}
\medskip

\noindent It is naturally expected that this behavior of absolute $\Xg$--distributions 
will not depend on the choice of the Dirac operator with exact lattice chiral symmetry.
We would also like to emphasize that the weaker statement, namely that the absolute
$\Xg$--distributions are strictly monotonic or constant on each chiral half, might hold
not only for low--lying modes but for all Dirac modes.

\subsection{Continuum Limit of Distributions at Fixed $\Lambda$}

Assuming that the dynamics of Dirac eigenmodes at fixed $\Lambda$ can be assigned a 
well--defined physical meaning, their dynamical polarization characteristics should have 
a proper continuum limit. One should point out though that whether this is true or not 
constitutes the test for the notion of eigenmodes at fixed $\Lambda$ rather than the test 
for applicability of absolute $\Xg$--distributions.

\begin{figure}[t]
\begin{center}
    \centerline{
    \hskip 0.00in
    \includegraphics[width=8.3truecm,angle=0]{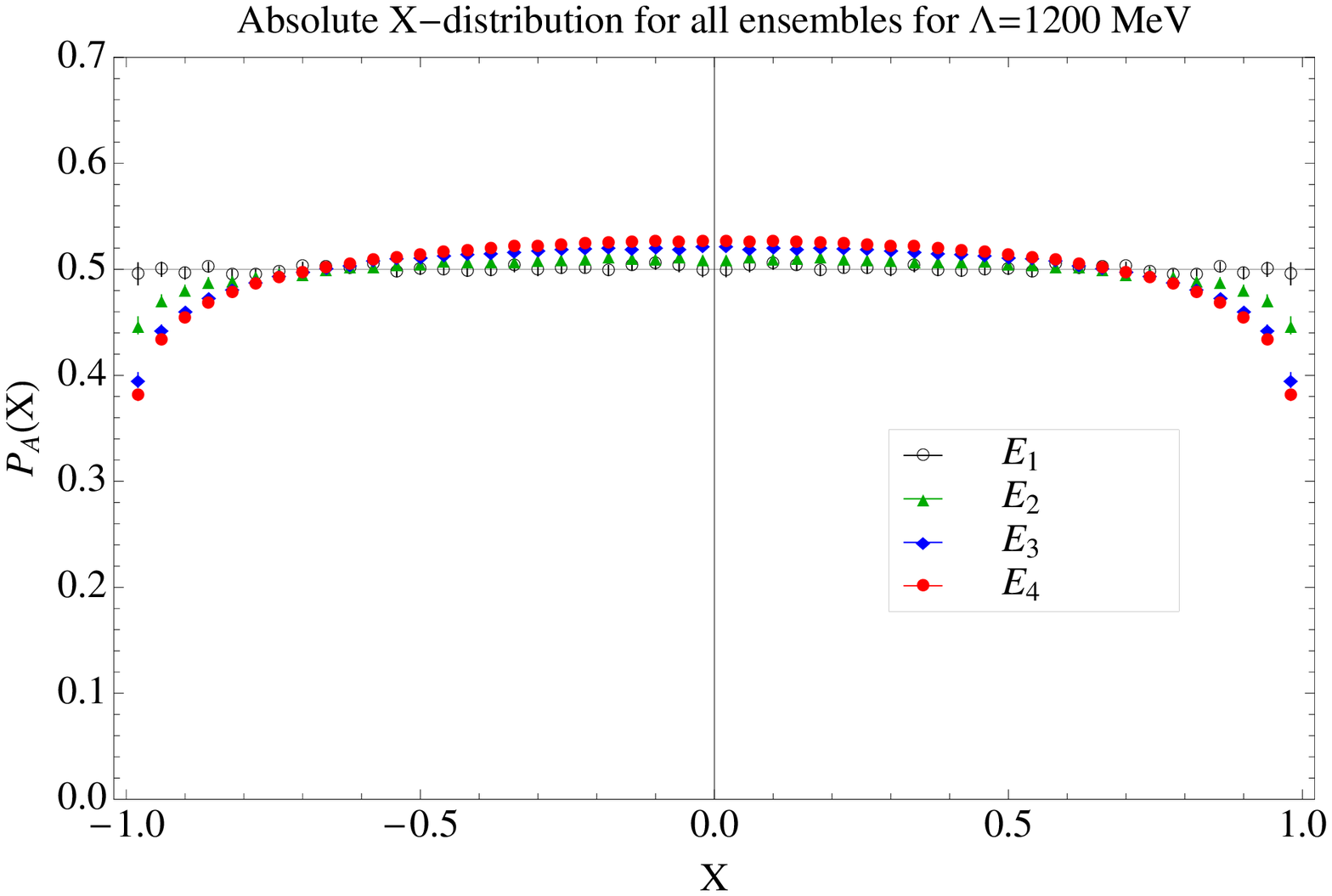}
    \hskip -0.0in
    \includegraphics[width=8.3truecm,angle=0]{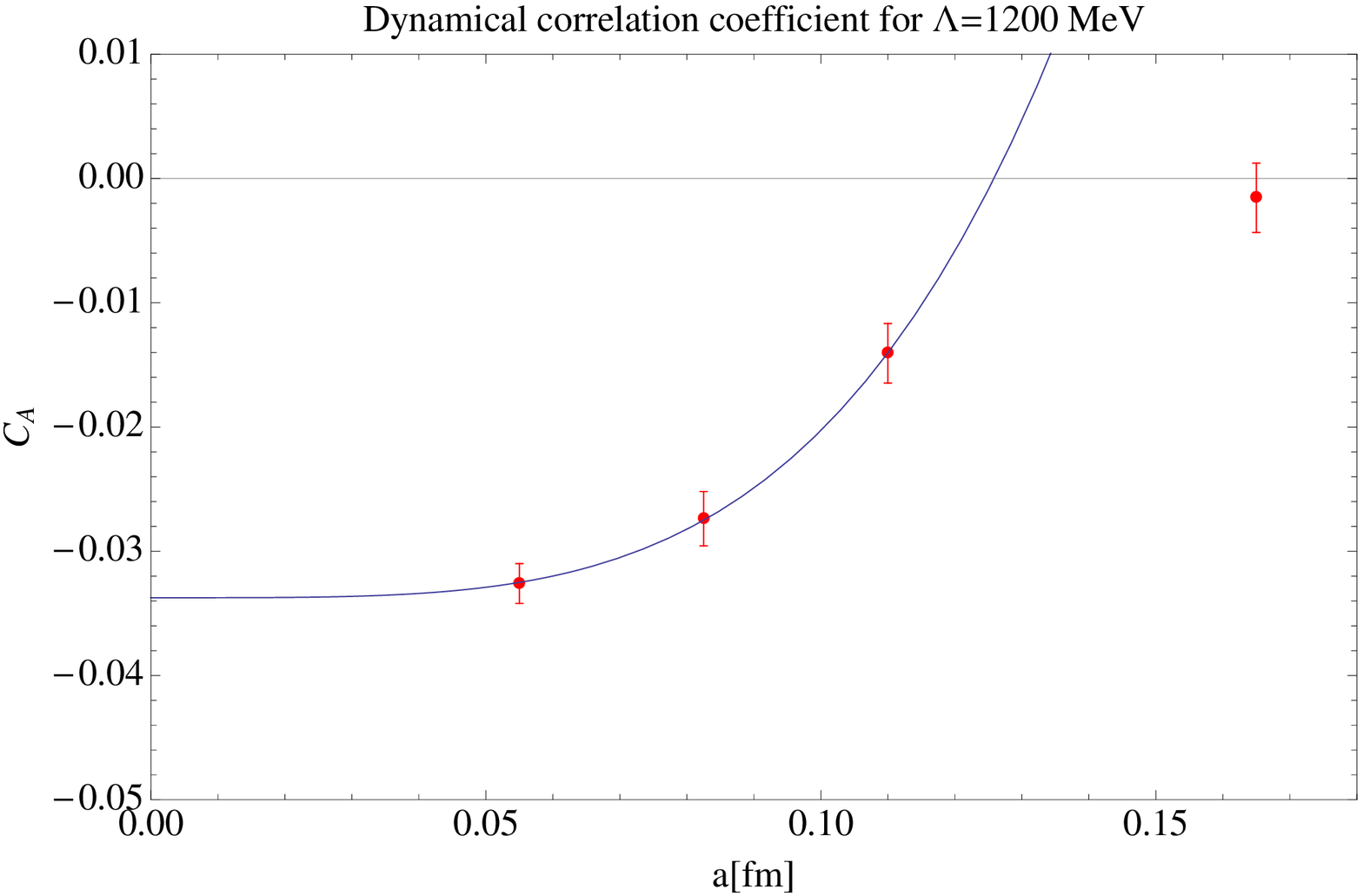}     
     }
     \vskip -0.00in
    \centerline{
    \hskip 0.00in
    \includegraphics[width=8.3truecm,angle=0]{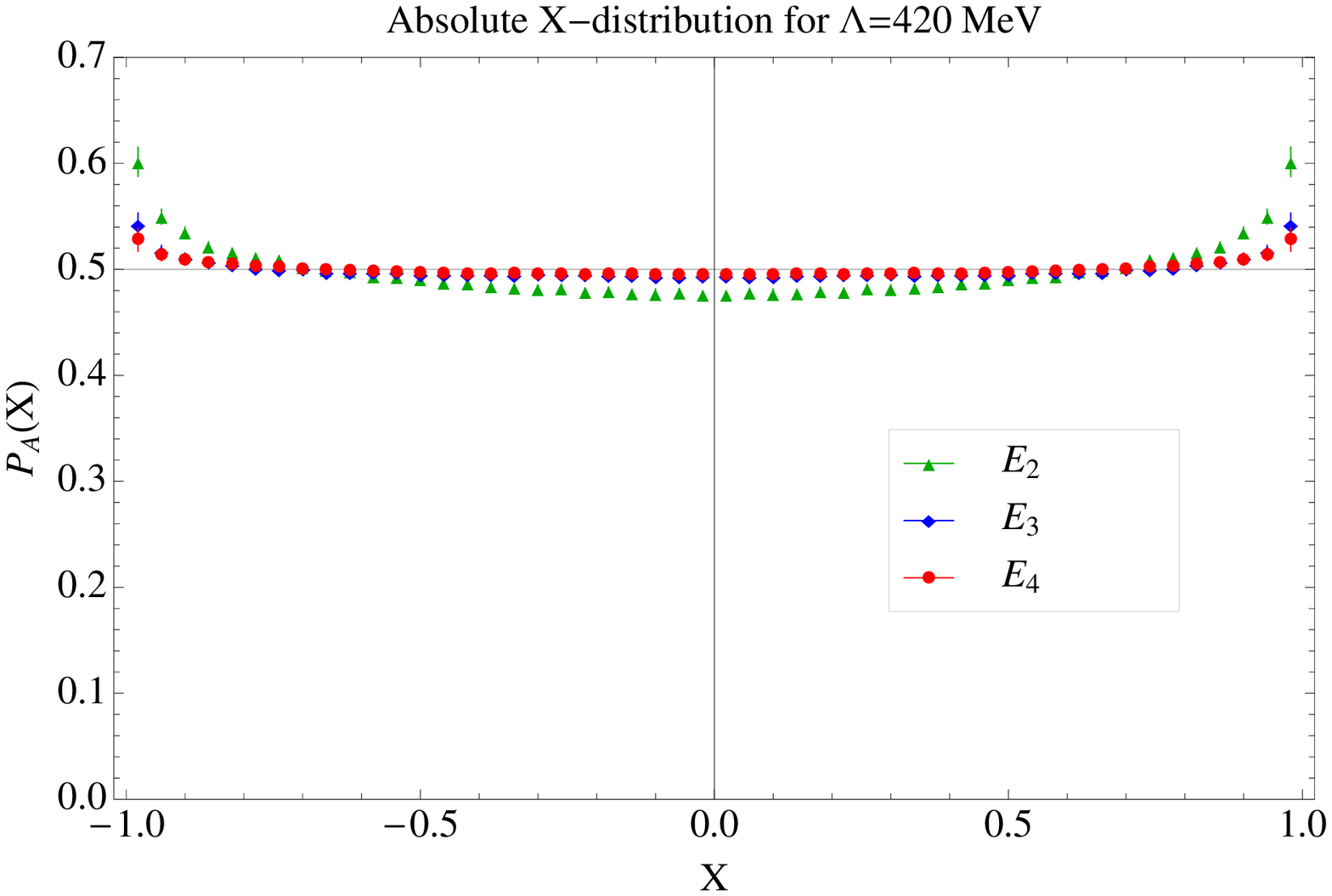}
    \hskip -0.0in
    \includegraphics[width=8.3truecm,angle=0]{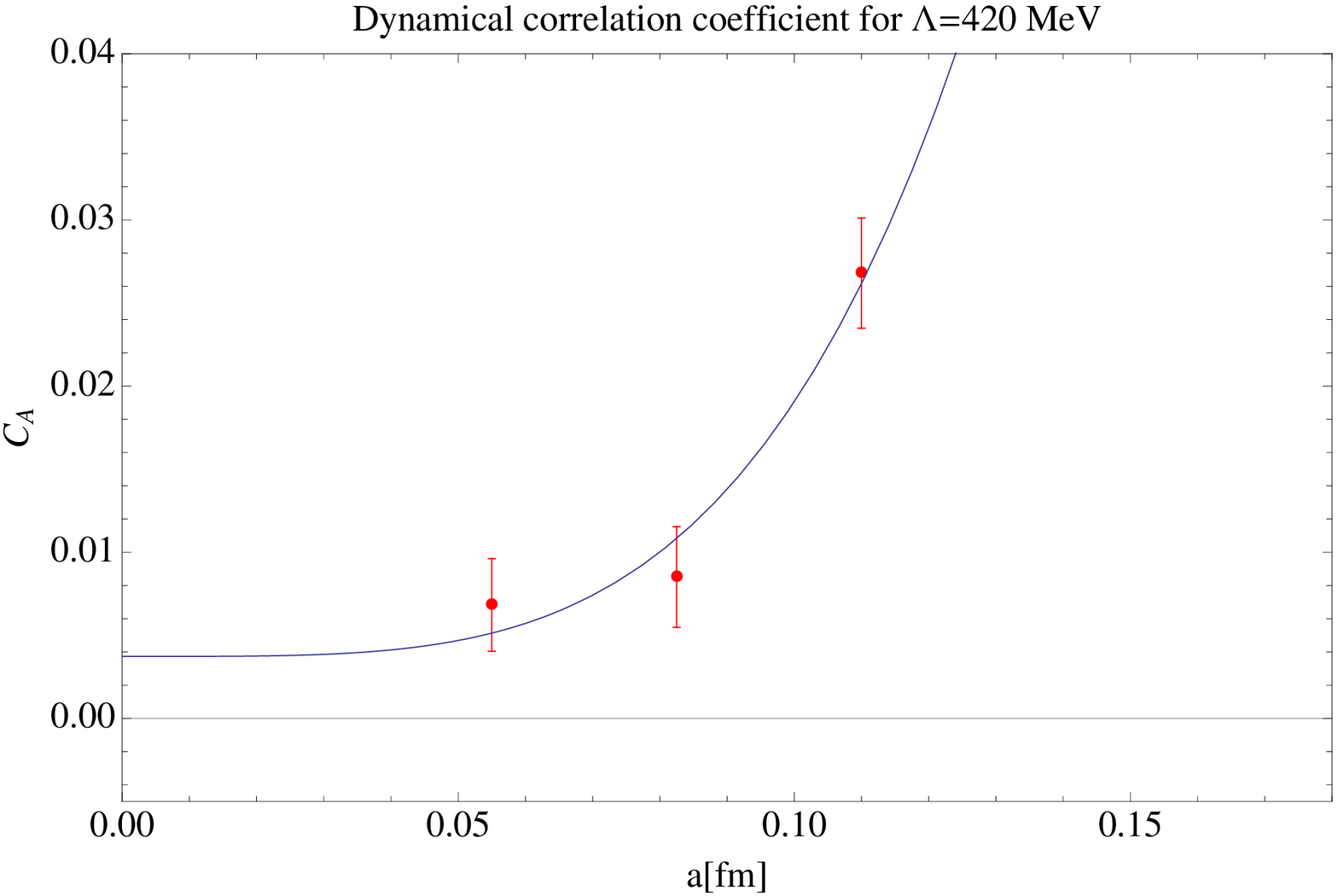}     
     }
     \vskip -0.00in
     \caption{Lattice spacing dependence of absolute $\Xg$--distributions for overlap Dirac 
              eigenmodes at $\Lambda=1200$ MeV (top) and $\Lambda=420$ MeV (bottom). 
              The corresponding correlation coefficients of polarization are shown 
              on the right.}
     \vskip -0.1in 
     \label{fig:pa-fixed-lam}
\end{center}
\end{figure}

Since our statistics are rather limited, we define the set of modes specifying the dynamics
at $\Lambda > 0$ in the following way. For each configuration from given gauge ensemble
we take two modes $\psi_{\lambda_i}$,  $\psi_{\lambda_{i+1}}$ such that 
$|\lambda_i| \le \Lambda < |\lambda_{i+1}|$. Here we have ordered the non--zero low--lying
modes by magnitude as $\lambda_1,\lambda_1^\star,\lambda_2,\lambda_2^\star, \ldots$. 
Since in practice one doesn't encounter any degeneracy of non--zero modes except for 
conjugate pairs, this definition uniquely 
picks out the two modes from the spectrum that are ``hugging'' the value $\Lambda$ assuming
that $\Lambda$ is sufficiently large 
($\Lambda \ge \Lambda_{LOW}^{MAX}$ from Table~\ref{tab:ensembles}).

In Fig.~\ref{fig:pa-fixed-lam} we show the convergence of absolute $\Xg$--distributions
for eigenmodes at 1200 MeV (top) and 420 MeV (bottom). Note that in case of the latter, 
the result for ensemble $E_1$ is not applicable since 420 MeV is less than 
$\Lambda_{LOW}^{MAX}$. This is not a problem because $E_1$ appears to be too coarse to be 
used in continuum extrapolations while our aim is to provide the example of the very 
low $\Lambda$. As one can see from plots on the left, the data quite clearly suggests 
point--wise convergence of absolute $\Xg$--distributions as the continuum limit is 
approached. To see this in a more coarse--grained manner, we show in the right column also 
the lattice spacing dependence of the moment of the distribution (correlation coefficient) 
in both cases. Their qualitative behavior suggests convergence as well. The lines shown
are fits to the form $c_1 + c_2 a^4$ to the points corresponding to $E_2$--$E_4$. 
While the power four is suggested by the data at 1200 MeV, the fit should only be viewed 
to guide the eye. Based on the above, we propose the following for further examination.

\medskip
\noindent {\em \underline{Proposition 2}\,(Continuum Limit):$\;$ 
           The absolute $\Xg$--distributions for overlap Dirac modes at scale
           $\Lambda$ have a continuum limit in SU(3) pure glue lattice gauge theory.}
\medskip


\subsection{Chiral Polarization Transition}

\begin{figure}
\begin{center}
    \centerline{
    \hskip 0.00in
    \includegraphics[width=16.6truecm,angle=0]{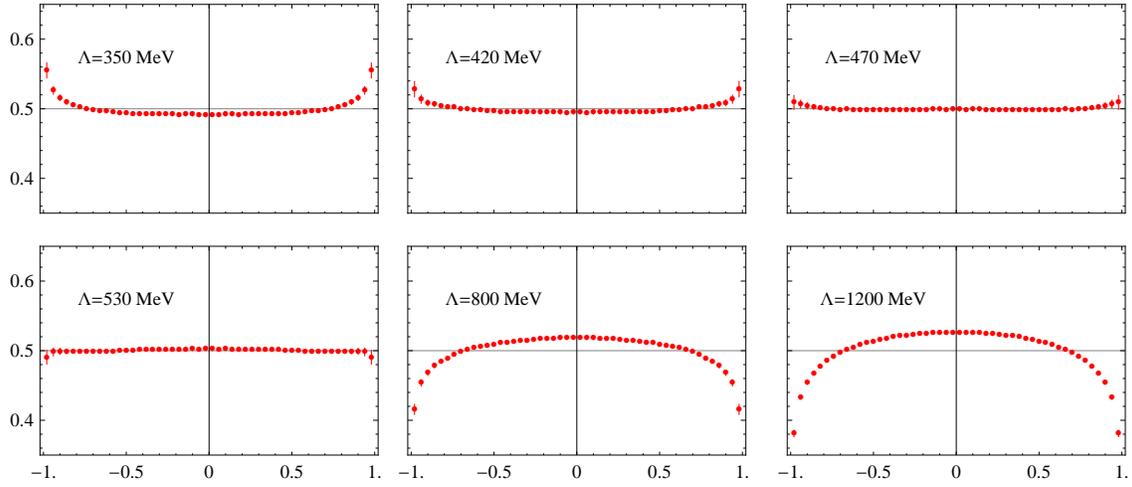}
     }
     \vskip -0.00in
     \caption{Absolute $\Xg$--distributions for gauge ensemble $E_4$ with changing scale 
              $\Lambda$ of the modes. Note the transition from convex to concave (positive
              correlation to negative correlation) between 470 MeV and 530 MeV.}
     \vskip -0.1in 
     \label{fig:pa-E4-alllam}
\end{center}
\end{figure}

The comparison of left and right columns in Fig.~\ref{fig:pa-low-high}, as well as 
the comparison of top and bottom parts in Fig.~\ref{fig:pa-fixed-lam} suggest 
a high degree of regularity in the dynamical polarization properties of the low--lying
modes. In particular, the dynamics of lowest--lying modes always seems to enhance 
chiral polarization while the dynamics of higher modes tends to suppress it.
To check if this is indeed a monotonic feature with respect to scale $\Lambda$ of the
modes, we show in Fig.~\ref{fig:pa-E4-alllam} how the absolute $\Xg$--distribution 
for gauge ensemble $E_4$ changes from low values upwards. As one can see, 
the transition from convex to concave indeed proceeds in a monotonic manner.  

The above behavior is qualitatively similar for all our ensembles and we are thus led 
to conclude that there exists a sharply defined transition point $\Lambda_T$ in 
the spectrum of overlap Dirac eigenmodes, where dynamical polarization properties 
of these modes qualitatively change. The QCD--inherited dynamics governing the lower 
part of the spectrum supports chiral polarization while in the upper part it produces
chiral anti--polarization. Apart from the mere existence of this dynamical feature, 
it is quite remarkable that it is not only the CCP that changes sign 
at $\Lambda_T$, but rather the whole polarization dynamics (absolute $\Xg$--distribution) 
appears to be indistinguishable (uniform) from statistical independence. 
This suggests that the existence of this {\em chiral polarization transition} represents
a robust dynamical property of QCD Dirac eigenmodes. 

\begin{figure}
\begin{center}
    \centerline{
    \hskip 0.00in
    \includegraphics[width=16.6truecm,angle=0]{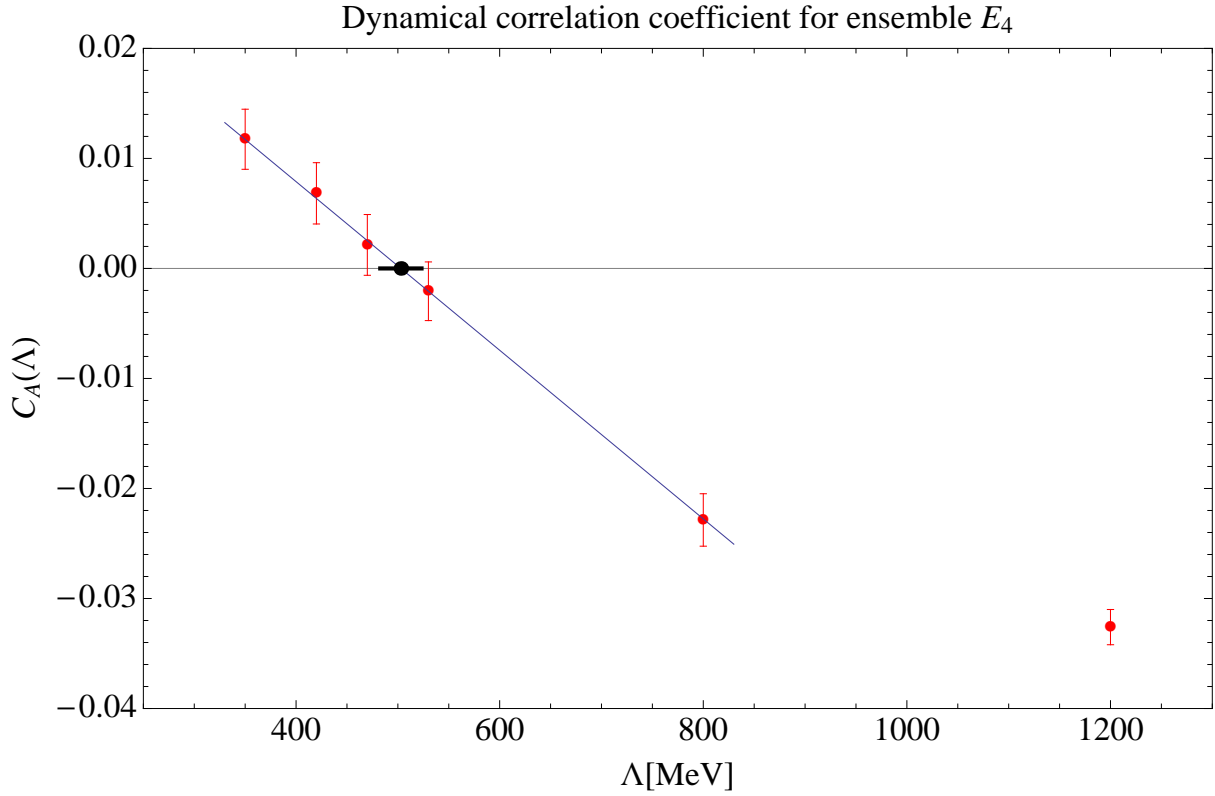}
     }
     \vskip -0.00in
     \caption{Determination of the chiral polarization transition point $\Lambda_T$ for 
              ensemble $E_4$.}
     \vskip -0.1in 
     \label{fig:ca-E4-alllam}
\end{center}
\end{figure}

In order to determine $\Lambda_T$ for our ensembles, we scanned the behavior of CCP
with respect to $\Lambda$ and looked for the behavior in the vicinity of the point 
where it changes sign. As shown in Fig.~\ref{fig:ca-E4-alllam} for gauge ensemble $E_4$, 
the data suggests linearity in the corresponding region. The transition point was thus
determined from the linear fit in this neighborhood. More details about this procedure can 
be found in Appendix~\ref{app:implement}.

\begin{figure}[t]
\begin{center}
    \centerline{
    \hskip 0.00in
    \includegraphics[width=16.6truecm,angle=0]{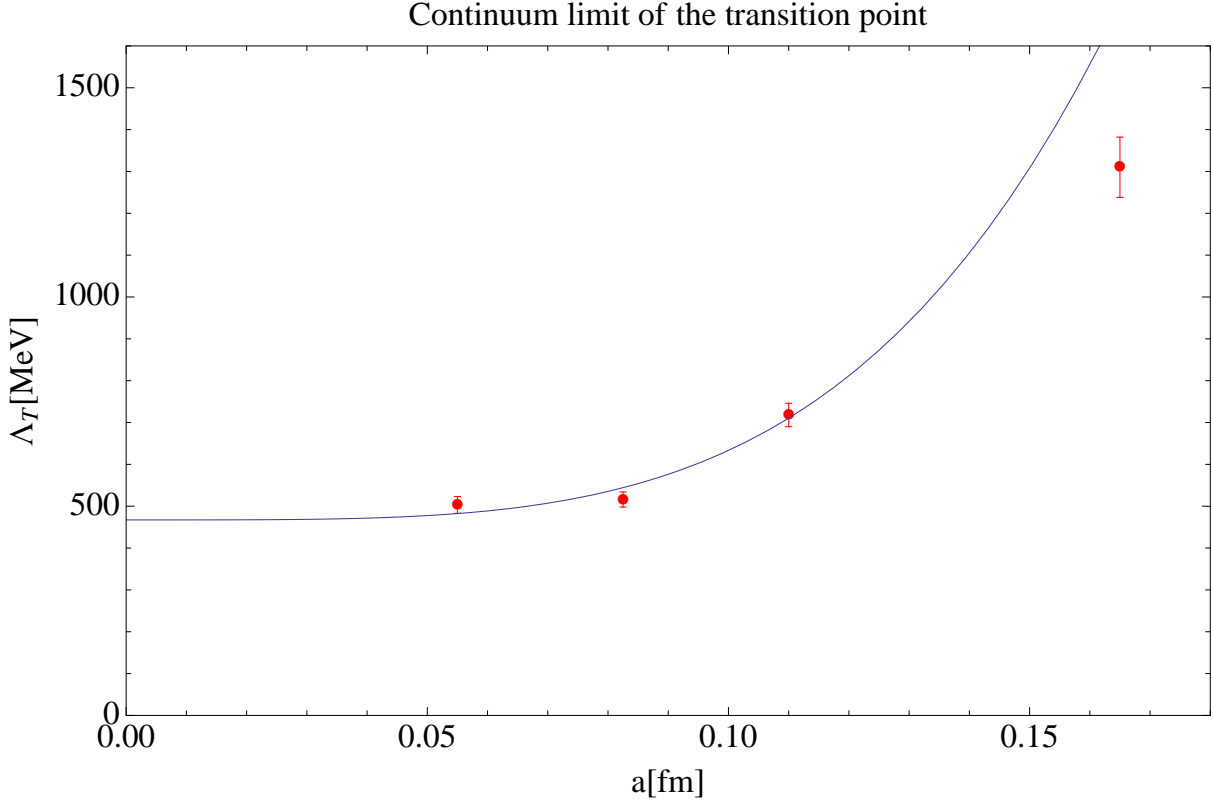}
     }
     \vskip -0.00in
     \caption{Dependence of chiral polarization point $\Lambda_T$ on the lattice
              spacing at fixed volume.}
     \vskip -0.1in 
     \label{fig:lambdaT-vs-a}
\end{center}
\end{figure}

Having performed the determination of $\Lambda_T$ for ensembles $E_1$--$E_4$, we show in 
Fig.~\ref{fig:lambdaT-vs-a} the dependence of $\Lambda_T$ on the lattice spacing. 
We emphasize that the ensembles in question have fixed physical volume. Our main 
interest here is to examine whether $\Lambda_T(V,a)$ has a finite continuum limit
at fixed $V$. In other words, whether there is a well--defined finite scale associated
polarization dynamics of QCD Dirac eigenmodes in finite volume. As one can see from 
the figure, our data clearly suggests that this is indeed the case. 
The fit to $c_1 + c_2 a^4$, with the coarsest ensemble $E_1$ excluded from the fit,
was again used to guide the eye. Based on our results, we propose the following for 
further investigation.

\medskip
\noindent {\em \underline{Proposition 3}\,(Chiral Polarization Scale):$\;$
           Consider pure glue SU(3) lattice gauge theory in finite physical 
           volume $V$. There exists a scale $\Lambda_T(V,a)$ in the spectrum of overlap Dirac 
           operator such that its eigenmodes are chirally polarized below $\Lambda_T(V,a)$
           and chirally anti-polarized above it. Moreover, the absolute $\Xg$--distribution
           changes from convex to concave and is thus uniform at the transition point.  
           The continuum limit $\lim_{a \to 0} \Lambda_T(V,a)$ exists and is non--zero.}
\medskip

\noindent It should be pointed out that the volume of our ensembles is rather small and 
it is not clear at this point whether the above {\em chiral polarization scale} remains
finite also in the infinite volume limit. We consider this an interesting question that 
should be studied in the future.

\section{Discussion and Conclusions}

Polarization tendencies, whether observed empirically in natural phenomena or studied 
theoretically in various model systems, are frequently of interest in physics. 
In fact, they are arguably the most basic attributes of dynamical behavior one can 
attempt to describe. For this reason they are also of interest for 
the {\em bottom--up} program of studying QCD vacuum structure~\cite{Hor06B}. Indeed, 
the major aspect of this approach is to systematically characterize the properties of 
the vacuum in a model--independent manner.

Polarization is a highly ``coarse--grained'' concept and, as such, offers a variety 
of descriptions that may be consistent with its intuitive meaning. However, such 
plurality is not necessarily a sign of redundancy. Rather, various descriptions may 
reflect large freedom in possible aspects of polarization that one could be interested 
in exploring. Once the intent of the description is meaningfully specified, suitable
quantifiers can emerge in a reasonably unique manner.
In this work we discussed polarization properties from a different angle. Indeed, rather 
than looking for description tailored for a specific physical context, our goal was
to construct a differential characteristic that is meaningful regardless of what 
the dynamical variable in question represents. As such, it could then be viewed as an 
attribute of the dynamics alone, a {\em dynamical polarization characteristic}.

To achieve the above, we considered a large set of descriptors consistent with
the qualitative notion of polarization, namely all possible $\Xg$--distributions labeled
by various polarization functions. If one imagines a rule of assignment 
associating a suitable description with any given physical context, then the set of 
$\Xg$--distributions would be the range of such map: a proper description 
is always found somewhere within this set. However, given our goals, one should not 
view different polarization functions as representing all possible physical contexts,
but rather as different ``reference frames'' in which any given dynamics can be looked 
at in terms of polarization. The $\Xg$--distribution associated with such a frame is 
kinematical in the same sense as velocity $\vec{v}$ in given inertial frame is 
kinematical for mechanical particle: it can be made arbitrary by choosing the reference 
frame appropriately. The necessary condition for a quantitative measure to be dynamical 
is its invariance under the change of reference frame, as exemplified by $d \vec{v}/d t$ 
and its Galilean invariance in case of Newtonian particle. One of our main points 
in this work is that, for situations involving polarization, the analog 
of acceleration is the absolute $\Xg$--distribution with its reparametrization invariance.

Similarly to the case of mechanical particle where one could consider e.g.\ 
higher time derivatives, it is possible to construct other $\Xg$--distributions 
with reparametrization invariance. Thus, frame independence is not sufficient to fix 
the dynamical polarization characteristic uniquely. The needed additional requirement 
we used was that the characteristic be ``correlational''.  Indeed, as emphasized 
throughout this article, the absolute $\Xg$--distribution represents the differential 
comparison of polarization in given dynamics relative to the case of statistically 
independent components. If one performs this comparison on average rather than
differentially, the correlation coefficient of polarization is obtained. 
This coefficient can thus be viewed as descending from absolute $\Xg$--distribution 
via a coarse--graining procedure. It has a well-defined statistical meaning and provides 
a suitable measure for ranking of possible dynamics in terms of their dynamical 
polarization properties. 

\begin{figure}
\begin{center}
    \centerline{
    \hskip 0.3in
    \includegraphics[width=9.3truecm,angle=0]{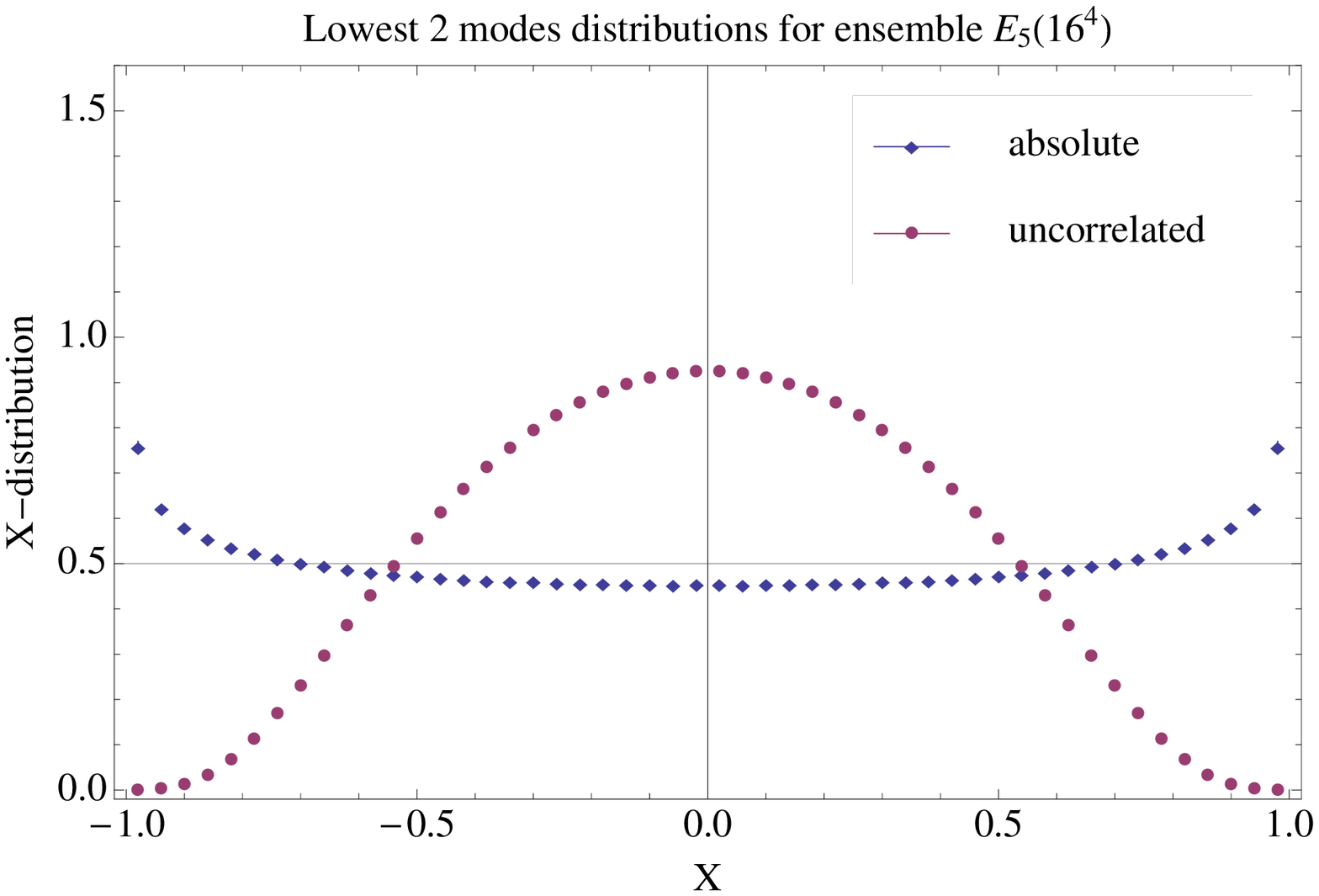}
    \hskip -0.4in
    \includegraphics[width=9.3truecm,angle=0]{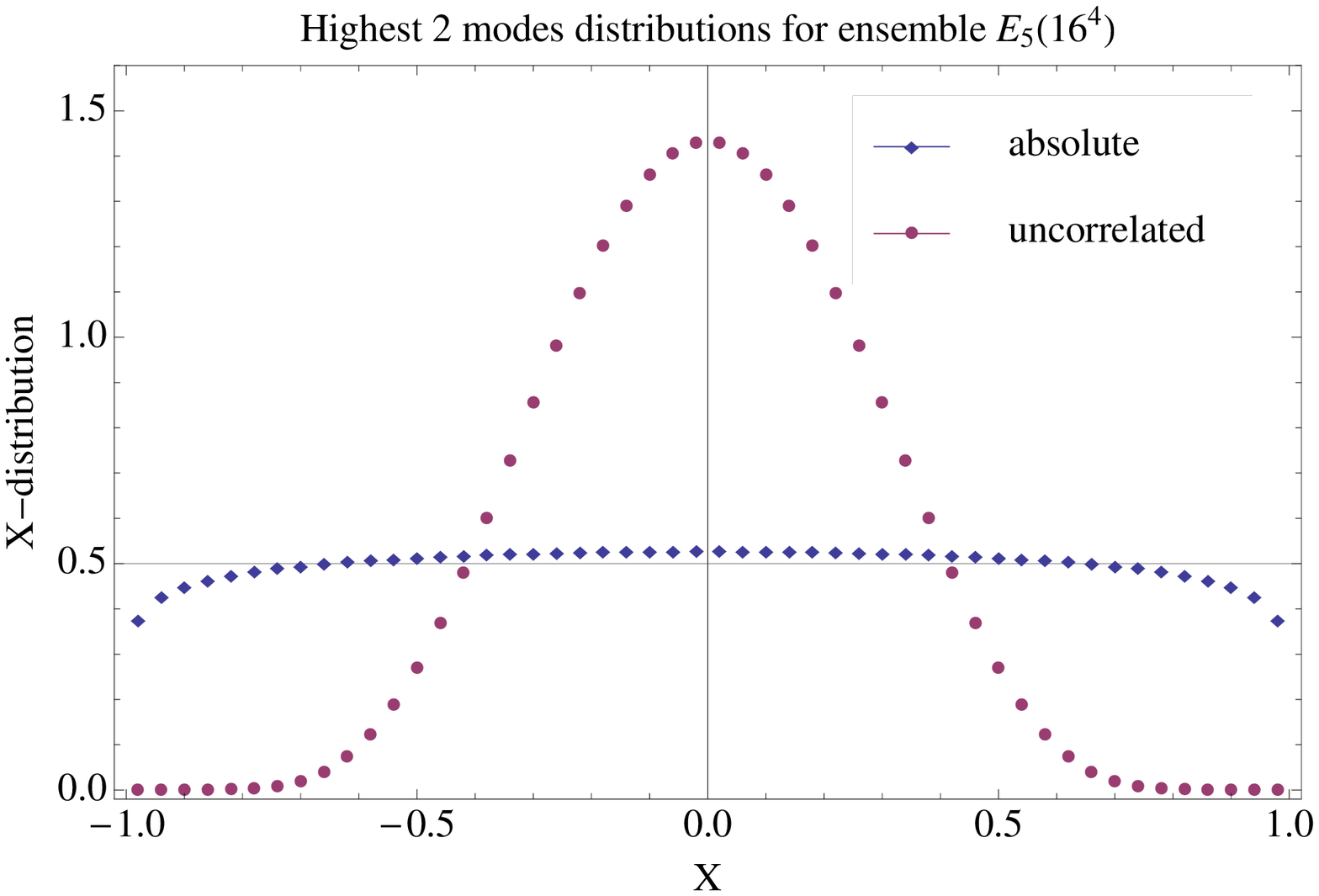}     
     }
     \vskip -0.00in
     \caption{Absolute (dynamical) and uncorrelated (kinematical) $\Xg$--distributions
     for gauge ensemble $E_5$ using the reference polarization coordinate for the latter.
     Results for lowest two non--zero pairs are shown on the left
     while the highest two pairs are shown on the right.}
     \vskip -0.1in 
     \label{fig:E5}
\end{center}
\end{figure}

Somewhat different perspective on our proposal is obtained when one views it 
as a reduction process wherein the full information on orthogonally decomposed 
variable $Q\equiv (Q_1,Q_2)$, described by the density function $\df(Q)$, is being 
restricted into the information relevant for describing its polarization properties
only. Integrating out unrelated 
information naturally leads to the notion of polarization dynamics or, equivalently, 
an $\Xg$--distribution $\xd(\Xg)$, but the associated non--uniqueness issues  
lead to the conclusion that this reduction is in fact insufficient. One should rather
consider the reduction to a pair
\begin{equation}
   \df(Q) \,\longrightarrow \, \db(q_1,q_2) \,\longrightarrow \,
   \xd(\Xg) \,,\, \xd^u(\Xg)
    \label{eq:405}
\end{equation}
where $\xd^u(\Xg)$ descends from $\df^u(Q)$, namely the distribution of statistically 
independent components $Q_1$ and $Q_2$. While both members of the above pair vary
with the choice of the polarization coordinate, there is an invariant 
information in the ``difference'' of the two, and we thus propose that the polarization
should rather be characterized by 
\begin{equation}
   \df(Q) \,\longrightarrow \, \xd_A(\Xg) \,,\, \xd^u(\Xg)
    \label{eq:410}
\end{equation}
where the first part is invariant (dynamical) and the second part is variant 
(kinematical).\footnote{Note that, as opposed to pair in (\ref{eq:405}), the arguments 
of these two $\Xg$--distributions have different meanings in this case.} 
Note that the kinematical part is still a useful ingredient in the description of
polarization. Indeed, if one stores the information on $\df(Q)$ in the above form with
standard choice of the polarization coordinate, such as $\xd_A(\Xg) \,,\, \dop^u(\rpc)$
for reference polarization coordinate, then it is possible to recover all other 
$\Xg$--distributions one might need in various physical contexts that this dynamics
could be modelling. As an example of such practice, we show in Fig.~\ref{fig:E5} the 
polarization characteristics for gauge ensemble $E_5$ and the group of lowest two (left) 
and highest two (right) non--zero modes computed. The above conceptual aspects as well 
as the theoretical structure underlying the method are discussed at length in 
Appendix~\ref{sec:general}. 


Taking a QCD perspective, it is interesting to apply the above techniques to Dirac 
eigenmodes and investigate their local chiral structure. This is where the related issues first 
arose~\cite{Hor01A}. To this end, we obtained and presented the basic picture of dynamical 
polarization properties in low--lying overlap Dirac eigenmodes for pure glue QCD. 
The most intriguing aspect of our findings relates to emergence of new 
sharply defined feature in the Dirac spectrum, namely that 
of a {\em chiral polarization transition}. Indeed, our data
indicate that, in finite physical volume, there exists a scale $\Lambda_T$ in the Dirac 
spectrum such that QCD dynamics leads to enhancement of local chirality in modes 
at lower scales, while the same dynamics works to suppress it in modes above 
the transition. In substantiating this, we focused on continuum limit at fixed physical 
volume but this basic feature appears to be robust when the volume is increased, 
as indicated by the results for ensemble $E_5$ shown in Fig.~\ref{fig:E5}.

Even though it has long been suspected that chirality properties of low--lying Dirac 
modes are relevant in the mechanism of spontaneous chiral symmetry breaking, it was not 
known before that QCD dynamics leads to such a clear--cut qualitative 
distinction in the behavior of low and high parts of the Dirac spectrum.
The above result not only confirms the conceptual value of the approach to 
polarization that we developed here, but also offers what could be a fruitful avenue 
for ``bottom--up'' investigation of spontaneous chiral symmetry breaking in QCD. 
An immediate point of interest in this regard is whether the chiral polarization
scale $\Lambda_T$ remains finite in the infinite volume. Indeed, the existence 
of such chirality--related scale, generated by QCD dynamics, would open an interesting 
new class of possibilities for the mechanism of spontaneous chiral symmetry breaking.

\bigskip

\noindent
{\bf Acknowledgments:}
Ivan Horv\'ath is indebted to Scott Hamann for his support and patience that was 
important for this work to materialize. Discussions with Robert Mendris are also 
appreciated. Andrei Alexandru is supported in part by U.S. Department of Energy 
under grant DE-FG02-95ER-40907. Terrence Draper is supported in part by 
U.S. Department of Energy under grant DE-FG05-84ER40154.
The computational resources for this project 
were provided in part by the Center for Computational Sciences at the University
of Kentucky, and in part by the George Washington University IMPACT initiative.

\bigskip

\begin{appendix}

\section{General Description of the Method}
\label{sec:general}

In this section we describe the absolute $\Xg$--distribution method in detail.
Since the formalism is applicable in a fairly generic setting, we carry this 
discussion out without reference to a specific physical situation in question.
This helps to bring the essential ingredients more to the forefront.
Consider a quantity $Q$ taking values in linear space with scalar product.
Assume further that there is some physically motivated decomposition of $Q$ into
a fixed pair of orthogonal subspaces in place, namely   
\begin{equation}
   Q = Q_1 + Q_2    \qquad\qquad  Q_1 \,\cdot \, Q_2 = 0
   \label{eq:10}  
\end{equation}
where ``$\,\cdot\,$'' denotes the scalar product. What we implicitly have in mind
here is that the orthogonal subspaces in question are ``equivalent'' e.g.\ of 
the same dimension, but the resulting methods can be generalized to include
non--symmetric cases.

What we mean by specifying the ``dynamics'' of $Q$ is selecting an arbitrary 
probability space with allowed values of $Q$ as its sample space. For practical 
considerations (and manipulations) we will assume however the existence of 
an underlying probability distribution (function or generalized function) 
$\df(Q)$ from which the probability of measurable events can be obtained 
via corresponding integrations. It is not 
important for this discussion whether and how this distribution descended from 
dynamics of a more complex or fundamental system. For example, in applications 
related to studying space--time structure of composite fields in QCD (QCD vacuum 
structure) this probability distribution would have ultimately descended from 
probability density of gauge configurations given by QCD 
action.

Since our goal is to describe polarization properties with respect 
to (\ref{eq:10}), it is convenient to consider dynamics in terms
of the polarization components, i.e.\ $\df(Q_1,Q_2)$. In addition 
to dynamics, the method we will be describing is built on the notion of 
a polarization function $\Fg(Q_1,Q_2)$. 
The role of $\Fg$ is to quantify the degree of polarization for 
any sample $(Q_1,Q_2)$. One can think of it in terms of a ``polarization meter'' 
accepting a sample as an input and providing its polarization value as an output. 
Different devices of this type are characterized by different polarization 
functions.
In what follows, we will carry out our discussion in the situation where
the dynamics are symmetric, namely $\df(Q_1,Q_2) = \df(Q_2,Q_1)$, and 
the polarization meters are also symmetric, namely
$|{\Fg}(Q_1,Q_2)|= |{\Fg}(Q_2,Q_1)|$. This is most relevant for our targeted 
applications. However, we will keep the framework and notation wide enough 
to incorporate future generalizations in this direction.  

Finally, we will be quantifying the tendency for polarization by relative 
magnitudes of $Q_1$ and $Q_2$. For example, by saying that $Q$ is 
strongly polarized in the first direction we mean that $|Q_1| \gg |Q_2|$. 
Consequently, the probability distribution of component magnitudes, namely
\begin{equation}
   \db (q_1,q_2) \equiv \int dR_1 dR_2 \,\df(R_1,R_2)\, 
   \delta(q_1-|R_1|) \;\delta(q_2-|R_2|)
   \qquad\qquad q_i \equiv |Q_i|
   \label{eq:20}  
\end{equation}
will be the object characterizing ``dynamics'' for our purposes, with the 
subscript ``$b$'' in $\db$ indicating that this is the representation of 
dynamics in base sample space variables ($q_1,q_2$).

\subsection{Polarization Functions}

To start with the $\Xg$-distribution method, one needs to select a suitable 
function $\Fg(Q_1,Q_2)$ serving as a measure of polarization on the 
``sample space''. Such {\em polarization function} will thus reflect 
the relative contribution of the two orthogonal subspaces in $Q$. 
Due to the linear structure involved, the values of $\Fg$ assigned 
to $Q$ and its arbitrary rescaling $\alpha Q$ ($\alpha \ne 0$) should be 
identical, i.e.\ we consider $Q$ and $\alpha Q$ to be equally polarized samples.
Restricting ourselves to characteristics $\Fg(q_1,q_2)$ that only depend on 
magnitudes $(q_1,q_2)$ this translates into the requirement that $\Fg(q_1,q_2)$ 
only depends on $t\equiv q_2/q_1$. 

Note that, in terms of $t$, ``being smaller'' indicates being more polarized with 
respect to the first direction while ``being larger'' signals larger polarization 
with respect to the second direction. Thus, even a simple choice 
$\Fg(q_1,q_2)=q_2/q_1$ has some basic features we are seeking. However, we wish 
to work with characteristics that are symmetric with respect to 
the two subspaces, and have a standardized compact range. 
To achieve this, we impose the following set of requirements on acceptable 
functions $\Fg$.
\medskip

\noindent {\em \underline{Definition 1}\,(Polarization Functions):$\;$ Consider 
real--valued functions $\Fg(q_1,q_2)$ on $\R_0^+\times \R_0^+$ where
$\R_0^+\equiv [0,\infty)$. The subset of such functions satisfying 
the following conditions:
\smallskip

\noindent $\,$ (a) $\; \Fg(q, t q)$ is non--decreasing function of $t \in \R_0^+$
                   for fixed $q>0$
\smallskip

\noindent $\,$ (b) $\; \Fg(0,q)\,=\,1  \quad$ for $\quad q>0$
\smallskip

\noindent $\,$ (c) $\; \Fg(q,t q)$ is independent of $q>0$ for fixed $t \in \R_0^+$
\smallskip

\noindent $\,$ (d) $\; \Fg(q_1,q_2)\,=\,-\Fg(q_2,q_1)$ 
\smallskip

\noindent will be referred to as a set of polarization functions. 
}

\medskip
\noindent Note that, with any polarization function, samples $(q_1,q_2)$ strictly 
polarized in the first direction $(q_1 > 0,q_2=0)$, strictly unpolarized 
$(q_1=q_2)$, and strictly polarized in the second direction $(q_1=0,q_2>0)$ 
are characterized by functional values $-1$, $0$ and $+1$ respectively. 
$\Fg$ serves as a comparator in a sense
that we say that sample $(q_1,q_2)$ is more polarized than sample $(r_1,r_2)$
if $|\Fg(q_1,q_2)| > |\Fg(r_1,r_2)|$. Similarly, given our conventions,
we consider sample $(q_1,q_2)$ more polarized in the first direction
than sample $(r_1,r_2)$ if $\Fg(q_1,q_2) < \Fg(r_1,r_2)$ while we consider
it more polarized in the second direction if $\Fg(q_1,q_2) > \Fg(r_1,r_2)$. 
There are obviously infinitely many such comparators (polarization functions).

Every polarization function vanishes at origin $(0,0)$ and is otherwise 
specified by the choice of non--decreasing 
$\Fg(q_2/q_1) \equiv \Fg(t)$ such that $\Fg(t)=-\Fg(1/t)$ and 
$\Fg(0)=-1$. Variable $t$ is in one to one correspondence with the polar 
angle $\varphi \equiv \tan^{-1}(t)$ in Cartesian plane, and it can thus be
useful to parametrize $\R_0^+ \times \R_0^+$ using polar coordinates 
$(\rho,\varphi)$. In this case condition {\em (c)} translates into 
$\Fg(\rho,\varphi) = \Fg(\varphi)$. Further simplifications arise by
``symmetrizing'' the angular variable $\varphi \in [0,\pi/2]$ 
via rescaled polar angle $\rpc$ of Eq.~(\ref{eq:n20}). 
In this case the remaining conditions on non--decreasing $\Fg(\rpc)$ translate 
into $\Fg(\rpc)=-\Fg(-\rpc)$ and $\Fg(1)=1$. In other words, the set of
polarization functions can be faithfully parametrized by odd, non--decreasing 
functions on $[-1,1]$ with unit maximal value. 

When viewed as functions of $\rpc$, then simple examples of polarization functions 
are odd integer powers and their generalizations to positive real powers, namely
\begin{equation}
    \Fg^C(\rpc;\alpha) \,=\, \sgn(\rpc) \; |\rpc|^{\alpha} \qquad\; \alpha > 0
    \label{eq:40}
\end{equation}
Note that $\Fg^C(\rpc;1)=\rpc$ corresponds to the original chiral orientation 
parameter of Ref.~\cite{Hor01A}. 
When viewed as functions of $t$, then another simple family
of polarization functions is specified by
\begin{equation}
    \Fg^R(t;\alpha) \,=\, \frac{t^\alpha - 1}{t^\alpha + 1}  
    \qquad\; \alpha > 0
    \label{eq:50}
\end{equation}
The special case of $\Fg^R(t;2)$ was considered in~\cite{Blu01A}. 
Another possibility is to work with
\begin{equation}
    \Fg^G(t;\alpha) \,=\, \frac{4}{\pi} \tan^{-1}(t^\alpha) - 1  
    \qquad\; \alpha > 0
    \label{eq:60}
\end{equation}
where $\alpha=1$ is the chiral orientation parameter of Ref.~\cite{Hor01A} 
and $\alpha=2$ was used in~\cite{Gat02A}.

\subsubsection{Set $\Fset$}

Let us denote the set of all odd non--decreasing real--valued  
functions on $[-1,1]$ with unit maximal value as $\Fset_0$. Each polarization
function is labeled by an element of $\Fset_0$ and, in fact, we will write
$\Fg \in \Fset_0$ regardless of whether we think of $\Fg$ in terms of 
a single variable $\Fg=\Fg(\rpc)$, or in terms of two variables 
$\Fg=\Fg(q_1,q_2)$. 

Before proceeding, it is useful to make a subtle distinction that will be 
needed later. Since functions in $\Fset_0$ are bounded and 
non--decreasing, they are continuous except possibly on a countable set 
of points where they have finite jumps. We will group the functions 
that only differ by values at points of discontinuity and treat them 
as a single measure for polarization. To describe the resulting equivalence 
classes as a new function set, we represent each class by a function
that is middle--valued for $|x|<1$.\footnote{Function $f$ will be 
referred to as middle--valued at $x$ if 
$f(x) = \lim_{\delta \to 0^+} (f(x+\delta) + f(x-\delta))/2$. Note that if
$f$ is continuous at $x$ then it is also middle--valued. At points of 
discontinuity with finite jumps, middle--valued function takes a value
in the middle of the jump.} We then have the following representation of 
polarization functions.

\medskip
\noindent {\em \underline{Definition 2}\,(Set $\Fset \,$):$\;$ 
 Set $\Fset$ is a subset of real--valued functions $f(x)$ on $[-1,1]$ 
 satisfying the following conditions:
 \smallskip

 \noindent $\,$ (1) $\; f(x)$ is non--decreasing 
 \smallskip

 \noindent $\,$ (2) $\; f(1)\,=\,1$ 
 \smallskip

 \noindent $\,$ (3) $\; f(x)$ is middle--valued at $|x|<1$
 \smallskip

 \noindent $\,$ (4) $\; f(-x) = -f(x)$
}

\medskip
\noindent It should be emphasized that even though $\Fset \subset \Fset_0$, 
it implies a very mild restriction on the set of polarization functions.
Indeed, for every element of $\Fset_0$ there is an element in $\Fset$ which
differs from it at most on a countable set in domain $[-1,1]$. The rationale
for this restriction is that polarization functions represented by elements
of $\Fset$ are in one to one correspondence with possible polarization 
dynamics as will be discussed shortly.

\subsection{Polarization Coordinates and X--Distributions}

When studying the polarization properties of dynamics $\db(q_1,q_2)$
using polarization function $\Fg(q_1,q_2)$, it is convenient
to express probability dependencies directly in terms of $\Fg$.
In particular, we define the following associated objects
\begin{eqnarray}
    \pr(\Xg,\rho) & = & 
      \int_0^\infty d q_1 \int_0^\infty d q_2 \, \db(q_1,q_2) \; 
      \delta\Bigl( \Xg - \Fg(q_1,q_2)\Bigr) \,
      \delta\Bigl( \rho - \sqrt{q_1^2+q_2^2} \, \Bigr)  
      \label{eq:73} \\
    \xd(\Xg) & = &  
      \int_0^\infty d q_1 \int_0^\infty d q_2 \,
      \db(q_1,q_2) \; \delta\Bigl( \Xg - \Fg(q_1,q_2) \Bigr)
      \label{eq:74} \\
    \plp & = &
      \int_0^\infty d q_1 \int_0^\infty d q_2 \,
      \db(q_1,q_2) \; |\, \Fg(q_1,q_2) \,|     
    \label{eq:75}
\end{eqnarray}
where $\Xg$ denotes a generic independent variable parametrizing 
the range of polarization functions, i.e.\ $\Xg \in [-1,1]$, and 
$\rho \in \R_0^+$ is the magnitude of sample $Q$. In this way, every 
polarization function $\Fg(q_1,q_2) \in \Fset$ defines a corresponding
``sample space coordinate'' $\Xg$ which will be referred to as 
the {\em polarization coordinate}. We emphasize that the constructs 
$\pr$,$\xd$ and $\plp$ depend both on the dynamics $\db$ and 
the polarization function $\Fg$. They can thus also be viewed as 
images of maps $\pr=\hat{\pr}[\db,\Fg]$, $\xd=\hat{\xd}[\db,\Fg]$ and 
$\plp = \hat{\plp}[\db,\Fg]$.

By construction, $\pr(\Xg,\rho)$ carries the most detailed information 
about the dynamics $\db(q_1,q_2)$ with respect to polarization properties 
defined by $\Fg$, and will be referred to as the {\em polarization 
representation}. The other two objects are related to it via the following 
``descent'' relations
\begin{equation}
   \pr(\Xg,\rho) \; \longrightarrow \;   
    \int_0^\infty d \rho \,\pr(\Xg,\rho)
    \,\equiv \, \xd(\Xg) \;\longrightarrow \;
    \int_{-1}^{1} d \Xg \, |\Xg| \, \xd(\Xg)
    \,\equiv \, \plp 
    \label{eq:77}
\end{equation}
i.e.\ $\xd$ is a marginal distribution of $\pr$ and $\plp$ is a moment 
of $\xd$ (and of $\pr$). In $\xd(\Xg)$ the information on dynamics 
unrelated to polarization is integrated out which corresponds
to the {\em $\Xg$--distribution} of Ref.~\cite{Hor01A} generalized to arbitrary 
$\Fg$. $\Xg$--distributions are central objects of interest in this work. 
The role of the averaged polarization characteristic $0 \le \plp \le 1$ will 
become clear in the upcoming sections.

If $\Fg(x) \in \Fset$ is a one to one map of $[-1,1]$ onto $[-1,1]$, thus 
entailing a faithful polarization coordinate on the sample space, then 
the dynamics specified by $\pr(\Xg,\rho)$ is fully equivalent to 
original $\db(q_1,q_2)$. The corresponding subset $\Fset^{f} \subset \Fset$ 
is comprised of odd, continuous, strictly increasing functions $\Fg(x)$ 
on $[-1,1]$ with unit maximal value. Moreover, utilizing such equivalent 
parametrizations is particularly efficient if they have a differentiable 
relationship to base coordinates $(q_1,q_2)$, since then the standard 
analytic manipulations in the underlying integrals can be performed. 
Thus, we also consider the subset $\Fset^{fd}$ containing those elements 
of $\Fset^{f}$ that are differentiable on their domain, 
i.e.\ $\Fset^{fd} \subset \Fset^{f} \subset \Fset$. 

Given the utility of polarization coordinates for our purposes, it is 
convenient to fix a reference polarization representation of dynamics 
$\db(q_1,q_2)$. 
Since ${\rpc}=\Fg(\rpc) \in \Fset^{fd}$, the rescaled polar coordinate 
(\ref{eq:n20}) is particularly suited to play this 
role.\footnote{There are two technical reasons to choose $\rpc$. First, 
the change of variables from base pair $(q_1,q_2)$ to polarization pair 
$(\rpc,\rho)$ doesn't involve $\rpc$--dependence of the Jacobian. 
Secondly, the specification of the set of polarization functions in terms 
of $\rpc$ is simple.} 
Changing variables from $(q_1,q_2)$ to $(\rpc,\rho)$, the associated 
{\em reference} polarization representation $\dr(\rpc,\rho)$ is explicitly 
specified by
\begin{equation}
  1 = \int_0^\infty d q_1 \int_0^\infty d q_2 \, \db(q_1,q_2) 
     =  \int_{-1}^{1} d\rpc \int_0^\infty d \rho \,
        \underbrace{  \frac{\pi}{4} \, \rho \, 
        \db\Bigl( q_1(\rpc,\rho),q_2(\rpc,\rho) 
           \Bigr)}_{\textstyle \dr(\rpc,\rho)} 
    \label{eq:85}
\end{equation}
with $q_1(\rpc,\rho) = \rho\, \cos\, (\frac{\pi}{4}(x + 1))$, 
$q_2(\rpc,\rho) = \rho\, \sin\, (\frac{\pi}{4}(x + 1))$. In terms of this
representation, equations (\ref{eq:73}-\ref{eq:75}) simplify to
\begin{eqnarray}
    \pr(\Xg,\rho) & = & 
      \int_{-1}^{1} \, d x \, \dr(\rpc,\rho) \; 
      \delta\Bigl( \Xg - \Fg(x)\Bigr ) 
      \label{eq:73a} \\
    \xd(\Xg) & = &  
      \int_{-1}^{1} \, d x \, \dop(\rpc) \; 
      \delta\Bigl( \Xg - \Fg(x) \Bigr )
      \label{eq:74a} \\
    \plp & = &
      \int_{-1}^{1} \, d x \, \dop(\rpc) \; 
      |\,\Fg(x) \,|
      \label{eq:75a}
\end{eqnarray}
where the restricted dynamics $\dop(\rpc) = \int_0^\infty d \rho \, \dr(\rpc,\rho)$ 
focuses on polarization only, and will thus be referred to as the reference
{\em polarization dynamics}. Note that, at the same time, $\dop$ can be thought of 
as the reference $\Xg$--distribution. 

To summarize, we think of all dynamics as being parametrized via 
distributions $\db(q_1,q_2)$ in base coordinates. However, when interested in 
characterizing polarization properties of $\db$, it is convenient to parametrize 
the same set of dynamics in the coordinate system that directly reflects 
the polarization of the sample, which is the role of $\dr(\rpc,\rho)$. 
Integrating out the coordinate orthogonal to reference polarization 
coordinate $\rpc$, we obtain a restricted dynamics $\dop(\rpc)$ describing 
polarization only, and we thus view $\dop$ as parametrizing possibilities
for polarization dynamics. While it is tempting to take $\dop(\rpc)$ to be a 
detailed polarization characteristic assigned to dynamics $\db$, the choice of 
$\rpc$ as a reference polarization coordinate is clearly arbitrary, and we could 
reparametrize using any polarization function $\Fg(\rpc) \in \Fset$.
Had we done so, $\dr(\rpc,\rho)$ would have been replaced by 
another polarization representation $\pr(\Xg,\rho)$, $\dop(\rpc)$
would have been replaced by $X$--distribution $\xd(\Xg)$, and the reference 
value of $\da$ would have changed to $\plp$. Thus, unlike $\dr$, $\dop$ and 
$\da$ which are assigned to a given dynamics, we view $\pr$, $\xd$ and $\plp$ as 
``floating objects'' reflecting dependence on $\Fg$ as well.

\subsection{Polarization Dynamics}
\label{subsec:pol_dynamics}

As emphasized before, the objects of our interest are various $\Xg$--distributions
$\xd$ that we can assign to given dynamics $\db$, since they carry the information 
on polarization properties. Given that the polarization dynamics $\dop$ associated
with $\db$ is itself an $\Xg$--distribution, it is not surprising that, 
as seen from Eq.~(\ref{eq:74a}), the full information contained in $\db(q_1,q_2)$ 
is not needed to study its $\Xg$--distributions. Rather, the polarization 
dynamics $\dop(\rpc)$ represents all we need. We can thus write 
$\xd = \hat{\xd}[\dop,\Fg]$ instead of $\xd = \hat{\xd}[\db,\Fg]$. 

Therefore, our analysis in what follows will focus on studying interrelations 
among the members of the triple $(\dop,\Fg,\xd)$. To do that, it is desirable
to be more specific about defining the set of possibilities for polarization 
dynamics. In particular, we would like to represent them in terms of 
a function set so that their description is on par with the representation
of polarization functions via set $\Fset$. In that regard, it is convenient
to use the information stored in $\dop(\rpc)$ in the form of the associated 
cumulative probability function
\begin{equation}
  S_r(\rpc) \equiv \int_{-1}^\rpc d y \, \dop(y)   
  \label{eq:93.10}
\end{equation}
which assigns probability to events $[-1,x]$. We will then consider 
polarization dynamics represented by $S_r \in \dopset$ where the function 
set $\dopset$ is defined as follows.
\medskip

\noindent {\em \underline{Definition 3}\,(Set $\dopset \,$):$\;$ 
 Set $\dopset$ is a subset of real--valued functions $f(x)$ on $[-1,1]$ 
 satisfying the following conditions
 \smallskip

 \noindent $\,$ (1) $\; f(x)$ is non--decreasing 
 \smallskip

 \noindent $\,$ (2) $\; f(1)\,=\,1$ 
 \smallskip

 \noindent $\,$ (3) $\; f(x)$ is right continuous at all $x<1$
 \smallskip

 \noindent $\,$ (4) $\; f(x) + f(-x) \,=\,1 + \Delta(x) \qquad 
              \quad x>-1 \,, \quad
              \Delta(x) \equiv 
              f(x) - \lim_{\delta \to 0} f(x-\delta) \,\ge\, 0$
}

\medskip
\noindent Condition {\em (4)} above expresses the symmetry of the underlying
dynamics in base variables. The function $\Delta(\rpc)$ detects 
``atoms'' in probability space associated with $f(x)$.
In particular, it represents the probability of a point--like event $\{x\}$, which 
is zero everywhere except possibly on a countable set of points (atoms) where it 
is positive. 

It should be noted that the description in terms of $S_r$ is in fact more general
than the description in terms of $\dop$ in the sense that the density 
representation doesn't exist for all possible cumulative probability functions. 
Probabilistic spaces with atoms have a discrete part that can be described via 
distributions ($\delta$--functions), but even in the case when $S_r$ is continuous,
it can be non--differentiable on a set of (Lebesgue) measure 
zero.\footnote{Continuous functions that are not monotonic can be non--differentiable
everywhere as exemplified by the famous Weierstrass's function. Monotonicity however 
greatly reduces options for such exotic behavior.} In some of such cases a 
{\em complete} description of probability space in terms of density simply doesn't 
exist. Nevertheless, we will interchangeably use both languages since we are not
aware of a situation where this distinction would be relevant physically. 
More importantly, all the constructs considered here can be defined solely in terms 
of $S_r$ for all elements of $\dopset$.~\footnote{This includes moments, e.g.\ 
$\plp$, as follows from properties of Stieltjes's extension to the Riemann integral.} 
Thus the notion of density doesn't even have to be invoked in principle. However, 
its convenience and physical utility certainly warrant its use.

\subsubsection{Polarization Functions -- Polarization Dynamics Correspondence}
\label{subsubsec:correspondence}

For considerations that will follow, it will be useful to single out polarization 
dynamics for which $S_r(\rpc)$ is a one to one map of $[-1,1]$ onto $[0,1]$. Such 
subset $\dopset^f \subset \dopset$ consists simply of continuous, strictly 
increasing functions with unit maximal value and satisfying $S_r(-x) + S_r(x) =1$.
Thus, a simple transformation
\begin{equation}
     S_r(\rpc)   \quad \longrightarrow \quad \Fg_r(\rpc) 
                  \equiv 2\,S_r(\rpc) - 1    
     \label{eq:94.40}
\end{equation} 
makes it into odd, continuous, strictly increasing function on $[-1,1]$ with 
unit maximal value, namely an element of $\Fset^f$. The construction obviously works
also in the reverse direction and (\ref{eq:94.40}) defines a natural correspondence 
between polarization dynamics and polarization functions -- a bijection between 
$\dopset^f$ and $\Fset^f$.

Polarization dynamics from $\dopset^f$ do not contain atoms or events 
$[\rpc_1,\rpc_2>\rpc_1]$ with zero assigned probability. 
The latter can be relaxed while the above correspondence still holds unchanged.
Formally, it is then a bijection between larger sets $\Fset^c$ ($\dopset^c$) of 
continuous polarization functions (dynamics). However, the extension to cases with 
atoms requires a minor modification. In particular, the problem with (\ref{eq:94.40}) 
then is that the image $\Fg_r(\rpc)$ is not odd at points of discontinuity due 
to property {\em (3)} of $S_r(\rpc)$, and thus not an element of $\Fset$. 
This is rectified by replacing the map (\ref{eq:94.40}) with  
\begin{equation}
     S_r(\rpc)   \quad \longrightarrow \quad \Fg_r(\rpc) 
                  \equiv \cases{
      \;  2\,S_r(\rpc) - 1 - \Delta(\rpc) \; &  \mbox{\rm for $|\rpc| < 1$} \cr
      \;  -1 &  \mbox{\rm for $\rpc=-1$} \cr
      \;  +1 &  \mbox{\rm for $\rpc=1$} \cr
       }   
     \label{eq:94.50}
\end{equation} 
which only differs from (\ref{eq:94.40}) for $S_r$ that are not elements 
of $\dopset^c$. Map (\ref{eq:94.50}) in fact defines a bijection between 
$\Fset$ and $\dopset$, and we can conclude that the freedom in choice of 
the polarization functions is as large as possibilities for polarization dynamics 
itself.

\subsection{Arbitrariness of Polarization Characteristics}
\label{subsec:arbitrary}

Viewing the constructs $\pr$, $\xd$ and $\plp$ as tools for characterizing 
the polarization properties of dynamics $\db$ with decreasing
level of detail, one has to ask how meaningful these characteristics can be 
since they depend on the choice of the polarization function. In this 
subsection we will clarify the extent of this arbitrariness in case of
$\Xg$--distributions. Specifically, fixing the polarization dynamics via 
$\dop(\rpc)$ (or $S_r(\rpc)$), we ask how much can $\Xg$--distributions 
$\xd(\Xg)$ (or $S(\Xg)$) change as we scan the set $\Fset$ of polarization 
functions.

\begin{figure}
\begin{center}
    \centerline{
    \hskip 0.00in
    \includegraphics[width=8.3truecm,angle=0]{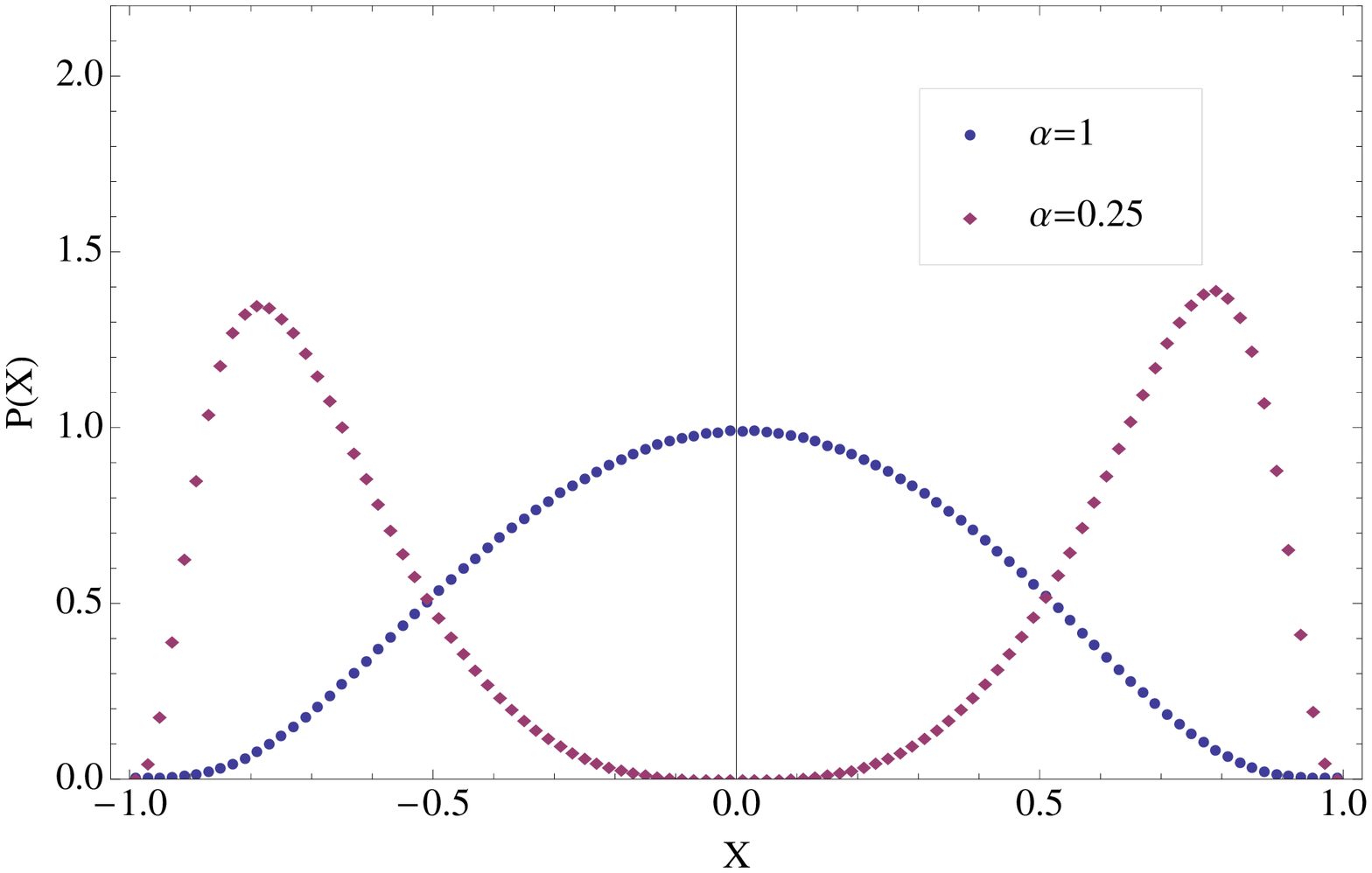}
    \hskip -0.0in
    \includegraphics[width=8.3truecm,angle=0]{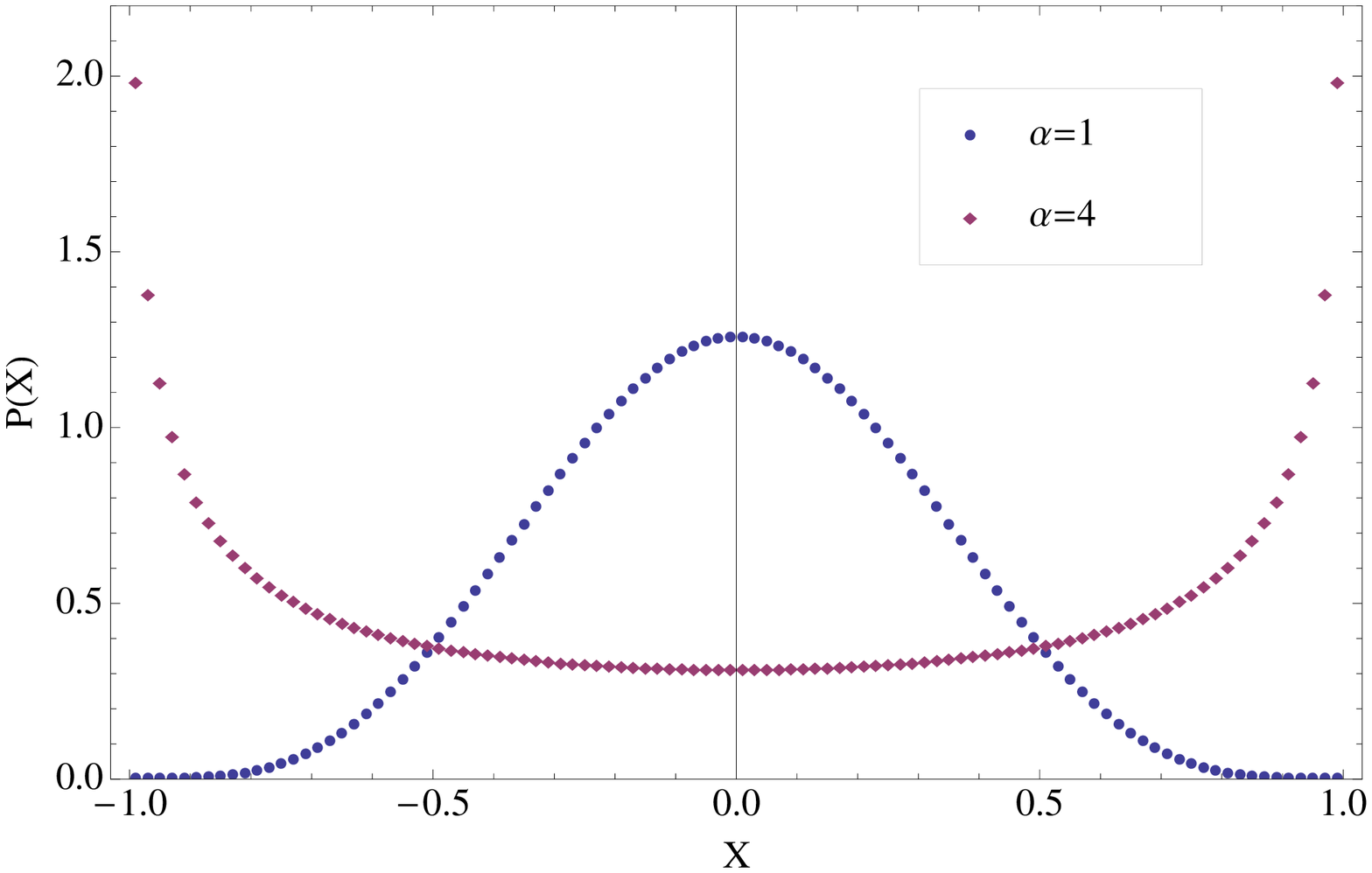}     
     }
     \vskip -0.00in
     \caption{Possible $\Xg$--distributions associated with fixed dynamics 
     can have qualitatively different behavior. In this example from lattice 
     QCD we selected two polarization functions from family $\Fg^C$ (left), 
     and two from family $\Fg^R$ (right).}
     \vskip -0.1in 
     \label{fig:arbitrary}
\end{center}
\end{figure}

To start with a relevant example, we show in Fig.~\ref{fig:arbitrary} several 
$\Xg$--distributions associated with dynamics induced by pure--glue 
lattice QCD ensemble $E_4$ in two pairs of eigenmodes with lowest non--zero 
eigenvalues (see Table~\ref{tab:ensembles} in Sec.~\ref{sec:qcd}). 
The left--right
decomposition of Dirac bispinor defines the polarization components,
and the details of related calculations are given in Sec.~\ref{sec:qcd}.
In the left panel we selected two polarization functions from the family
$\Fg^C$ of Eq.~(\ref{eq:40}), with values of $\alpha$ as shown. 
One can see that the qualitative behavior of the $\Xg$--distribution 
differs dramatically for the two choices with $\alpha=1$ case suggesting 
a double--peaked (highly polarized) behavior while $\alpha=1/4$ case 
exhibiting the opposite (highly unpolarized) tendencies. On the right 
panel we show a similar example for functions $\Fg^R$ specified
by Eq.~(\ref{eq:50}).
 
To understand the variability of $\Xg$--distributions, it is easiest to 
vary polarization functions within $\Fg(\rpc) \in \Fset^{fd}$, 
i.e.\ to consider the subset of differentiable faithful polarization 
coordinates. Then we can make the definition (\ref{eq:74a}) of $\xd(\Xg)$ 
more explicit via analytic change of variable to obtain 
\begin{equation}
   \xd(\Xg) \,=\, 
      \frac{1}{ \Fg'\biggl(\Fg^{-1}(\Xg)\biggr) } \;
      \dop \biggl( \Fg^{-1}(\Xg)\biggr)
      \qquad\qquad \Fg \in \Fset^{fd}
  \label{eq:90}
\end{equation}
where $\Fg'(\rpc) \equiv d\Fg(\rpc)/d\rpc$. The form of this result explains 
why we can obtain wildly varied $\Xg$--distributions for $\dop(\rpc)$ by 
changing the polarization function $\Fg(\rpc)$. 

For more general analysis, it is convenient to express the underlying 
relations in terms of cumulative probability functions. In that case 
the defining condition for $\Xg$--distribution $S(\Xg)$ 
associated with $S_r(\rpc)$ and $\Fg(\rpc)$ is that it is the probability
of the event $\{\rpc \,|\, \Fg(\rpc) \le \Xg\}$ in the probability space
specified by $S_r(\rpc)$. To make this more explicit, notice that 
if $\Xg$ belongs to the range ${\eusm R}[\Fg]\subset [-1,1]$ of $\Fg$, then 
$\{\rpc \,|\, \Fg(\rpc) \le \Xg\} = [-1,\bar{\Fg}(\Xg)]$, a closed interval, 
while if $\Xg\not\in {\eusm R}[\Fg]$ then it is an open interval 
$[-1,\bar{\Fg}(\Xg))$. Here the generalized inverse
$\bar{\Fg}(\Xg)$ of polarization function $\Fg(\rpc)$ is defined via 
the relevant least upper bound, namely
\begin{equation}
   \bar{\Fg}(\Xg) \,\equiv\, \sup \{ \rpc \,|\, \Fg(\rpc) \le \Xg\}
   \label{eq:92}
\end{equation}
and it coincides with the standard inverse $\Fg^{-1}(\Xg)$ for $\Fg \in \Fset^f$.
We then have explicitly
\begin{equation}
     S(\Xg) \equiv \cases{
      \;  S_r\Bigl(\bar{\Fg}(\Xg)\Bigr)  \; &  
          \mbox{\rm if $\Xg \in {\eusm R}[\Fg]$} \cr
      \;  S_r\Bigl(\bar{\Fg}(\Xg)^-\Bigr)  \; &  
          \mbox{\rm if $\Xg \not\in {\eusm R}[\Fg]$} \cr
       }   
     \label{eq:92.10}
\end{equation} 
where $S_r(\rpc^-)$ is the left limit of $S_r$ at $\rpc$. The above expression 
assigns the $\Xg$--distribution $S(\Xg) \in \dopset$ to arbitrary 
$S_r \in \dopset$ and $\Fg \in \Fset$, and is the analog of equation 
(\ref{eq:74a}). Taking it as a starting point, one can straightforwardly 
obtain special cases of interest. For example, considering faithful polarization 
coordinates, i.e.\ $\Fg \in \Fset^f$, we have
\begin{equation}
   S(\Xg)  \,=\, S_r\Bigl(\Fg^{-1}(\Xg)\Bigr) \qquad
   \mbox{\rm or} \qquad 
   S_r(\rpc) \,=\, S\Bigl(\Fg(\rpc)\Bigr)  \qquad\qquad
   \Fg \in \Fset^f
   \label{eq:94.15}
\end{equation}
One use of the above relations is that they show, in an explicit manner, 
that fixing an invertible polarization function defines a one to one map of 
$\dopset$ onto itself, namely a bijection between polarization dynamics 
and $\Xg$--distributions. Let us 
also restrict the polarization dynamics to be from $\dopset^f \subset \dopset$ 
(see section \ref{subsubsec:correspondence}) for which the corresponding 
$S_r(\rpc)$ is a one to one map of $[-1,1]$ onto $[0,1]$, and thus invertible. 
From (\ref{eq:94.15}) it then follows that $S(\Xg)$ assigned to $\Fg \in \Fset^f$ 
is an element of $\dopset^f$, and thus also invertible. We then have
\begin{equation}
   \Fg(\rpc) \,=\, S^{-1}\Bigl(S_r(\rpc)\Bigr)
   \qquad\qquad   \Fg \in \Fset^f, \quad S_r \in \dopset^f  
   \label{eq:94.20}
\end{equation}
The above equation establishes that, given a fixed polarization dynamics
$S_r$ from $\dopset^f$, we can obtain an arbitrary $\Xg$--distribution
$S(\Xg)$ from $\dopset^f$ by adjusting the polarization function 
$\Fg \in \Fset^f$ according to (\ref{eq:94.20}). In other words,
scanning through all polarization functions in $\Fset^f$, the constructed
$\Xg$--distributions span the whole $\dopset^f$. One can continue further
and make this result even more general, but we will not go into additional 
details. The conclusion is that, for $S_r \in \dopset^f$, the arbitrariness 
in its $\Xg$--distributions is essentially as large as the full variety of 
polarization dynamics themselves.

The above analysis reveals the simplicity of the structure governing 
the interrelations in possible triples $(S_r,\Fg,S)$ when $S_r$ and $\Fg$ 
are faithful. In particular, the $\Xg$--distribution $S$ doesn't have to be 
thought of as a ``dependent variable'' determined by $S_r$ and $\Fg$, but
can be also viewed as an independent object from $\dopset^f$. In other
words, we have equivalent functional dependencies $S = \hat{S}[S_r,\Fg]$, 
$S_r = \hat{S_r}[S,\Fg]$, and $\Fg = \hat{\Fg}[S_r,S]$ whose explicit forms 
are given above. In what follows, we will implicitly assume the above
restriction, i.e.\ that polarization dynamics are chosen from $\dopset^f$, 
and that polarization functions are chosen from $\Fset^f$ so that 
the above structure readily holds. Moreover, when our discussion involves
probability densities ($\dop$ or $\xd$), the reader can simply assume that 
we are working in the framework of $\Fset^{fd}$ and $\dopset^{fd}$. However, 
we emphasize again that everything can be straightforwardly extended 
to the $\dopset^{f}$, $\Fset^{f}$ combination in that case. Most situations 
of practical interest are covered in this way but when the departure from 
$\dopset^f$, $\Fset^f$ is necessary, we will explicitly point it out.

\subsection{Motivation for Relative Measures}
\label{subsec:rel_measures}

Considerations of the previous section suggest that, in order to meaningfully 
characterize the dynamics of polarization via $\Xg$--distributions, one should 
proceed in one or both of the following ways. (i) Fix the polarization function 
$\Fg$ using well-motivated physics considerations, i.e.\ select polarization 
function that has direct physical meaning. Whether this can be done depends on 
the nature of quantity $Q$ in question. (ii) Use only those aspects of 
$\Xg$--distributions that do not depend on the choice of the polarization 
function. We emphasize that (i) and (ii) really represent conceptually 
different approaches to the issue. The purpose of this paper is to construct 
universal characteristics of the dynamics in the spirit of the latter. 
The following considerations turn out to be relevant for achieving that goal.

The feature that puts all polarization functions $\Fg \in \Fset^f$ on the same 
footing is that they are equally good comparators of polarization on the sample 
space due to strict monotonicity. In other words, if sample $(q_1,q_2)$ was 
established to be more polarized than $(t_1,t_2)$ using polarization function 
$\Fg_0$, then this would hold also if any other $\Fg \in \Fset^f$ was used 
to relate them. 
Thus, any characteristic that may be defined via statistics of such comparisons 
will be invariant under the choice of the polarization function. To see a simple 
and useful construct of this type, let us compare the dynamics (distribution) 
$\db^{(1)}$ to dynamics $\db^{(2)}$ via the probability $\plp_R$ that a sample 
drawn from $\db^{(1)}$ is more polarized than a sample independently drawn 
from $\db^{(2)}$. While one can approach the determination of $\plp_R$ using 
different polarization functions, the answer will always be the same. Indeed, 
this invariance becomes obvious if one pictures evaluation of $\plp_R$ 
by producing simultaneous draws from $\db^{(1)}$ and $\db^{(2)}$, and recording 
the double--sequence of corresponding reference polarization 
coordinates $\{(\rpc^{(1)}_i,\rpc^{(2)}_i), i=1,2,\ldots\}$.
The relative frequency of encountering $|\rpc^{(1)}_i| > |\rpc^{(2)}_i|$ in this 
process determines $\plp_R$, and the result would not change had 
we tested whether $|\Fg(\rpc^{(1)}_i)| > |\Fg(\rpc^{(2)}_i)|$ instead. Obviously,
the information stored in $\dop^{(1)}$, $\dop^{(2)}$ is sufficient to carry out 
the above process and in terms of the associated $\Xg$--distributions we have 
formally
\begin{equation}
   \plp_R(\dop^{(1)},\dop^{(2)},\Fg) \,\equiv\, 
   \int_{|Y| < |\Xg|} d\Xg dY \, \xd(\Xg;\dop^{(1)},\Fg) \, 
                                 \xd(Y;  \dop^{(2)},\Fg) 
   \,=\, \plp_R(\dop^{(1)},\dop^{(2)})
   \label{eq:100}
\end{equation}
where we have made the dependence of various objects on the dynamics and 
the polarization function explicit everywhere. For obvious reasons,
$\plp_R$ will be referred to as the {\em probability of larger polarization} (PLP) 
in $\db^{(1)}$ relative to $\db^{(2)}$.

More detailed invariant objects can be defined using analogous reasoning. 
One example relevant for our purposes involves the probability $\bar{S}_R(s)$ 
that a sample drawn from $\db^{(1)}$ is more polarized in the first
direction than fraction $1-s$ ($0 \le s \le 1$) of the population drawn from
$\db^{(2)}$. This is invariant with respect to the choice of the polarization
function because we can compute $\bar{S}_R(s)$ through the following process. 
Selecting a reference polarization coordinate as our measure for polarization, 
we again generate a double--sequence of draws from $\db^{(1)}$ and $\db^{(2)}$. 
From the sequence of length $N$ of such draws, namely  
$\{(\rpc^{(1)}_i,\rpc^{(2)}_i), i=1,2,\ldots N\}$, 
we can obtain an estimate $\bar{S}_R(s,N)$ of $\bar{S}_R(s)$ by first sorting 
$\{\rpc^{(2)}_i\}$ in ascending order obtaining a sequence 
$\{ \rpc^{(2)}_{j(i)} \}$ of monotonically increasing samples, and setting 
$\rpc_0 \equiv \rpc^{(2)}_{j(i_0)}$, where 
$i_0$ is the smallest index such that  $i_0/N > s$. Note that $\rpc_0$ 
approximates the separation point where the population from $\db^{(2)}$ splits
into fraction $s$ of samples most polarized in the first direction, and the rest.
The estimate $\bar{S}_R(s,N)$ is then obtained by counting the relative frequency
of occurrences with $\rpc^{(1)}_i < \rpc_0$. 
It is clear that we could have used arbitrary $\Fg(\rpc) \in \Fset^f$ to make 
all the above comparisons with the same result. Thus, both $\bar{S}_R(s,N)$ and 
$\bar{S}_R(s) = \lim_{N \to \infty} \bar{S}_R(s,N)$ are invariant under the choice
of the polarization function for all $s$. To make the contact with 
$\Xg$--distributions, we can change the variable and transit from  $\bar{S}_R(s)$ 
to $S_R(X) \equiv \bar{S}_R(1/2 + X/2)$. This resulting object again only 
depends on the polarization dynamics involved, i.e.\ we have
$S_R(X;\dop^{(1)},\dop^{(2)})$. We will refer to it as 
{\em relative $\Xg$--distribution} of $\db^{(1)}$ relative to $\db^{(2)}$.
Note that, by construction, 
if  $\db^{(1)} = \db^{(2)}$, i.e.\ when comparing given dynamics to itself,
we have $\bar{S}(s)=s$ and $S_R(X)=1/2 + X/2$.

\subsection{Relative Measures and their Reparametrization Invariance}

Discussion of the previous section fully defines two new objects, namely 
the probability of larger polarization $\plp_R$ and the relative 
$\Xg$--distribution $S_R$, both assigned to a pair of polarization dynamics 
from $\dopset^f$. However, in order to connect them to previously defined 
$\plp$ and $S$, we need to convert the constructive definitions into explicit 
ones. Let us start by a closer look at equation (\ref{eq:100}) representing $\plp_R$
which can be written as
\begin{eqnarray}
   \plp_R(\dop^{(1)},\dop^{(2)})  & = & 
      \int_{-1}^{1} \, d \rpc \, \dop^{(1)}(\rpc) \, 
      \int_{-|\rpc|}^{|\rpc|} \, d y \, \dop^{(2)}(y) \;=\; 
      \int_{-1}^{1} \, d \rpc \, \dop^{(1)}(\rpc) \, 
      |\, 2 S_r^{(2)}(\rpc) - 1 \,| \;=\;
      \nonumber \\
      & = &  
      \int_{-1}^{1} \, d \rpc \, \dop^{(1)}(\rpc) \, |\,\Fg_r^{(2)}(\rpc) \,|
      \;=\; \plp(\dop^{(1)},\Fg_r^{(2)})
      \label{eq:105}
\end{eqnarray}
where $\Fg_r^{(2)}$ is the polarization function assigned to dynamics $S_r^{(2)}$
(or $\dop^{(2)}$) via natural map (\ref{eq:94.40}). Thus the PLP in $\dop^{(1)}$
relative to $\dop^{(2)}$ is the average polarization $\plp$ of $\dop^{(1)}$
in polarization function $\Fg_r^{(2)}$, as defined by Eq.~(\ref{eq:75a}).

To express the relations of the above type, it is convenient to use the 
``hat notation'' for functional prescriptions involving domains 
of maps (functions) themselves. For example, the functional prescriptions for maps 
$\hat{S} \,:\, \dopset \times \Fset^f \mapsto \dopset$ and
$\hat{\Fg} \,:\, \dopset^f \times \dopset^f \mapsto \Fset^f$,   
introduced via relations (\ref{eq:94.15}) and (\ref{eq:94.20}) respectively, 
are given by
\begin{equation}
   \hat{S}\,[\,S,\Fg\,] \,\equiv\, S \circ \Fg^{-1}  \qquad  \mbox{\rm and} \qquad
   \hat{\Fg}\,[\,S_{(1)},S_{(2)}\,] \,\equiv\, S_{(2)}^{-1} \circ S_{(1)} 
   \label{eq:109}
\end{equation}
where the arguments are to be thought of merely as generic elements of the function 
sets involved. It is further useful to express the natural map (\ref{eq:94.40}) as
\begin{equation}
    S \equiv \hat{S}_\star[\,\Fg\,] \,=\, S_\star \circ \Fg  
    \qquad  \mbox{where} \qquad   
    S_\star(\Xg) \equiv \frac{1 + \Xg}{2}   
    \label{eq:111}
\end{equation}
Note that $S_\star \in \dopset^f$ represents polarization dynamics associated  
with the uniform distribution $P_\star(X)=1/2$, and that $S_r = \hat{S}_\star[\,\Fgr\,]$. 
According to Eq.~(\ref{eq:105}) 
we then have, in terms of cumulative probability distributions, 
that $\plp_R = \hat{\plp}_R[\,S_r^{(1)}, S_r^{(2)}\,]$ with
\begin{equation}
   \hat{\plp}_R [\,S_{(1)}, S_{(2)}\,] \,\equiv\, 
   \hat{\plp} [\, S_{(1)}, S_\star^{-1} \circ S_{(2)} \,]
   \label{eq:113}
\end{equation}
Similarly, one can inspect that the construction of the relative $\Xg$--distribution
$S_R$ corresponds to setting $S_R = \hat{S}_R[\,S_r^{(1)}, S_r^{(2)}\,]$ with
\begin{equation}
   \hat{S}_R [\,S_{(1)}, S_{(2)}\,] \,\equiv\, 
   \hat{S} [\, S_{(1)}, S_\star^{-1} \circ S_{(2)} \,] \,=\, 
   S_{(1)} \circ S_{(2)}^{-1} \circ S_\star
   \label{eq:116}
\end{equation}
Notice that since 
$\hat{S}[\,S_r,\Fg_r \,] = \hat{S}[\,S_r,S_\star^{-1} \circ S_r\,] = S_\star$, 
the relative construct $\hat{S}_R[\,S_{(1)}, S_{(2)}\,]$ represents a familiar
$\Xg$--distribution for dynamics $S_{(1)}$ with respect to the polarization 
function for which the $\Xg$--distribution of $S_{(2)}$ is uniform. 

An important feature of the above definitions, discussed in the previous section, 
is that they are invariant under the choice of the polarization coordinate. In other 
words, had we chosen a polarization coordinate associated with polarization function 
$\Fg(\rpc)$ instead of our reference polarization coordinate $\rpc$, then the 
polarization dynamics described by $S_r^{(i)}$ would have been described by its 
$\Xg$--distribution $\hat{S}[\,S_r^{(i)},\Fg\,] = S_r^{(i)} \circ \Fg^{-1}$ 
instead. As is obvious from definition (\ref{eq:116}), the relative $\Xg$--distribution
$S_R$ is invariant under such reparametrization. Consequently, being a moment 
of associated $P_R$, PLP is invariant as well.

The basic property of PLP is that
\begin{equation}
    \hat{\plp}_R [\,S_{(1)}, S_{(2)}\,] \,=\,
     1 \,-\, \hat{\plp}_R [\,S_{(2)}, S_{(1)}\,]
   \label{eq:124}
\end{equation}
as is obvious from its statistical interpretation described
in Sec.~\ref{subsec:rel_measures}, and can be easily derived analytically following 
the definition given above. Similarly, one can immediately see from (\ref{eq:116})
that the relative $\Xg$--distribution satisfies
\begin{equation}
   \hat{S}_R [\,S_{(2)}, S_{(1)}\,] \,=\,
   S_\star \circ  \Bigl( \hat{S}_R [\,S_{(1)}, S_{(2)}\,] \Bigr)^{-1} \circ S_\star
   \label{eq:126}
\end{equation}
Thus, up to some necessary modifications by the $S_\star$ operation, 
$\hat{S}_R [\,S_{(2)}, S_{(1)}\,]$ and $\hat{S}_R [\,S_{(1)}, S_{(2)}\,]$ are inverses
of one another, as one can easily understand from the constructive definition of 
relative $\Xg$--distribution.
We emphasize again that the above relations hold for $S_{(i)} \in \dopset^f$.

\subsubsection{Implementation}

Since the definition of absolute $\Xg$--distribution, to be discussed in 
the next section, is based on the relative $\Xg$--distribution, it is worth 
commenting on the implementation of the latter. The procedure is straightforward 
if the polarization dynamics $S_{(1)}$, $S_{(2)}$ (or $P_{(1)}$, $P_{(2)}$) are 
explicitly given. Indeed, in that case one simply implements the evaluation of
\begin{equation}
   S_R(X) \,=\, S_{(1)}\Bigl(  S_{(2)}^{-1}\, 
                            (\, 1/2 + X/2\,) \Bigr)
   \label{eq:130}
\end{equation}
or in terms of probability density
\begin{equation}
   P_R(X) \,=\, \frac{1}{2} \, 
                \frac{P_{(1)}\Bigl(  S_{(2)}^{-1}\,(\, 1/2 + X/2\,) \Bigr)}
                     {P_{(2)}\Bigl(  S_{(2)}^{-1}\,(\, 1/2 + X/2\,) \Bigr)}
   \label{eq:132}
\end{equation}

However, the explicit forms of relevant distributions are frequently not known
in practice. Rather, the dynamics involved are only specified implicitly, 
e.g.~in terms of some reduction process applied to underlying more complex  
degrees of freedom. 
In such situations one typically has an access to the dynamics through
a finite probabilistic chain of samples known to be generated from
the associated distribution. One option to implement the relative 
$\Xg$--distribution in that case is to use the constructive 
definition given in Sec.~\ref{subsec:rel_measures}. An equivalent approach 
is to simply follow the original ``histogram method'' 
of Ref.~\cite{Hor01A}, and apply it to the probabilistic chain for $P_{(1)}$
and the polarization function 
\begin{equation}
   \Fg(\rpc) \equiv 2 \int_{-1}^\rpc d y \,P_{(2)}(y) - 1
   \label{eq:134}   
\end{equation}
The specifics of the implementation then boil down to the selection
of a procedure for approximating $\Fg(\rpc)$ from a finite probabilistic chain
for $P_{(2)}$.

\subsection{Absolute $\Xg$--Distribution}

The introduction of relative $\Xg$--distribution doesn't change the arbitrariness
of it as a tool characterizing polarization properties of given dynamics. 
Indeed, it just moves the arbitrariness in the choice of polarization function 
into the arbitrariness of the dynamics serving as a comparative standard. However, 
it does introduce a different angle into the problem that turns out to be useful. 
Indeed, contrary to the question 
``Which polarization function should be used?'' the question 
``Which dynamics should we compare to?'' does evoke a meaningful answer. 

The main idea is as follows. When seeking a characteristic describing a {\em dynamical 
tendency for polarization}, what we have in mind is the information on how the projection
component behaves in relation to the complementary one, rather than what does it do
``in isolation'', or independently of it. In other words, we wish to compare 
to the dynamics that produces identical answers for any questions involving components 
``in isolation'', but no effects attributable to the interplay between them. 
For given dynamics $\db$, such dynamics $\db^u$ is uniquely specified, and represents 
the behavior of statistically independent components with the same distributions
of components themselves (marginal distributions). Specifically, denoting the 
marginal distribution associated with $\db(q_1,q_2)$ as $p(q)$, i.e.
\begin{equation}
    p(q) \, \equiv \, \int_0^\infty d q_2 \, \db(q,q_2) \,=\,
                      \int_0^\infty d q_1 \, \db(q_1,q)
   \label{eq:140}
\end{equation}
the corresponding ``uncorrelated distribution'' $\db^u$ is given by
\begin{equation}
    \db^u(q_1,q_2) \,\equiv\, p(q_1)\, p(q_2) 
   \label{eq:150}
\end{equation}
Note that the superscript ``$u$'' will be used also in other related constructs, i.e.\ 
the polarization dynamics associated with $\db^u$ will be denoted as 
$\dop^u$ (or $S_r^u$). We then define the {\em absolute $\Xg$--distribution} associated
wit dynamics $\db$ (in its cumulative form) as
\begin{equation}
   S_A \,=\, \hat{S}_A[\,\db\,] \, \equiv \, \hat{S}_R\,[\,S_r,S_r^u \,]
   \,=\,  S_r \circ (S_r^u)^{-1} \circ S_*
   \label{eq:160}
\end{equation}
namely as $\Xg$--distribution of $S_r$ relative to $S_r^u$. There are several 
remarks regarding this definition that might be useful to emphasize.
\smallskip

\noindent {\em (i)} The feature that makes the above construction qualitatively 
different from the old $\Xg$--distribution approach is that it discards the notion
of preferred fixed polarization function. This is obvious from definition 
(\ref{eq:160}) corresponding to the choice of the polarization function
$\Fg(x) \,=\, 2 S_r^u(x) - 1$, which depends on the dynamics $\db$ itself. More
precisely, the associated polarization function will be the same for members of 
the equivalence class consisting of all dynamics with fixed marginal distribution 
in base variables. The members of the same class will be distinguished by their 
dynamical tendencies for polarization of course, namely by their absolute 
$\Xg$--distributions.
\smallskip

\noindent {\em (ii)} Note that the information in polarization dynamics $S_r$ is not 
sufficient for constructing the absolute $\Xg$--distribution for dynamics $\db$.
Indeed, the marginal distribution $p(q)$ (and thus $S_r^u$) cannot be reconstructed 
from $S_r$ alone. Thus, we cannot scale down from $\hat{S}_A[\db]$ to $\hat{S}_A[S_r]$.
This is related to point {\em (i)} in that for the old ``fixed polarization function''
approach this is always possible.
\smallskip

\noindent {\em (iii)} While the notion of absolute polarization function is not viable 
for characterizing dynamical tendency for polarization, the notion of absolute 
$\Xg$--distribution remains. Indeed, the construction (\ref{eq:160}) is invariant with 
respect to the choice of the polarization function to implement it. In other words,
we could have started with other polarization representation leading to $\Xg$--distributions
$S$ and $S^u$ for $\db$, but we would nevertheless end up with the same absolute 
$\Xg$--distribution $S_A$. At the technical level, this follows from reparametrization 
invariance of relative $\Xg$--distributions, and is made obvious by the discussion in 
subsection~\ref{subsec:rel_measures}.

\subsection{Correlation Coefficient of Polarization} 

The absolute PLP is defined in the same way as the absolute $\Xg$--distribution, namely
\begin{equation}
   \plp_A \,=\, \hat{\plp}_A[\,\db\,] \, \equiv \, \hat{\plp}_R\,[\,S_r,S_r^u \,]
   \label{eq:180}
\end{equation}
and can be computed as the moment of the latter (see Eq.~(\ref{eq:77})). Since $\plp_A$
is an average characteristic based on comparison to statistical independence, 
it can be thought of as a correlation measure specifically designed for polarization. 
To be used as such, we define the {\em correlation coefficient of polarization} (CCP) via
\begin{equation}
   \cop_A \,\equiv\, 2 \plp_A - 1 \,=\, S_\star^{-1}(\plp_A)    
   \label{eq:200}
\end{equation}
Note that $\cop_A \in [-1,1]$. The positive correlation ($\plp_A > 1/2$) signals 
that dynamics is acting to enhance polarization, while negative correlation 
($\plp_A < 1/2$) indicates that dynamics works to suppress it. Due to its 
invariant nature, it quantifies an inherent property of $\db$, namely its 
overall dynamical tendency for polarization.

A natural question to ask in this regard is why the standard Pearson correlation
coefficient is not suitable for this purpose. This is easy to see. Consider 
the dynamics $\db(q_1,q_2)$ with support only on line $q_1=q_2$. While the Pearson
correlation will be perfect ($r=1$) in this case, it is obvious that there is 
no dynamical tendency for polarization in such dynamics. In fact, it is just the
opposite with dynamics inhibiting any possibility of polarized samples. 
Indeed, the absolute PLP is obviously zero in this situation for all dynamics 
with non--singular marginal distributions. These dynamics are thus perfectly
antipolarized and characterized by $\cop_A=-1$ as appropriate.

\subsection{Singular Dynamics} 

At this point we wish to briefly comment on including in the formalism of
absolute $\Xg$--distributions the cases involving dynamics $\db$ 
for which the associated polarization dynamics $S_r$ is not an element 
of $\dopset^f$. Let us recall that there were no restrictions on the notion of 
standard $\Xg$--distribution $S(\Xg)$ which can be uniquely assigned to arbitrary 
$S_r \in \dopset$ and $\Fg \in \Fset$. However, it was natural to focus on 
the case of bijective maps (elements of $\dopset^f$, $\Fset^f$), especially 
as we transitioned to the notion of relative $\Xg$--distributions.
This is reflected in our formulas for $\hat{S}_R [\,S_{(1)}, S_{(2)}\,]$, such
as (\ref{eq:116}) and (\ref{eq:130}), by the presence of $S_{(2)}^{-1}$. In fact,
everything discussed in the context of relative $\Xg$--distributions applies 
to arbitrary $S_{(1)} \in \dopset$ and $S_{(2)} \in \dopset^f$. While this covers
most cases of practical interest, it does not afford the construction of absolute
$\Xg$--distributions for $\db$ with $S_r \not\in \dopset^f$ since in that case 
$S_r^u$ is typically not an element of $\dopset^f$ either.

The main point about the extension from $S_{(2)} \in \dopset^f$ to 
$S_{(2)} \in \dopset$ in the context of relative $\Xg$--distributions is that 
it may not simultaneously retain the features coexisting in the former case. 
In particular, with faithful polarization comparators $S_{(2)} \in \dopset^f$ 
we have the interpretation of $\hat{S}_R [\,S_{(1)}, S_{(2)}\,]$ both in terms 
of standard $\Xg$--distribution with particular polarization 
function, and in terms of statistical comparisons between the two dynamics, 
as described in constructive definition of Sec.~\ref{subsec:rel_measures}. 
In fact, one can easily see that this dual view of the concept extends to 
the case when $S_{(2)} \in \dopset^c$. Indeed, the object 
$S_\star^{-1} \circ S_{(2)}$ needed in the definition
$\hat{S}_R [\,S_{(1)}, S_{(2)}\,] \equiv 
\hat{S} [\, S_{(1)}, S_\star^{-1} \circ S_{(2)} \,]$ is a valid polarization
function in this case, as discussed in Sec.~\ref{subsubsec:correspondence}. 
At the same time, this definition is equivalent to the comparative one 
in unchanged form.

However, the situation is different when $S_{(2)}$ is discontinuous and thus
represents dynamics with probabilistic atoms. The extensions in that case are
non--unique in both of the two approaches that become non--equivalent in 
general. If one follows the rationale of polarization functions, then it is 
natural to simply extend the $S_\star^{-1}$ operation using Eq.~(\ref{eq:94.50})
and thus define $\hat{S}_R$ via canonical polarization dynamics -- polarization
function bijection of Sec.~\ref{subsubsec:correspondence}. When utilizing  
this definition, then absolute $\Xg$--distribution assigned to dynamics that
one generally thinks of as maximally polarized, i.e.\ those with 
$\xd_{r,p}(\rpc) \equiv \half\delta(\rpc+1) + \half\delta(\rpc-1)$,  
is $\xd_A^p(\Xg)=\xd_{r,p}(\Xg)$ and the correlation coefficient is $\cop_A=1$. 
However, if one prefers to focus on the concept of statistical comparisons, 
then one is led to an alternative path which we now briefly describe.

The root cause of ambiguity in the comparative definitions of 
Sec.~\ref{subsec:rel_measures} is that for distributions with atoms,
comparisons using ``$>$'' and ``$\ge$'' can become non--equivalent, and one
has to fix the treatment of the ``$=$'' case. To see the associated issues,
let us again consider the polarization dynamics $\xd_{r,p}$, but 
assume in addition that the underlying dynamics $\db$ has a marginal distribution
$p(q)=\half \delta(q) + \half \delta(q-1)$. This implies that its uncorrelated
polarization dynamics is 
$\xd_{r,p}^u(\rpc) \equiv \quarter\delta(\rpc+1) + 
 \half\delta(\rpc) + \quarter\delta(\rpc-1)$. Adopting the comparative scheme
with sharp inequalities (as we did in Sec.~\ref{subsec:rel_measures}), one 
immediately obtains that $\hat{\plp}_R[\xd_{r,p},\xd_{r,p}^u]=1/2$ and 
$\hat{\plp}_R[\xd_{r,p}^u,\xd_{r,p}]=0$. On the other hand, admitting 
the equality always as a positive outcome, we obtain 
$\hat{\plp}_R[\xd_{r,p},\xd_{r,p}^u]=1$ and 
$\hat{\plp}_R[\xd_{r,p}^u,\xd_{r,p}]=1/2$ instead. Note that in both cases
the equation (\ref{eq:124}) reflecting conservation of probability is not 
satisfied. In the former case, some probability is ``lost'' by ignoring 
the equal outcomes, while in the latter case it is ``created'' via 
double counting. The natural approach is to insist that Eq.~(\ref{eq:124})
be satisfied and that the two dynamics being compared are treated on equal
footing. This forces us to assign half of the equal outcomes in favor of first 
dynamics and the other half to the second dynamics. That way we obtain 
$\hat{\plp}_R[\xd_{r,p},\xd_{r,p}^u]=3/4$ and
$\hat{\plp}_R[\xd_{r,p}^u,\xd_{r,p}]=1/4$, implying that $\cop_A=1/2$ in this
case. We emphasize that while we are focusing on a specific example, the above 
rule uniquely fixes the assignment of PLP and the correlation coefficient for 
all possible dynamics.

\begin{figure}
\begin{center}
    \centerline{
    \hskip 0.00in
    \includegraphics[width=8.3truecm,angle=0]{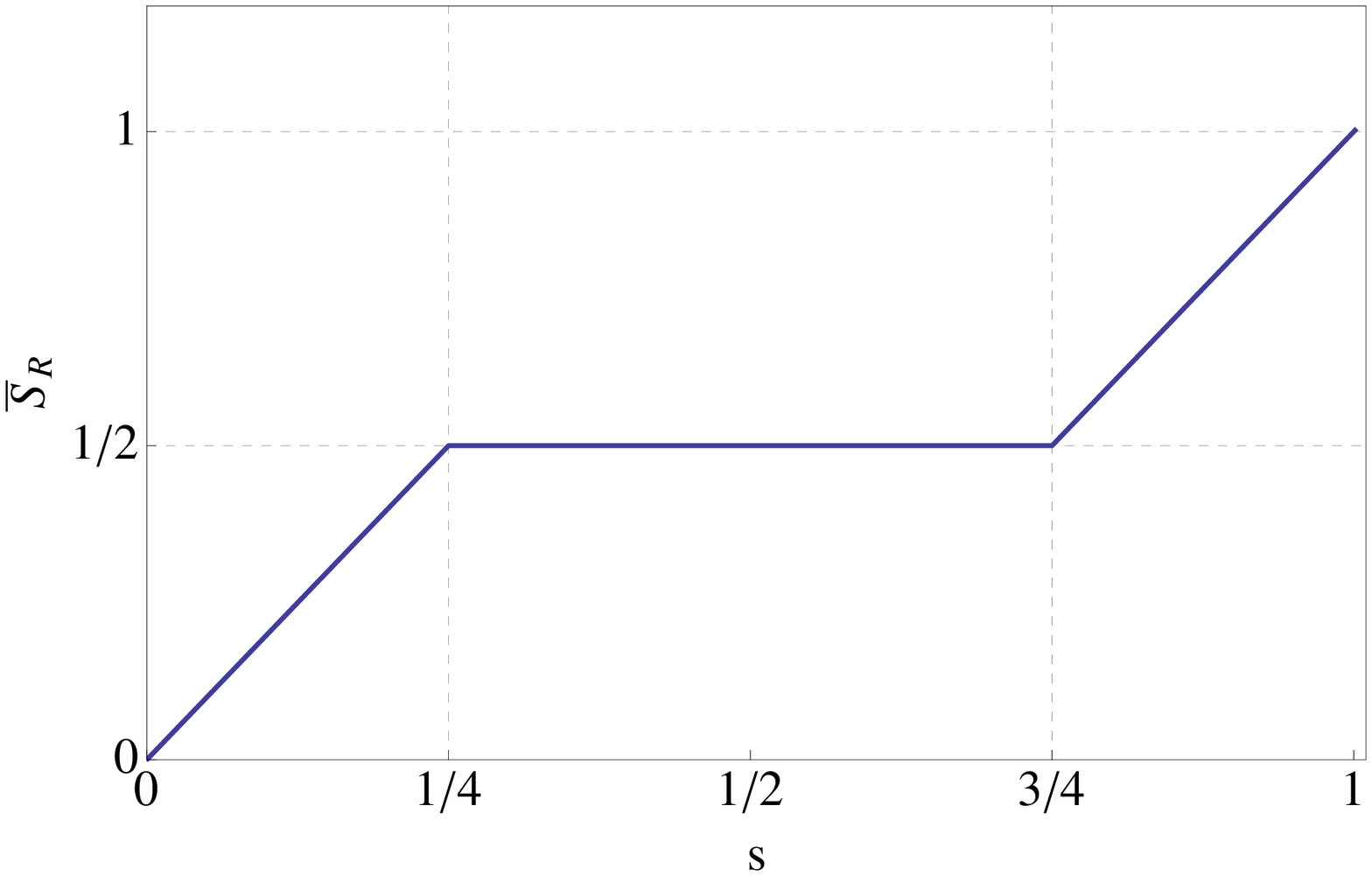}
    \hskip -0.0in
    \includegraphics[width=8.3truecm,angle=0]{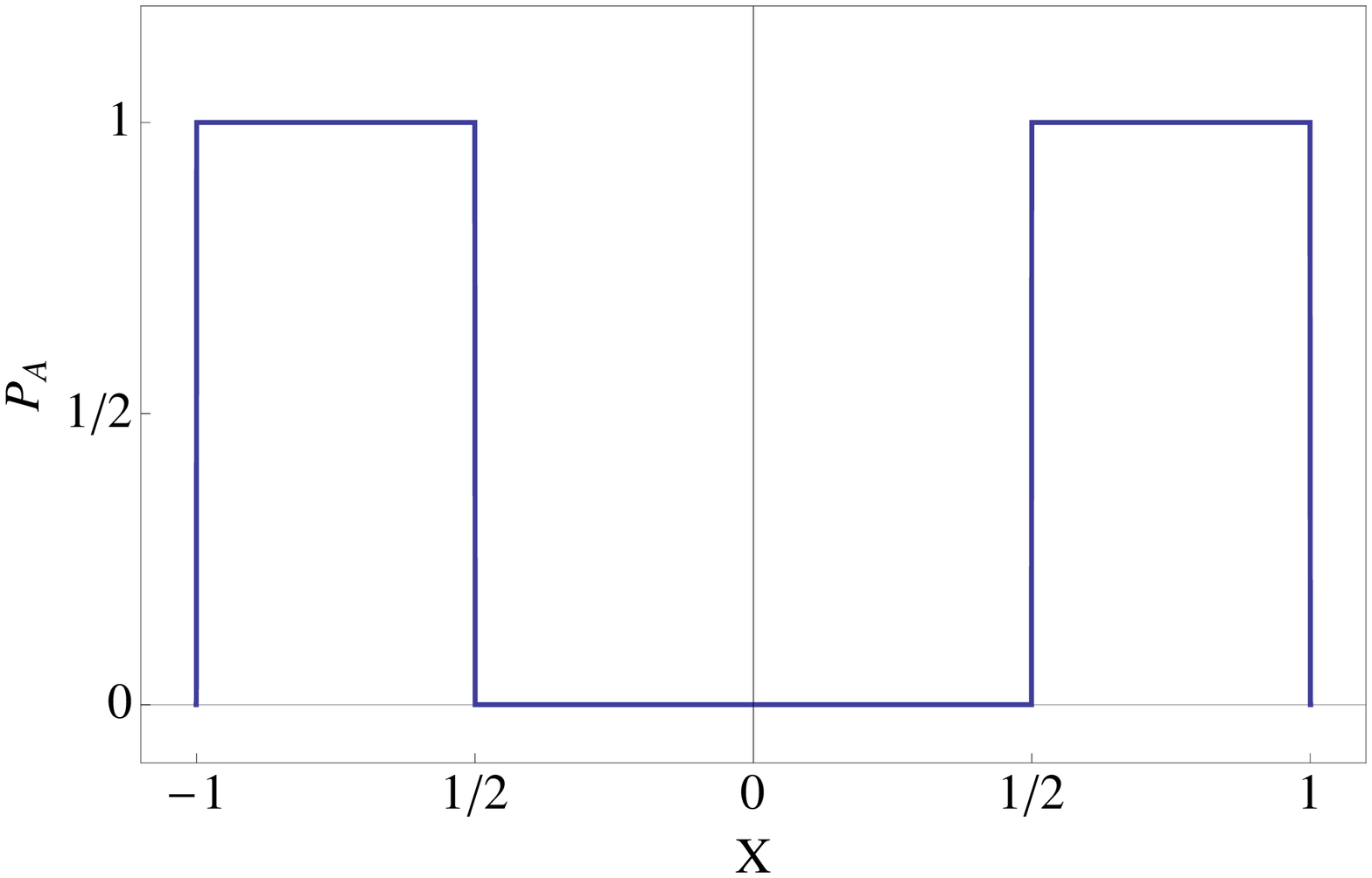}     
     }
     \vskip -0.00in
     \caption{Construction of absolute $\Xg$--distribution for singular
     dynamics characterized by $\xd_{r,p}(\rpc)$ (see text) in the context
     of statistical comparisons. The intermediate step, $\bar{S}_R(s)$, is 
     shown on the left while $P_A(\Xg)$ is on the right.}
     \vskip -0.2in 
     \label{fig:strict}
\end{center}
\end{figure}

The above approach can be consistently extended to the level of 
$\Xg$--distributions as well. As one can easily check, the construction of
Sec.~\ref{subsec:rel_measures} for relative $\Xg$--distributions only becomes 
ambiguous when both dynamics have an atom at the same point $\rpc_0$.
The proper prescription is that the segment of cumulative probability function 
associated with such points be linear. Illustrating this again on the case of 
$\xd_{r,p}$ and $\xd_{r,p}^u$, we show in Fig.~\ref{fig:strict} the construction 
of $\bar{S}_R$ for $\xd_{r,p}$ relative to $\xd_{r,p}^u$, and the associated 
absolute $\Xg$--distribution $P_A$. Note that 
$\int d\Xg \, P_A(\Xg) \, |\Xg| = 3/4 \equiv \plp_A$ consistently with the above 
prescription for PLP.

\subsection{More on the Invariance of Absolute Measures} 

Apart from reparametrization invariance, absolute measures possess additional 
symmetry that we would like to mention. To do that, it is instructive to think
of reparametrization invariance in the following way. Imagine that the generator
based on the dynamics $\db(q_1,q_2)$ produces the sequence of samples
$\{(q_1, q_2)_i\}$, 
distributed accordingly. The operator of the device broadcasts the sequence 
to various distant receivers, each of which has a polarization
meter at his disposal. When fed a sample $(q_1, q_2)_i$,  the polarization 
meter outputs its degree of polarization $\Xg_i$, thus allowing each receiver 
to analyze the polarization properties of $\db$. 
While the operation of each existing polarization meter is based on a valid 
polarization function $\Fg \in \Fset^f$, there is no standard choice adopted 
by the manufacturers. Thus, each receiver obtains a different sequence $\{\Xg_i\}$ 
to work on in general. This would seem to preclude the possibility of producing 
equivalent polarization characteristics across the spectrum of receivers. 

However, every receiver is required to follow special instructions,
identical for all, on how to use his polarization meter to determine 
the {\em dynamical} polarization properties of the transmitted dynamics. According
to this recipe, in addition to running $\{(q_1, q_2)_i\}$ through the polarization 
meter and producing $\{\Xg_i\}$, he is also required to form an independent
sequence~\footnote{What we mean by ``independent'' in this case is independent with 
respect to the ordering of the elements.}
$\{(q_1, q_2)_i^u\}$, in which the received components 
are assigned to one another randomly rather than as dictated by the dynamics. 
The receiver is then to run this randomized sequence through his polarization 
meter to obtain the associated $\{\Xg_i^u\}$. Finally, he is instructed to
perform the statistics of comparisons between $\Xg_i$ and $\Xg_i^u$ to yield 
the desired dynamical measure of polarization (absolute $\Xg$--distribution or 
correlation coefficient). Reparametrization invariance expresses the fact that, 
assuming they follow the above recipe, all receivers will end up with exactly 
the same dynamical polarization characteristic. 

Thus, by virtue of providing the above instructions, the sender doesn't have to worry
about the fact that receivers use various kinds of polarization meters. The additional
symmetry of absolute measures pertains to the fact that, to some extent, the same is also
true in reverse. In particular, by virtue of the same prescription, the receivers are 
protected from the fact that the sender could have used various kinds of generators. 
Indeed, recall
that the purpose of the formalism that we developed was to extract from the full dynamics
of quantity $Q$ the information on polarization. In doing that, we postulated that 
the information on dynamics of magnitudes $q_1$, $q_2$ of the projections, 
i.e.\ $\db(q_1,q_2)$, is sufficient for this since the relevant measures are based
on the ratios of these magnitudes. However, we could have used the distribution
of $(q_1^\alpha, q_2^\alpha)$ ($\alpha > 0$) associated with the same dynamics of $Q$ for 
this purpose. Thus, different senders communicating information on the same dynamics
could have used different kinds of generators and sending the sequence of samples
$\{(q_1^\alpha, q_2^\alpha)_i\}$, instead of $\{(q_1, q_2)_i\}$. As one can easily
see though, the above recipe for constructing absolute measures guarantees that 
the result is insensitive to this 
freedom because the statistics of required comparisons will not change. Formally,
these measures are invariant under 
\begin{equation}
      \db(q_1,q_2) \;\longrightarrow \; \pr_b^{(\alpha)} \,\equiv \,
      \frac{1}{\alpha^2} \, q_1^{\frac{1}{\alpha}-1} q_2^{\frac{1}{\alpha}-1} \,
      \db(q_1^{\frac{1}{\alpha}}, q_2^{\frac{1}{\alpha}})
      \label{eq:220}
\end{equation}
It is due to simultaneous validity of both types of above invariances that we view 
the constructs discussed here as universal polarization characteristics truly 
characterizing the dynamics itself, rather than representing our choices for description 
of polarization.

\section{Elements of Implementation}
\label{app:implement}

In this Appendix we will discuss certain numerical aspects of the results 
presented in Sec.~\ref{sec:qcd}. While the implementation described below was applied 
specifically to the case of overlap Dirac eigenmodes, one can of course use these 
strategies in more general settings. We will pay most attention to the computation 
of absolute $\Xg$--distributions, but will also discuss the calculation of correlation 
coefficient $C_{A}$ and the procedure for the determination of chiral polarization scale 
$\Lambda_T$.

Before we start, it is useful to recall that the input for our calculations is certain
dynamics $\db(q_{1},q_{2})$, but we only have a restricted statistical knowledge of it 
from relevant numerical simulation. In other words, what we have is a finite population 
of pairs $\{\, (q_{1}^{(i)},q_{2}^{(i)}) \;|\; i=1,\ldots,N \,\}$, whose statistics 
is governed by $\db(q_{1},q_{2})$. By virtue of this finite population, the latter can 
be thought of as being ``approximated'' by a discrete distribution
\begin{eqnarray}
  \nonumber
  {\cal P}_{b}(q_{1},q_{2}) & \longrightarrow & \tilde{\cal P}_{b}(q_{1}, q_{2}) \equiv 
  \frac{1}{N}\sum_{i=1}^{N} \delta(q_{1}-q_{1}^{(i)}) \, \delta(q_{2}-q_{2}^{(i)}), \\
  {\cal P}_{b}^{u}(q_{1},q_{2}) & \longrightarrow & \tilde{\cal P}_{b}^{u}(q_{1}, q_{2})
  \equiv \frac{1}{N} \sum_{i=1}^{N} \delta(q_{1}-q_{1}^{(i)}) \, \frac{1}{N} 
                   \sum_{j=1}^{N}\delta(q_{2}-q_{2}^{(j)})
  \label{eq:a20}
\end{eqnarray}
where the second equation relates to the ``uncorrelated'' counterpart 
$\db^{u}(q_{1},q_{2})$ of $\db(q_{1},q_{2})$. In case of Dirac eigenmodes the population 
comes from local values in the selected group of modes, such as those at scale $\Lambda$ 
in given gauge ensemble, with pairs representing the magnitudes of right and left chiral 
components. 

Our implementation followed the primer calculation described in Sec.~\ref{sec:primer}.
Technical details not specified there have to do with the finiteness 
of the population representing $\db(q_1,q_2)$ and the fact that, while one wishes to have
its size $N$ as large as possible in order to represent the dynamics adequately, 
the computational complexity involved in handling the complete information as expressed
in Eq.~(\ref{eq:a20}) grows as $N^2$ in case of uncorrelated distribution. 
Since $N$ can be quite large even with relatively modest size of gauge ensembles,
i.e.\ we have $N\approx 6\times 10^{7}$ for ensemble $E_{4}$ with only two eigenmodes 
per configuration contributing, it is computationally advantageous to do some 
coarse--graining as explained below. 
\medskip

{\bf Absolute $\Xg$--distribution.} Given our statistics, we have decided to evaluate 
$P_A(\Xg)$ at $N_b=50$ equidistant values based on the information stored in the associated
$N_b=50$ bins of probability. This allows us to plot the distribution with enough resolution 
to capture relevant details while it also allows for enough statistics in each bin 
so that the statistical errors are reasonably small. In order to reduce the computational 
effort, we organized the calculation as follows:

\medskip

\begin{itemize}
\item[1.] Compute the reference polarization coordinates $x_{1},...,x_{N_{b}-1}$ representing 
the boundaries of each bin. To avoid the numerical evaluation of inverse tangent and using the
fact that our construction doesn't depend on the parametrization of the polar angle, we actually
work directly with slopes $t_{1},...,t_{N_{b}-1}$ ($t \equiv q_2/q_1$). These slopes separate 
the sample space quadrant into segments containing equal number of samples drawn from 
the uncorrelated distribution ${\cal P}_{b}^{u}$. In Fig.~\ref{fig:dyn_E1} we indicated with 
solid gray lines these boundaries when the quadrant is partitioned into $10$ such bins.

\item[2.] Count how many samples $(q_{1}^{(i)},q_{2}^{(i)})$ drawn from the correlated 
distribution ${\cal P}_{b}$ fall into each bin by comparing their slope with the boundaries 
determined in the previous step.
\end{itemize}

Since we only need to inspect each sample once to determine which bin it belongs to,
the second step has a complexity $O(N)$, which is negligible compared to the first step
which is $O(N^{2})$ in the naive implementation. One possibility to deal with the first step 
is to use a randomly drawn sub--population from the corresponding $N^{2}$ samples 
to approximate the uncorrelated distribution. The problem with this approach is that it 
interferes with the error analysis. 
However, one can organize the calculation carefully and compute the first step using a method 
that has a complexity only $O(N \log N)$.

The key to this is the ability to compute $S_{r}^{u}(t)$ efficiently. Indeed, the problem of
finding the boundary is equivalent to solving the equation $S_{r}^{u}(t_{i})=i/N_{b}$, and 
one can solve this equation using a bisection method which can determine $t_{i}$ to one part 
in a million with only $20$ evaluations of $S_{r}^{u}(t)$.
To compute $S_{r}^{u}(t)$ we sort $\{\, q_{2}^{(j)} \,\}$ in ascending order, a task of 
complexity $O(N\log N)$. For every $q_{1}^{(i)}$ we then compute the fraction $f_{i}$ of 
elements from $\{\, q_{2}^{(j)} \,\}$ smaller than $ t \, q_{1}^{(i)}$. 
Since $\{\, q_{2}^{(j)} \,\}$ is sorted, one can use a binary search which has a $O(\log N)$ 
complexity. Computing the fraction $f_{i}$ for all $N$ values of $q_{1}^{(i)}$
to get the answer $S_{r}^{u}(x) = \frac{1}{N} \sum_{i} f_{i}$, the total complexity for this 
task is $O(N\log N)$ as advertised.

While the above strategy makes the calculation of $\Xg$--distributions manageable, it can 
still take a considerable amount of time. To speed it up, we sort the values 
$q_{1}^{(i)}$ and separate them into bins with equal number of samples. Each bin is then 
replaced by a single representative value equal to the average of the samples it contains. 
For the smallest ensemble $E_{1}$ where one can comfortably carry out the calculation 
without binning, we compared the results of the binned version using 1000 elements per bin 
with the exact calculation. We found that the differences are much smaller than the error 
bars. For larger ensembles, the approximation is expected to be even better since 
the number of samples is increased. The binned version was thus used in our evaluations.

The result of the above calculation determines the fraction of samples drawn from 
${\cal P}_{b}$ belonging to each of the $N_b$ bins. This information allows us to plot 
the distribution (histogram) of $dS_{r}/dS_{r}^{u}$ as a function of $S_{r}^{u}$, which 
can then easily be scaled into the absolute $\Xg$--distribution $\xd_{A}(\Xg)$. 
To estimate the errors for each value of argument $\Xg$ (or bin), we used single 
elimination jackknife method with new samples generated by removing one configuration 
from the gauge ensemble.
\medskip

{\bf The Correlation Coefficient.} The evaluation of correlation coefficient $\cop_A$ is 
equivalent to the calculation of $\plp_A$, namely the probability of larger polarization. 
While $\plp_{A}$ can be evaluated independently of the absolute $\Xg$--distribution,
we follow its definition as a moment of $\xd_{A}(\Xg)$ since this streamlines the error 
analysis. Every jackknife sample $(\xd_A)^j$ in the process of calculating the absolute 
$\Xg$--distribution produces the jackknife sample $\plp_A^j$. In particular, if 
$(\xd_A)^j_i$ is the value of the estimate for bin $i$, then we have explicitly
\begin{equation}
   \plp_{A}^j \,=\, \sum_{i=1}^{N_{b}} \frac{|\Xg_{i}+\Xg_{i-1}|}{2} \, (P_{A})^j_{i}
   \label{eq:a30}  
\end{equation}
where $X_{i}$ are the boundaries of the bins that divide the domain $[-1,1]$ into
$N_b$ equal size bins, and thus $\Xg_{0}=-1$ and $\Xg_{N_{b}}=1$. 
The jackknife analysis of the above estimates then determines the mean value of 
the correlation coefficient and its error.
\medskip

{\bf Chiral Polarization Scale.} One of the main results in this work is that 
the correlation coefficient $C_{A}$ changes sign at a specific scale $\Lambda_T$
of the eigenmodes. If we denote by $C_{A}(\Lambda)$, the correlation coefficient 
evaluated using a pair of eigenmodes $\lambda_{1,2}$ ``bracketing'' the scale 
$\Lambda$, i.e.\ $\lambda_{1} \le \Lambda < \lambda_{2}$, then the chiral polarization
transition point is defined via $C_{A}(\Lambda_{T})=0$. 
To determine this scale, we first crudely map out the behavior of $C_{A}(\Lambda)$ 
by evaluating the correlation coefficient on a grid with steps of about $200$ MeV.
Note that, using the spectral information from Table~\ref{tab:ensembles}, we choose 
the overall position of these grid points such that all gauge configurations 
in the ensemble contribute modes for the calculation. Determining the transition 
region this way, we then evaluate $C_{A}(\Lambda)$ at few more points in 
the estimated vicinity of the transition point. Using this information, several 
values $\Lambda_{i}$ closest to the transition are finally selected.
The correlation coefficient is then evaluated for each of the jackknife samples 
and the estimate of the correlation matrix for $C_{A}(\Lambda_{i})$, using a jackknife 
method, is formed. The final result for $\Lambda_{T}$ and its error are evaluated 
using a linear regression based on this correlation matrix.
\medskip

{\bf Calculation of Eigenmodes.} We determine the eigenvalues of the zero-mass overlap 
operator using our own implementation of the implicitly restarted Arnoldi algorithm 
with deflation~\cite{ira}. To speedup the calculation and to reduce the amount of
memory required, we compute the eigenvalues of $D^{\dagger}D = D^{\dagger}+D$ which
is Hermitian and commutes with $\gamma_{5}$. This allows us to project on a chiral 
subspace and compute only one member of the chiral pair. The eigenvectors 
and eigenvalues of the overlap operator are then straightforwardly constructed using 
the basic properties of the overlap Dirac matrix.

\end{appendix}

\end{document}
\bye